\documentclass[11pt,a4paper]{article}
\usepackage{color}
\usepackage{amsmath}
\usepackage{amssymb}
\usepackage{esint}

\usepackage{jcappub}

\usepackage{graphicx}
\usepackage{caption}
\usepackage{subcaption}
\usepackage{slashed}
\usepackage{amsfonts}
\usepackage{enumerate}
\usepackage{color}
\usepackage{bm}

\usepackage{marginnote}
\newcommand{\mk}[1]{#1}

\newcommand{\mc}[1]{\mathcal{#1}}

\title{Slow-roll corrections in multi-field inflation: a separate universes approach}

\author[a]{Mindaugas Kar\v{c}iauskas,}
\emailAdd{mindaugas.m.karciauskas@jyu.fi}

\author[b,c,d]{Kazunori Kohri,}
\emailAdd{kohri@post.kek.jp}

\author[b,c]{Taro Mori}
\emailAdd{moritaro@post.kek.jp}

\author[b]{\\ and Jonathan White}
\emailAdd{jwhite@post.kek.jp}

\affiliation[a]{University of Jyvaskyla, Department of Physics, P.O.Box 35 (YFL), FI-40014\\ University of Jyv\"{a}skyl\"{a}, Finland}
\affiliation[b]{Theory Center, KEK, Tsukuba 305-0801, Japan}
\affiliation[c]{The Graduate University for Advanced Studies (Sokendai)\\Department of Particle and Nuclear Physics, Tsukuba 305-0801, Japan}
\affiliation[d]{Rudolf Peierls Centre for Theoretical Physics, The University of
Oxford, 1 Keble Road, Oxford OX1 3NP, UK}

\abstract{In view of cosmological parameters being measured to ever higher precision, theoretical predictions must also be computed to an equally high level of precision. In this work we investigate the impact on such predictions of relaxing some of the simplifying assumptions often used in these computations. In particular, we investigate the importance of slow-roll corrections in the computation of multi-field inflation observables, such as the amplitude of the scalar spectrum $P_\zeta$, its spectral tilt $n_s$, the tensor-to-scalar ratio $r$ and the non-Gaussianity parameter $f_{NL}$. To this end we use the separate universes approach and $\delta N$ formalism, which allows us to consider slow-roll corrections to the non-Gaussianity of the primordial curvature perturbation as well as corrections to its two-point statistics. In the context of the $\delta N$ expansion, we divide slow-roll corrections into two categories: those associated with calculating the correlation functions of the field perturbations on the initial flat hypersurface and those associated with determining the derivatives of the e-folding number with respect to the field values on the initial flat hypersurface.  Using the results of Nakamura \& Stewart '96, corrections of the first kind can be written in a compact form.  Corrections of the second kind arise from using different levels of slow-roll approximation in solving for the super-horizon evolution, which in turn corresponds to using different levels of slow-roll approximation in the background equations of motion.  We consider four different levels of approximation and apply the results to a few example models.  The various approximations are also compared to exact numerical solutions.}

\begin{document}
 \global\long\def\idx#1{\textsc{{\tiny\ensuremath{\left(#1\right)}}}}

 \global\long\def\nc{N_{\mathrm{c}}}

 \global\long\def\rc{\rho_{\mathrm{c}}}

 \global\long\def\i{\mathrm{i}}

 \global\long\def\f{\mathrm{f}}

 \global\long\def\s{\mathrm{S}}

 \global\long\def\l{\mathrm{L}}


\maketitle

\section{Introduction and outline}

A period of inflation in the early Universe is now widely accepted as
being responsible for generating the primordial density perturbations
that seed all large-scale structure (LSS) in the Universe as well as the
temperature fluctuations observed in the cosmic microwave background
(CMB). Constraints on the amplitude and scale dependence of the
primordial spectrum are already very precise, 
at 68\% confidence level they are 
given as \cite{Ade:2015lrj}\footnote{See
Section~\ref{Sec:corr} for definitions of $A_s$ and $n_s$.}
\begin{gather}\label{obs1}
\ln(10^{10}A_{s})=3.062\pm0.029 \qquad \mbox{and}\qquad n_{s}=0.968\pm0.006.
\end{gather}
Another key property of the spectrum is its deviation from Gaussianity.
In the so-called squeezed limit this can be quantified in terms of the
parameter $f_{NL}^{\rm local}$, and current constraints are still
consistent with this being zero \cite{Ade:2015ava}:
\mk{
\begin{align}
f_{NL}^{{\rm local}}=0.8\pm5.0
\end{align}
}
at the $68\%$ confidence level. While this constraint is not yet as tight as those on $A_s$ and $n_s$,
future LSS observations are forecast to improve constraints on
non-Gaussianity to the level $\sigma (f_{NL}^{\rm local })\sim 0.1$ \cite{Takahashi:2015zqa,Yamauchi:2016wuc}.  

\mk{}In addition to density perturbations, inflation is also expected to give
rise to gravitational waves.  At present the properties of the
gravitational wave spectrum are less well constrained, with an upper
limit on the amplitude essentially being given as
\cite{Ade:2015lrj,Array:2015xqh,Aghanim:2016cps}
\mk{
\begin{equation}
r<0.1\,.
\end{equation}
}
at the $95\%$ confidence level. However, CMB polarization observations are expected to improve this
constraint to the level $\sigma (r)\sim 10^{-3}$ \mk{\cite{Errard:2015cxa}}.

The high-precision nature of current and forecast constraints
necessitates that theoretical predictions used to compare candidate
inflation models with the data must be equally as precise, and any
approximations made in deriving these predictions must therefore be
carefully scrutinised.  One such approximation often made is the
so-called slow-roll approximation, whereby the Hubble flow parameters
defined as
\begin{align}
\epsilon  \equiv-\frac{\dot{H}}{H^{2}}\qquad\mbox{and}\qquad
\eta  \equiv\frac{\dot{\epsilon}}{\epsilon H}\label{eps-eta-def}
\end{align}
are assumed to be small, namely $\epsilon,\eta\ll 1$.  These two
requirements are very generic: the first is necessary for
(quasi-)exponential inflation, while the second ensures that inflation
lasts for long enough.  In the case of single-field inflation, on making
an expansion in the two parameters $\epsilon$ and $\eta$ we find that
the leading-order expression for $n_s$ is given as
 \begin{equation}\label{sftilt}
  n_s -1= -2\epsilon_k -\eta_k,
 \end{equation}
 where the subscript $k$ denotes that the quantities should be evaluated
 at the time at which the scale under consideration, $k$, left the
 horizon during inflation.  Barring a cancellation between the two terms
 on the right hand side of \eqref{sftilt}, from the observed value of
 $n_s$ given in \eqref{obs1} we conclude that $\epsilon,\eta\lesssim
 \mathcal O(10^{-2})$.  As such, in the case of single-field inflation
 we see that any slow-roll corrections of order $\epsilon$ or $\eta$ will
 likely represent percent-level corrections, and should thus be included
 if we desire percent-level precision in our theoretical predictions.

While current observations are perfectly consistent with single-field
inflation, in the context of high-energy unifying theories one naively
expects the presence of many fields.  It is thus important that we are
able to precisely determine the predictions of multi-field inflation
models in order that they can be constrained using current and future
data.  Perhaps of particular interest in the context of multi-field
inflation is the possibility of obtaining an observably large
non-Gaussianity.  In the case of single-field inflation it is known that
the three-point function in the squeezed limit is unobservably small
\cite{Maldacena:2002vr}.  This suggests that if a non-negligible
three-point function were to be observed then this would be a strong
\mk{indication of} 
the presence of multiple fields during inflation.  Even if
not observed, however, it is still important to determine what the
implications of this are for multi-field models of inflation.  Indeed,
while it is possible to obtain a large non-Gaussianity in such models,
it is by no means a generic prediction.  When considering slow-roll
corrections in multi-field inflation the situation is somewhat more
complicated than in the single-field case.  In addition to assuming
$\epsilon,\eta\ll 1$, the slow-roll approximation is often taken to
imply the smallness of a larger set of expansion parameters that we
label $\eta_{IJ}$, with $I,J=1,..., M$ and $M$ being the number of
fields (see Section~\ref{Sec:background} for details).  One then finds
that observational constraints on $n_s$ do not require all components of
$\eta_{IJ}$ to be $\lesssim \mathcal O(10^{-2})$, such that
slow-roll corrections of order $\eta_{IJ}$ may represent
larger-than-percent-level corrections.  If this is the case, the
inclusion of next-to-leading order corrections will be crucial, and it
is plausible that next-to-next-to-leading order corrections may also
become important.  Indeed, slow-roll corrections to the two-point
statistics of scalar and tensor perturbations in multi-field inflation
models have already been considered in the literature
\cite{Byrnes:2006fr,Avgoustidis:2011em},
and in some cases corrections were shown to be on the order of $20\%$.

In this paper we revisit the issue of slow-roll corrections in
multi-field inflation.  In particular, we are interested in the
slow-roll corrections to predictions for the curvature perturbation on
constant density slices, $\zeta$, and quantities related to its
correlation functions.  Our analysis makes use of the separate \mk{universes}
approach and $\delta N$ formalism, which allows us to consider slow-roll
corrections to the non-Gaussianity of $\zeta$ as well as corrections to
its two-point statistics.  As discussed above, precise predictions for
the non-Gaussianity will be key in constraining multi-field models of
inflation with current and future data.  The paper is structured as
follows.  In Section \ref{Sec:background} we introduce the class of
models under consideration and present the relevant background field
equations.  The slow-roll approximation and associated expansion
parameters are also introduced, and the background equations of motion
are expanded up to next-to-leading order in the slow-roll expansion,
with details of the derivation being given in Appendix
\ref{sec:SR-aprx}. In Section~\ref{sec:sepuniv} we review the separate
universes approach and $\delta N$ formalism on which our analysis is
based.  Rather than using a finite-difference method, as is often used,
our method is based on the transport techniques introduced and developed
in
\cite{Yokoyama:2007uu,Yokoyama:2007dw,Mulryne:2009kh,Mulryne:2010rp,Anderson:2012em,Seery:2012vj,Mulryne:2013uka}.
In this section we outline how these techniques can be applied under
various levels of slow-roll approximation.  In Section~\ref{Sec:corr} we
turn to the correlation functions of $\zeta$ as given in terms of the
correlation functions of field perturbations on an initial flat slice at
horizon crossing.  We find that slow-roll corrections arising from
slow-roll corrections to the initial field perturbations can be
expressed in a compact form and in terms of quantities that in principle
are observable.  In Section~\ref{Sec:examples} we apply the results of
the earlier sections to a few example models, focussing on a comparison
between the various levels of slow-roll approximation.  We finally
summarise our results in \mk{Section~\ref{Sec:conclusions}}.  In this work we
use the geometrized units $c=\hbar=M_{\mathrm{Pl}}=1$, where
$M_{\mathrm{Pl}}=\left(\hbar c/8\pi G\right)^{-1/2}$ is the reduced
Planck mass.

\section{Background dynamics and the slow-roll approximation}
\label{Sec:background}

We consider the class of multi-field models of inflation with an action
\begin{equation}
S=\int d^{4}x\sqrt{-g}\left(\frac{R}{2}-\frac{1}{2}\delta_{IJ}g^{\mu\nu}\partial_{\mu}\phi^{I}\partial_{\nu}\phi^{J}-V(\vec{\phi})\right),\label{S-def}
\end{equation}
where $R$ is the Ricci scalar associated with the spacetime metric
$g_{\mu\nu}$, $g$ is the determinant of $g_{\mu\nu}$,
$\phi^{I},\,I=1,..., M$ are $M$ scalar fields and we have used the
notation $\vec{\phi}=\left(\phi^{1},\ldots,\phi^{M}\right)$.
$V(\vec{\phi})$ in the above action is some $\phi^{I}$-dependent
potential. Note that we take the metric in field space to be Euclidean,
so the distinction between upper and lower indices as in $\phi^{I}$ and
$\phi_{I}$ is immaterial. We will use both of the notations
interchangeably and summation is implied over repeated indices, unless
stated otherwise.

At background level we take the metric $g_{\mu\nu}$ to be of the
Friedmann-Lemaitre-Robertson-Walker (FLRW) type, i.e. $g_{\mu\nu}={\rm
diag}\left(-1,a^{2},a^{2},a^{2}\right)$.  The dynamics is then fully
determined by the equations of motion for the $M$ scalar fields and the
Friedmann equation, which are given as
\begin{equation}
\ddot{\phi}_{I}+3H\dot{\phi}_{I}+V_{,I}=0,\label{EoMt}
\end{equation}
\begin{equation}
3H^{2}=\rho = \frac{1}{2}\dot\phi_I\dot\phi_I + V,\label{Frdmt}
\end{equation}
where $H = \dot a/a$, $\rho$ is the total energy density of the scalar
fields and we have used the notation $V_{,I}\equiv\partial
V/\partial\phi_{I}$.  Taking the time derivative of \eqref{Frdmt} and
using \eqref{EoMt} we are able to find an expression for $\epsilon$ as
defined in \eqref{eps-eta-def}, which can in turn be used to determine
$\eta$.  The resulting expressions are given as
\begin{equation}
\epsilon = \frac{1}{2}\frac{\dot\phi_I\dot\phi_I}{H^2},\qquad \eta = 2\frac{\ddot\phi_I\dot\phi_I}{H\dot\phi_J\dot\phi_J} +2\epsilon.  \label{eps-eta-dphidt}
\end{equation}

For the purpose of this work it is convenient to use the e-folding
number $N$, rather than $t$, as the time parameter, with $N$ being
defined as
\begin{equation}
N\equiv\ln\left(\frac{a_{\f}}{a_{\i}}\right)=\intop_{t_{\i}}^{t_{\f}}H\mathrm{d}t,\label{N-def}
\end{equation}
where the indices `$\i$' and `$\f$' denote initial and final values.  In
terms of $N$, eqs.~\eqref{EoMt}, \eqref{Frdmt} and
\eqref{eps-eta-dphidt} take the form
\begin{gather}
\phi_{I}''+\left(3-\epsilon\right)\left(\phi_{I}'+\frac{V_{,I}}{V}\right)=0,\label{EoMN}\\
 H^{2}=\frac{V}{3-\epsilon}=\frac{\rho}{3},\label{FrdmN}\\
 \epsilon = \frac{1}{2}\phi_{I}'\phi_{I}',\qquad \eta = 2\frac{\phi^{\prime\prime}_I\phi^\prime_I}{\phi^\prime_J\phi^\prime_J}\label{eps-eta-dphidN}
\end{gather}
where a prime denotes a derivative with respect to $N$,
e.g. $\phi_{I}'\equiv\mathrm{d}\phi_{I}/\mathrm{d}N$.

As discussed in the introduction, in the context of inflation one often
makes the so-called slow-roll approximation.  At the most fundamental
level this approximation corresponds to assuming $\epsilon,\eta\ll 1$,
which ensures that we obtain quasi-exponential inflation that lasts for
long enough.\footnote{While $\epsilon$ is positive definite, $\eta$ may
be negative, so strictly speaking we should perhaps require $|\eta|\ll
1$.  However, if $\eta$ is negative, meaning that $\epsilon$ is
decreasing with time, then the concern that inflation may not last long
enough is no longer an issue.  Instead, one will need to ensure that
there is a suitable mechanism to end inflation.}  In both the single-
and multi-field cases the assumption $\epsilon\ll 1$ allows us to
approximate the Friedmann equation \eqref{FrdmN} as
\begin{equation}
 H^{2}\simeq \frac{V}{3}\equiv\frac{\rho^{(0)}}{3}\equiv \left.H^{(0)}\right.^2.\label{FrdmN-SR}
\end{equation}
In the single-field case, we see from \eqref{eps-eta-dphidt} and
\eqref{eps-eta-dphidN} that the requirement $\eta\ll 1$ is equivalent to
the requirement $\ddot\phi/(H\dot\phi)\ll 1$, or
$\phi^{\prime\prime}/\phi^\prime\ll 1$, which allows us to approximate
\eqref{EoMN} as
\begin{gather}
\phi'+\frac{V_{,\phi}}{V}=0.\label{SFEoM-SR}
\end{gather}
We thus find that the slow-roll approximation is equivalent to assuming
that an attractor regime has been reached, in which the field velocity
is no longer an independent degree of freedom but is given as a function
of the field value \cite{Liddle:1994dx}.  In the multi-field case,
however, the situation is more complicated, as the condition $\eta\ll 1$
only constrains the component of $\phi''_I$ that lies along the
background trajectory, i.e. it does not necessarily imply
$|\phi''_I/\phi'_I|\ll 1$ (or equivalently
$|\ddot\phi_I/(H\dot\phi_I)|\ll 1$) for all $I$ (note that there is no
summation over $I$ in these expressions).  Nevertheless, by the
Cauchy-Schwarz inequality, the constraint $\eta\ll 1$ will be satisfied
if we assume that the magnitude of the field acceleration vector is
smaller than the magnitude of the field velocity vector, namely $|\vec
\phi''|/|\vec \phi'|\ll 1$.  If we then further assume that $\phi'_I\sim
\mathcal O(|\vec\phi'|)$ for all $I$ -- that is, we assume the field
basis is not aligned in such a way that some of the field velocity
components vanish or are considerably smaller than $|\vec\phi'|$ -- then
we can take the slow-roll approximation to mean that
$\phi''_I/\phi'_I\ll 1$ for all $I$, such that \eqref{EoMN} can be
approximated as
\begin{gather}
\phi_I'+\frac{V_{,I}}{V}=0.\label{EoM00}
\end{gather} 
In this case, we again find ourselves in an attractor regime where all
field velocities are given as functions of the field positions.  For
later convenience we rewrite \eqref{EoM00} as
\begin{eqnarray}
\phi_{I}' & \simeq & U_{,I}^{\left(0\right)}, \qquad \mbox{where}\qquad U^{(0)} = -\ln V.\label{EoM0}
\end{eqnarray}
Using \eqref{EoM0} we are then able to derive consistency conditions for
the first and second derivatives of the potential.  Explicitly, the
conditions $\epsilon,\eta\ll 1$ become
\begin{gather}
 \epsilon \simeq  \frac{1}{2}\frac{V_{,I}V_{,I}}{V^2} \equiv \epsilon^{(0)} \ll 1,\label{pot_epssrparam1}\\
 \eta \simeq  2\frac{U^{(0)}_{,I}U^{(0)}_{,IJ}U^{(0)}_{,J}}{U^{(0)}_{,K}U^{(0)}_{,K}} =
 4\epsilon^{(0)} -\frac{1}{\epsilon^{(0)}}\frac{V_{,JK}V_{,J}V_{,K}}{V^3}
\equiv \eta^{(0)} \ll 1. 
\end{gather}
The more stringent constraint $|\vec \phi''|/|\vec \phi'|\ll 1$ reads
\begin{align}
 4\left(\epsilon^{(0)}\right)^2 -2 \frac{V_{,I}V_{,IJ}V_{,J}}{V^3} + \frac{V_{,I}V_{,IJ}V_{,JK}V_{,K}}{V^2 V_{,L}V_{,L}} \ll 1.\label{stringsrc}
\end{align}
Given that $\epsilon^{(0)}\ll 1$, the requirement $\eta\ll 1$ gives us
the condition $V_{,J}V_{,JK}V_{,K}/V^3 \ll \epsilon^{(0)}$.  This in
turn means that the second term in eq.~\eqref{stringsrc} is negligible,
such that the final constraint reduces to
$V_{,I}V_{,IJ}V_{,JK}V_{,K}/(V^2 V_{,L}V_{,L})\ll 1$.  Introducing the notation
\begin{equation}\label{etaIJdef}
 \eta_{IJ} = \frac{V_{,IJ}}{V},
\end{equation}
both of the
constraints on the second derivatives of $V$ can be satisfied if the
eigenvalues of $\eta_{IJ}$ are small.  This in turn will
be the case if we assume\footnote{For large $M$, i.e. a large number of
fields, the more stringent constraint $\eta_{IJ}\ll 1/M$ would be
appropriate.}
\begin{align}
 \eta_{IJ}\ll 1,\label{pot_epssrparam2}
\end{align}
for all $I$ and $J$.  The conditions $\epsilon^{(0)},\eta_{IJ}\ll 1$
thus constitute the slow-roll assumptions that we will make in this
paper.

While \eqref{EoM0} represents the lowest-order slow-roll equations of
motion, in this paper we will be interested in considering
next-to-leading order corrections, and we thus require the equations of
motion valid to next-to-leading order in the slow-roll approximation.
Details of how these can be calculated are given in
Appendix~\ref{sec:SR-aprx}, but the final result can be written in a
form almost identical to \eqref{EoM0} as
\begin{align}
\phi_{I}' & \simeq U_{,I}^{\left(1\right)},\label{EoM1}
\end{align}
with
\begin{equation}
U^{\left(1\right)}\equiv U^{\left(0\right)}-\frac{1}{6}U_{,I}^{\left(0\right)}U_{,I}^{\left(0\right)}.\label{U1-def}
\end{equation}
The Friedmann equation at next-to-leading order is given as
\begin{align}
 H^2 \simeq  \frac{V}{3-\epsilon^{(0)}}\equiv \frac{\rho^{(1)}}{3}\equiv \left.H^{(1)}\right.^2 \simeq \frac{V}{3}\left(1+\frac{\epsilon^{(0)}}{3}\right),\label{Frdm1}
\end{align}
and the next-to-leading order expressions for $\epsilon$ and $\eta$,
denoted $\epsilon^{(1)}$ and $\eta^{(1)}$ respectively, can be
determined by substituting \eqref{EoM1} into the full expressions
\eqref{eps-eta-dphidN}.

Before moving on, it will turn out to be convenient later on to
re-express the full equations of motion \eqref{EoMN} in a more compact
form that mirrors the form of eqs. \eqref{EoM0} and \eqref{EoM1}.  In
order to do so, we introduce the notation
\begin{equation}
\varphi_{i}\equiv\left(\phi_{I},\phi_{I}'\right),\label{varj-def}
\end{equation}
where the index $i$ runs from $1$ to $2M$.  This allows us to write
\eqref{EoMN} as
\begin{equation}
\varphi'_i=\mathbb{U}_{i},\label{EoM-j}
\end{equation}
where
\begin{align}
\mathbb{U}_{i}\equiv\phi_{i}' \qquad & \text{ for }  1\leq i\leq M\label{U-M-def}\\
\mathbb{U}_{i}\equiv-\left(3-\epsilon\right)\left(\phi_{i-M}'+\frac{V_{,i-M}}{V}\right)\qquad & \text{ for }  M+1\leq i\leq2M.\label{U-2M-def}
\end{align}

\section{The curvature perturbation $\zeta$\label{sec:sepuniv}}

In this section we outline the method we use to determine the curvature
perturbation on constant density slices, $\zeta$, which is based on the
separate universes approach and $\delta N$ formalism
\cite{Sasaki:1995aw,Sasaki:1998ug,Wands:2000dp,Lyth:2004gb,Sugiyama:2012tj,Starobinsky:1982ee,Starobinsky:1986fxa}.

\subsection{The separate universes approach and $\delta N$ formula}

 The separate universes approach corresponds to the lowest-order
approximation in a gradient expansion \cite{Lyth:2004gb,Sugiyama:2012tj}. First,
one performs a smoothing of the Universe on some scale $L$ that is
larger than the Hubble scale $1/H$, namely $LH\gg1$. Associating the
small parameter $\xi=1/(LH)$ with spatial gradients, one then expands
the relevant dynamical equations in powers of $\xi$, and to the lowest
order neglects terms of order $\mathcal{O}(\xi^{2})$ or
greater. Consequently, one finds that individual $L$-sized patches
behave as separate universes, evolving independently and according to
the background field equations.  On scales larger than $L$, the
differences between neighbouring patches simply result from differences
in initial conditions.  If we are interested in the comoving scale with
wavenumber $k$ then we have $\xi = k/(aH)$.  During inflation this
parameter will be decreasing exponentially with time, and the separate
universes approach will be applicable after the Horizon-crossing time,
which is defined as the time at which $k = aH$.

Ultimately we are interested in computing the curvature perturbation on constant density slices,
$\zeta$, and in the context of the separate universes approach it can be shown that $\zeta$ is given by the number of e-foldings of expansion between a flat and constant density slice, $\delta N$.  To see why this is, let us consider the spatial metric, which can be written as
\begin{equation}
g_{ij}\left(t,\boldsymbol{x}\right)=a^{2}\left(t\right)\mathrm{e}^{2\psi\left(t,\boldsymbol{x}\right)}\gamma_{ij}\left(\boldsymbol{x}\right),\label{gij}
\end{equation}
where $a(t)$ is the fiducial background scale factor, $\psi(t,\bm x)$ is the curvature perturbation and $\gamma_{ij}$, satisfying $\mathrm{det}\gamma_{ij}=1$, contains gravitational waves.  Here we follow \cite{Sugiyama:2012tj} and fix the threading such that the scalar and vector parts of $\gamma_{ij}$ are set to zero.  Note that the fiducial background universe can be associated with some given location $\bm x_0$, namely $g_{ij}(t,\bm x_0) = a^2(t)\delta_{ij}$.  We can then consider \mk{$\tilde a(t,\boldsymbol{x})\equiv a(t)\, {\rm{e}}^{\psi(t,\boldsymbol{x})}$} as the effective scale factor at a given location $\bm x$, and the associated e-folding number is given as\footnote{The validity of this relies on the shift vector being negligible on large scales, which has been shown to be the case in e.g. \cite{Sugiyama:2012tj,Lyth:2004gb}.}
\begin{align}
 \tilde N(t_{\rm f},t_{\rm i},\bm x) = \ln\left( \frac{\tilde a(t_{\rm f},\bm x)}{\tilde a(t_{\rm i},\bm x)}\right) = \psi(t_{\rm f},\bm x) -\psi(t_{\rm i},\bm x) + N(t_{\rm f},t_{\rm i}). 
\end{align}
If we were to consider flat slices, where $\psi(t,\bm x) = 0$, then we
 see that the scale factor reduces to that of the fiducial background,
 namely $\tilde a(t,\bm x) = a(t)$, and correspondingly the e-folding
 number remains unperturbed, i.e. $\tilde N(t_{\rm f},t_{\rm i},\bm x) =
 N(t_{\rm f},t_{\rm i})$.\mk{}
On
 the other hand, if we were to consider moving between a flat slice at
 $t_{\rm i}$, with $\psi(t_{\rm i},\bm x) = 0$ such that $\tilde
 a(t_{\rm i},\bm x) = a(t_{\rm i})$, and a constant density slice at
 $t_{\rm f}$, on which $\delta\rho(t_{\rm f},\bm x) = 0$ and $\psi(t_{\rm f},\bm x) = \zeta(t_{\rm f},\bm x)$ such that $\tilde a(t_{\rm
 f},\bm x) = a(t_{\rm f})\rm e^{\zeta(t_{\rm f},\bm x)}$, then we find
\begin{gather}
\zeta(t_{\rm f},\bm x) = \delta N(t_{\rm f},\bm x) = \tilde N(t_{\rm f},t_{\rm i},\bm x) - N(t_{\rm f},t_{\rm i}).\label{dNz}
\end{gather}
Note that the left-hand side of the above equation is independent of $t_{\rm i}$, i.e. it is independent of when we take our initial flat slice.  This reflects the fact that the number of e-foldings between any two flat slices is homogeneous, and therefore does not contribute to $\delta N(t,\bm x)$.

If we consider the case $t_{\rm i} = t_{\rm f} - \delta t$, then $\delta
N(t,\bm x)$ given in \eqref{dNz} corresponds to the perturbation in the
number of e-foldings that results from making a gauge transformation
between the flat and constant-density slices at the final time $t_f$.
In this case we can derive a simple relation between $\delta N(t,\bm x)$
and the density perturbation on the flat hypersurface.  As the e-folding
number is unperturbed on flat hypersurfaces, it is useful to use this as
a time parameter.  We can then decompose the density on the flat
hypersurface as $\rho(N,\bm x) = \rho(N) + \delta \rho(N,\bm x)$, where
$\rho(N)$ is the density of the fiducial background trajectory
associated with the location $\bm x_0$, namely $\rho(N,\bm x_0) =
\rho(N)$. 
Next we consider the shift in $N$ required at each location $\bm x$ such
that $\rho(N + \delta N,\bm x) = \rho(N)$, that is, the time-shift
required to move from the flat slice to the constant density
slice~\cite{Yokoyama:2007dw,Yokoyama:2007uu}.  Assuming
$\delta\rho(N,\bm x)$ to be small, expanding $\rho(N+\delta N,\bm x)$ up
to second order in $\delta N$ and solving iteratively one
obtains\footnote{Note that because we are interested in computing the
bispectrum of $\zeta$, it is enough to keep only terms up to second
order in the expansion.}

\begin{equation}
\delta N(N,\bm x) = \frac{1}{6H^2\epsilon}\left[\delta\rho(N,\bm x) + \frac{1}{6H^2\epsilon}\delta\left(\frac{d\rho(N,\bm x)}{dN}\right)\delta\rho(N,\bm x)+\frac{1}{2}\frac{\left(2\epsilon - \eta\right)}{6H^2\epsilon}\delta\rho^2(N,\bm x)\right],\label{dN-rho}
\end{equation}
where $\delta\left(\frac{\mathrm{d}\rho}{\mathrm{d} N}\right)\equiv\left.\frac{\mathrm{d}\rho}{\mathrm{d}N}\right|_{\boldsymbol{x}}-\left.\frac{\mathrm{d}\rho}{\mathrm{d}N}\right|_{\boldsymbol{x}_{0}}$ and we have used
 \begin{align}
\left.\frac{d\rho}{dN}\right. =
 -6H^2\epsilon,\qquad \left.\frac{d^2\rho}{dN^2}\right. = 6H^2\epsilon\left(2\epsilon
 - \eta\right),\label{drhodN} 
 \end{align}
 which can be determined from the fact that $3H^2 = \rho$.  This result
 is in agreement with that obtained in e.g.~\cite{Dias:2014msa}.

In the context of inflation, where we assume the dynamics to be
determined by multiple scalar fields $\phi_{I}$, the validity of the
\mk{separate universes} picture has been confirmed explicitly to all orders in
perturbation theory~\cite{Sugiyama:2012tj}. In this case, the
perturbation $\delta\rho$ can be expressed in terms of perturbations in
the fields $\phi_{I}$ and their velocities $\phi_{I}'$ on the flat
slicing.  Using the compact notation introduced in \eqref{varj-def}, we
can expand $\delta \rho$ as
\begin{align}
 \delta\rho(N,\bm x) = \rho_{,i}\delta\varphi^i(N,\bm x) + \frac{1}{2}\rho_{,ij}\delta\varphi^i(N,\bm x)\delta\varphi^j(N,\bm x),\label{drho}
\end{align}
which on substituting into eq.~\eqref{dN-rho}
gives an expansion of the form 
\begin{equation}
\delta N\left(N,\boldsymbol{x}\right)\simeq N_{,i}\left(N\right)\delta\varphi^{i}\left(N,\boldsymbol{x}\right)+\frac{1}{2}N_{,ij}\left(N\right)\delta\varphi^{i}\left(N,\boldsymbol{x}\right)\delta\varphi^{j}\left(N,\boldsymbol{x}\right).\label{dNdphif}
\end{equation}
In the above equation we used the same notation as in eq.~\eqref{EoMt},
where the subscripts in coefficients $N_{,i}$ and $N_{,ij}$ denote
differentiation with respect to unperturbed fields $\varphi_{i}$.  The
explicit form of these coefficients will be given below.

While the expansion \eqref{dNdphif} is in terms of the field
perturbations at the final time $N$, in the context of the separate
universes approach we know that the field values and velocities on the
flat slicing evaluated at a given time $N$ and smoothed over a
superhorizon size patch at some spatial coordinate $\boldsymbol{x}$,
$\varphi^i(N,\bm x)$, are simply solutions to the background equations
of motion with initial conditions given at some earlier time $N_0$ that
are local to that patch, namely
 \begin{align}
  \varphi^i(N,\bm x) = \varphi^i(N,\varphi^a(N_0,\bm x)).
 \end{align}
Following e.g. \cite{Anderson:2012em,Elliston:2012gr,Seery:2012vj}, let us introduce
notation in which different types of indices are used to denote
quantities evaluated at different times. In particular, indices from the
beginning of the Latin alphabet will be used to denote quantities
evaluated at the initial time $N_{0}$,
e.g. $\varphi_{a}\left(\boldsymbol{x}\right)\equiv\varphi_{a}\left(N_{0},\boldsymbol{x}\right)$,
while indices from the end of the alphabet will be used to denote
quantities at the final time $N$,
e.g. $\varphi_{z}\left(\boldsymbol{x}\right)\equiv\varphi_{z}\left(N,\boldsymbol{x}\right)$.
This allows us to write
 \begin{equation}
\varphi^{z}\left(\boldsymbol{x}\right)=\varphi^{z}+\varphi_{,a}^{z}\delta\varphi^{a}\left(\boldsymbol{x}\right)+\frac{1}{2}\varphi_{,ab}^{z}\delta\varphi^{a}\left(\boldsymbol{x}\right)\delta\varphi^{b}\left(\boldsymbol{x}\right)+\ldots\label{jz}
\end{equation}
where $\varphi_{,a}^{z}\equiv\partial\varphi^{z}/\partial\varphi^{a}$
and
$\varphi_{,ab}^{z}\equiv\partial^{2}\varphi^{z}/\partial\varphi^{a}\partial\varphi^{b}$. In
this expression $\varphi^{z}\left(\boldsymbol{x}\right)$ are the values
of fields smoothed on superhorizon patch at $\boldsymbol{x}$ and
$\varphi^{z}$ (without an argument) is some fiducial trajectory in field
space, which can be associated with the location $\bm
x_0$. 

\mk{}$\delta\varphi^{a}\left(\boldsymbol{x}\right)$ are the
perturbations of initial conditions.  Thus we have an expansion of
$\delta\varphi^z(\bm x)$ in terms of $\delta \varphi^a(\bm x)$, which on
substituting into \eqref{dNdphif} gives an expansion of the form
\begin{equation}
\delta N\left(N,\boldsymbol{x}\right)=N_{,z}\varphi_{,a}^{z}\delta\varphi^{a}\left(\boldsymbol{x}\right)+\frac{1}{2}\left(N_{,z}\varphi_{,ab}^{z}+N_{,yz}\varphi_{,a}^{y}\varphi_{,b}^{z}\right)\delta\varphi^{a}\left(\boldsymbol{x}\right)\delta\varphi^{b}\left(\boldsymbol{x}\right),\label{z-phi}
\end{equation}
Note that for the same reason as discussed above, the choice of $N_0$ in
evaluating \eqref{z-phi} is arbitrary.  This can be used to our
advantage, as the \mk{$\delta\varphi$'s} are most easily computed soon after
the scales under consideration leave the horizon, corresponding to
$N_{0}\simeq N_{k}$.  In this case, we are able to make contact with the
perhaps more familiar form of the $\delta N$ expansion, where $\delta N$
is expanded in terms of the field perturbations shortly after horizon
crossing.  Explicitly, the derivatives of $N$ with respect to the field
values and velocities around horizon crossing are given as
\begin{align}
 N_{,a} = N_{,z}\varphi^z_{,a},\qquad N_{,ab} = N_{,z}\varphi^z_{,ab} + N_{,yz}\varphi^y_{,a}\varphi^z_{,b}.
\end{align}
The choice of $N$ in evaluating eq.~\eqref{z-phi}, on the other hand, is crucial.
To find the final value of $\zeta$, $N$ should be chosen to be some
\mk{time} after an adiabatic limit has been reached and $\zeta$ freezes in.
In general this can happen long after the end of inflation.

We see that there are three key elements that are required in order to
determine $\zeta(N,\boldsymbol{x})=\delta
N\left(N,\boldsymbol{x}\right)$: the derivatives of $N$ with respect to
$\varphi_{z}$, the derivatives of the final field values $\varphi_{z}$
with respect to the initial conditions $\varphi_{a}$ and the
perturbations in the initial conditions themselves,
$\delta\varphi_{a}\left(\boldsymbol{x}\right)$. The remainder of this
section we will be dedicated to computing these three elements.

\subsection{The derivatives of $N$}

The quantities $N_{,z}$ are determined by combining eqs. \eqref{dN-rho}
and \eqref{drho}, and for explicit expressions we need to determine
explicit expressions for $\rho_{,z}$ and $\rho_{,zy}$.  The final
results will depend on the level of slow-roll approximation that we
make, and we thus consider the following three cases in turn: no
slow-roll approximation, leading order slow-roll approximation and
next-to-leading order slow roll approximation.

\subsubsection{No slow-roll approximation}

In the case that no slow-roll approximation is used, the energy density
$\rho(N,\bm x)$ is given in terms of the fields and their velocities as
in eq.~\eqref{FrdmN}.  Perturbing this expression up to second order in
$\delta\phi_I$ and $\delta\phi'_I$ we obtain
\begin{align}\nonumber
\delta\rho&=\rho\left(\frac{V_{,I}}{V}\delta\phi_{I}+\frac{\phi_{I}'}{3-\epsilon}\delta\phi_{I}'+\frac{V_{,IJ}}{2V}\delta\phi_{I}\delta\phi_{J}\right.\\ &\hspace{2cm}\left.+\frac{\phi'_IV_{,J}}{(3-\epsilon)V}\delta\phi'_I\delta\phi_J + \frac{1}{2\left(3-\epsilon\right)}\left[\delta_{IJ} + 2\frac{\phi_I'\phi_J'}{3-\epsilon}\right]\delta\phi_{I}'\delta\phi_{J}'\right).\label{drho-exp}
\end{align}
Similarly, perturbing the first of \eqref{drhodN} we obtain
\begin{equation}
\delta\left(\frac{d\rho(N,\bm x)}{dN}\right) = -\delta\rho(N,\bm x)\phi'_I\phi'_I - 2\rho(N)\phi'_I\delta\phi'_I(N,\bm x).\label{ddNdrho}
\end{equation}
Plugging eqs.~\eqref{drho-exp} and \eqref{ddNdrho} into eq.~\eqref{dN-rho}
and comparing with eq.~\eqref{dNdphif} we can read off the coefficients
$N_{,i}$ and $N_{,ij}$
\begin{align}
N_{,Z} & =\frac{1}{2\epsilon}\frac{V_{,Z}}{V},\label{NI-rez1}\\
N_{,Z'} & =\frac{1}{2\epsilon}\frac{\phi_{Z}'}{3-\epsilon},\label{NdI-rez1}\\
N_{,YZ} & =\frac{1}{2\epsilon}\left(\frac{V_{,YZ}}{V}-\frac{\epsilon+\eta/2}{\epsilon}\frac{V_{,Y}}{V}\frac{V_{,Z}}{V}\right),\\
N_{,YZ'} & =-\frac{3-\epsilon+\eta/2}{2\epsilon^{2}\left(3-\epsilon\right)}\frac{V_{,Y}\phi_{Z}'}{V},\\
N_{,Y'Z'} & =\frac{1}{2\epsilon}\frac{1}{3-\epsilon}\left(\delta_{YZ}-\frac{6-3\epsilon+\eta/2}{\epsilon\left(3-\epsilon\right)}\phi_{Y}'\phi_{Z}'\right),\label{NdIdJ-rez1}
\end{align}
where $N_{,Z'}\equiv\partial N/\partial\phi_{Z}'$. It is worth reiterating
that while the above result is only valid up to second order in field perturbations
and their temporal derivatives, no slow-roll assumption was made.

While the current form of the coefficients in
eqs.~\eqref{NI-rez1}-\eqref{NdIdJ-rez1} is perfectly acceptable, it is
possible to simplify them by recognising that there are certain
combinations of perturbations that decay on super-horizon scales.  This
observation is closely related to the relation between the
constant-density and comoving surfaces in phase space.  The comoving
condition is defined as the requirement $T^0{}_i=0$, where $T^\mu{}_\nu$
is the the energy momentum tensor associated with the multiple scalar
fields.  Using $N$ as the time coordinate we find that in the flat gauge
and on super-horizon scales we have
\begin{align}
  T^0{}_i(\bm x) = - H^2(\bm x)\phi'_I(\bm x)\partial_i\phi_I(\bm x) + \mc O(\xi^2).\label{T0i}
\end{align}
As discussed in \cite{Sasaki:1998ug}, the condition $T^0{}_i=0$ with $T^0{}_i$ as
given in \eqref{T0i} does not in general define a surface in phase
space, except at linear order in perturbations.  However, one finds that
 \begin{align}
  - H^2(\bm x)\phi'_I(\bm x)\partial_i\phi_I(\bm x) = \partial_i\left(\frac{\rho(\bm x)}{3}\right)- H(\bm x)B_i(\bm x),\label{drho-dT0i}
 \end{align}
where
\begin{equation}
B_{i}\left(\boldsymbol{x}\right)\equiv\frac{H\left(\boldsymbol{x}\right)}{3-\epsilon\left(\boldsymbol{x}\right)}\left[\phi_{I}'\left(\boldsymbol{x}\right)\partial_{i}\phi_{I}'\left(\boldsymbol{x}\right)-\phi_{I}''\left(\boldsymbol{x}\right)\partial_{i}\phi_{I}\left(\boldsymbol{x}\right)\right].\label{Bi}
\end{equation}
This means that the comoving and constant
density conditions differ by the term $H(\bm x)B_i(\bm x)$.
As was pointed out by Sasaki \& Tanaka in \cite{Sasaki:1998ug}, and later by \mk{Sugiyama}
\textit{et al.} in \cite{Sugiyama:2012tj}, if we take the
spatial gradient of the equations of motion \eqref{EoMN} and then
contract this with $\phi_{I}'(\boldsymbol{x})$, we find that the
quantity $B_{i}$ satisfies
\begin{equation}
B_{i}'+3B_{i}=0,
\end{equation}
from which we deduce that $B_{i}$ decays as $a^{-3}$.  As such, we find
that even at non-linear order the comoving and constant-density
conditions coincide on super-horizon scales.  This result is well known
at linear order in perturbation theory, and is usually shown by
combining the energy and momentum constraints, see
e.g. \cite{Bassett:2005xm}.  Here, on the other hand, we simply made use
of the equations of motion for the scalar fields.\footnote{Striclty
speaking, we have relied on Einstein's equations to confirm that the
lapse function $\beta^i$ and the time dependence of $\gamma_{ij}$ decay
on large scales.}

Neglecting the decaying $H(\bm x)B_i(\bm x)$ term in \eqref{drho-dT0i}
we are able to find a relation between $\delta\rho(\bm x)$ and
$\delta\phi_I(\bm x)$ and $\delta\phi'_I(\bm x)$ that is simpler than
the one given in \eqref{drho-exp}.  Decomposing $\phi_I(\bm x) = \phi_I +
\delta\phi_I(\bm x)$ and similarly for $\phi'_I(\bm x)$ and $\rho(\bm
x)$, eq.~\eqref{drho-dT0i} becomes
\begin{align}
 \partial_i\delta\rho(\bm x) \simeq -\left(\rho + \delta\rho(\bm
 x)\right)\left(\phi'_I + \delta\phi'_I(\bm
 x)\right)\partial_i\delta\phi_I(\bm x).
\end{align}
At linear order the right-hand side of this expression can be written as
a pure gradient, such that we obtain $\delta\rho(\bm x) =
-\rho\phi'_I\delta\phi_I(\bm x)$.  Making use of this result we then
find that to second order we obtain
\begin{align}
 \delta\rho(\bm x) = \rho\left[- \phi'_I\delta\phi_I(\bm
 x)+\frac{1}{2}\left(\phi'_I\delta\phi_I(\bm x)\right)^2 -
 \frac{1}{2}\delta\phi'_I(\bm x)\delta\phi_I(\bm x)\right] -
 \frac{3H}{2}\partial^{-2}\partial_i D_i(\bm x),\label{drho-simp}
\end{align}
where
\begin{align}
D_{i}\left(\boldsymbol{x}\right) =
H\left[\delta\phi_{I}'\left(\boldsymbol{x}\right)\partial_{i}\delta\phi_{I}\left(\boldsymbol{x}\right)-\partial_{i}\delta\phi_{I}'\left(\boldsymbol{x}\right)\delta\phi_{I}\left(\boldsymbol{x}\right)\right].\label{Di}
\end{align}
Making use of \eqref{ddNdrho} and substituting \eqref{drho-simp} into
\eqref{dN-rho}, we obtain
\begin{align}
 \delta N =
 \frac{1}{2\epsilon}\left[-\phi'_Z\delta\phi_Z(\bm x) -
 \frac{1}{2}\left(\delta_{YZ} -
 2\frac{\phi'_Y\phi'_Z}{\epsilon}\right)\delta\phi_Y(\bm
 x)\delta\phi'_Z(\bm
 x)-\frac{\eta}{4\epsilon}\left(\phi'_Z\delta\phi_Z(\bm
 x)\right)^2\right]- \frac{\partial^{-2}\partial_i D_i(\bm x)}{4\epsilon H},\label{dN-NL}
\end{align}
which can be shown to be in agreement with the expression for
$\zeta^{{\rm simple}}$ in ref.~\cite{Dias:2014msa}.
As such, we see that at second order the use of \eqref{drho-simp} has
introduced a non-local term into the expansion of $\delta N$. However,
taking the derivative of $D_{i}$ with respect to $N$ we find that it
satisfies
\begin{equation}
a^{-3}\left(a^{3}D_{i}\right)'=\phi_{I}'\delta\phi_{I}'\partial_{i}B^{(1)}-\phi_{I}'\partial_{i}\delta\phi_{I}'B^{(1)},
\end{equation}
where $B^{(1)}$ is such that at linear order in perturbations we have
$B_i = \partial_i B^{(1)}$, from which we deduce that $B^{(1)}$ is
decaying as $1/a^3$.  
Provided $|\phi_{I}'|$ and $|\delta\phi_{I}'|$ do not grow too fast after horizon exit, we thus find that $D_i$ also decays.
As such, neglecting all decaying terms in the $\delta N$ expansion we find
that the second order expression for $\delta N$ remains local, and we
finally obtain
\begin{align}
\label{NI-rez2} N_{,Z} & \simeq-\frac{1}{2\epsilon}\phi_{Z}'\\
N_{,Z'} & \simeq0\\
N_{,YZ} & \simeq-\frac{\eta}{4\epsilon^{2}}\\
N_{,YZ'} & \simeq-\frac{1}{4\epsilon}\left(\delta_{YZ}-\frac{2}{\epsilon}\phi_{Y}'\phi_{Z}'\right)\\
\label{NdIdJ-rez2} N_{,Y'Z'} & \simeq0.
\end{align}
The fact that the non-local terms decay on super-horizon scales was
shown for the two-field case in \cite{Langlois:2006vv}, and here we have
generalised this result to the case of more than two fields.

\subsubsection{Leading and next-to-leading order slow roll approximations}

In the case that we make the slow-roll approximation we can use the same
method as above but now we obtain our expansion of $\delta\rho$ in terms
of $\delta\phi_I$ from the expressions for $\rho^{(0)}$ given in
\eqref{FrdmN-SR} and $\rho^{(1)}$ given \eqref{Frdm1} for the leading
and next-to-leading order cases respectively.  Note that the velocities
$\phi'_I$ are no-longer independent degrees of freedom, so that the
expansion of $\delta N$ is only in terms of the field fluctuations
$\delta\phi_I$.

Perhaps a slightly quicker way to obtain the necessary results, however,
is to realise that in the slow-roll case the quantities $B_i$ and $D_i$
introduced above exactly vanish.  Considering $B_i$ first, in the
slow-roll case we have
\begin{align}
 \phi'_I(\bm x)= U^{(i)}_{,I}(\bm x) \qquad\rightarrow\qquad \begin{array}{l}
  \phi''_I(\bm x) = U^{(i)}_{,IJ}(\bm x)U^{(i)}_{,J}(\bm x), \\ \partial_i\phi'_I(\bm x) = U^{(i)}_{,IJ}(\bm x)\partial_i\phi_J(\bm x), \end{array} 
\end{align}
where here the superscript $i$ takes the values $0$ or $1$ to denote the
leading or next-to-leading order cases respectively.  Substituting these
results into \eqref{Bi} we find that indeed $B_i$ vanishes.  Similarly,
turning to $D_i$, at linear order we have
\begin{align}
 \delta\phi'_I(\bm x) = U^{(i)}_{,IJ}\delta\phi_J(\bm x)\qquad
 \rightarrow \qquad \partial_i\delta\phi'_I =
 U^{(i)}_{,IJ}\partial_i\delta\phi_{J}(\bm x),
\end{align}
which on substituting into \eqref{Di} gives $D_i=0$.  As such, the
expression for $\delta N$ given in \eqref{dN-NL} with the last term set
to zero becomes exact, in the sense that the decaying terms involving
$B_i$ and $D_i$ are exactly zero in the slow-roll case rather than just
decaying.  On expressing $\phi'_I$ and $\delta\phi'_I(\bm x)$ in terms
of $U^{(i)}$ and its derivatives, we thus obtain
\mk{
\begin{align}
 N_{,Z} &= -\frac{1}{2\epsilon^{(i)}}U^{(i)}_{,Z},\label{Nz-SR}\\
 N_{,YZ}&=\frac{1}{2\left(\epsilon^{(i)}\right)^2}\left[-\epsilon^{(i)}
 U^{(i)}_{,YZ} + U^{(i)}_{,Y}U^{(i)}_{,X}U^{(i)}_{,XZ} +
 U^{(i)}_{,Z}U^{(i)}_{,X}U^{(i)}_{,XY} -
 \frac{\eta^{(i)}}{2}U^{(i)}_{,Y}U^{(i)}_{,Z}\right].\label{Nzy-SR}
\end{align}
}
Recall that our use of indices from the end of the alphabet here indicates that these are the derivatives of $N$ with respect to field values at the final time at which we wish to evaluate $\zeta$ using eq.~\eqref{z-phi}.
For the leading-order case, i.e. taking $i=0$ and substituting the
relevant expressions for $U^{(0)}$, $\epsilon^{(0)}$ and $\eta^{(0)}$,
we obtain results that are in agreement with \cite{Yokoyama:2007uu}.  As
far as we are aware, the next-to-leading order case, with $i=1$, has not
been considered in the literature.

\subsection{The derivatives of $\varphi_{z}$}

Having found the coefficients $N_{,z}$ and $N_{,yz}$ that appear in
eq.~\eqref{z-phi} we next compute the quantities $\varphi_{,a}^{z}$ and
$\varphi_{,ab}^{z}$.  Once again it is possible to do this under various
levels of slow-roll approximation.

\subsubsection{No slow-roll approximation}

Recall that the quantities $\varphi_{,a}^{z}$ and $\varphi_{,ab}^{z}$
appear in the expansion of $\varphi^z(\bm x)$ about some fiducial
trajectory as in \eqref{jz}.  Next\mk{,} let us recall that
as a consequence of the separate universes approach, the equations of
motion for $\phi_{I}\left(\boldsymbol{x}\right)$ are of exactly the same
form as those for the unperturbed fields given in eq.~\eqref{EoMN},
which we then conveniently re-expressed in terms of $\varphi_i$ as in
\eqref{EoM-j}.  Perturbing eq.~\eqref{EoM-j} up to second order we find
\begin{equation}
{\delta\varphi^{z}}'(\bm x)=\mathbb{U}_{,y}^{z}\delta\varphi^{y}(\bm x)+\frac{1}{2}\mathbb{U}_{,yx}^{z}\delta\varphi^{y}(\bm x)\delta\varphi^{x}(\bm x).\label{ddvphidN}
\end{equation}
Then, plugging eq.~\eqref{jz} into the above result we find that
$\varphi_{,a}^{z}$ and $\varphi_{,ab}^{z}$ satisfy the equations
\begin{align}
{\varphi_{,a}^{z}}' & =\mathbb{U}_{,y}^{z}\varphi_{,a}^{y},\label{djzdaa}\\
{\varphi_{,ab}^{z}}' & =\mathbb{U}_{,y}^{z}\varphi_{,ab}^{y}+\mathbb{U}_{,yx}^{z}\varphi_{,a}^{y}\varphi_{,b}^{x},\label{djzdab}
\end{align}
with the initial conditions $\varphi^z_{,a}(N_0) = \delta^z_a$ and
$\varphi^z_{,ab}(N_0) = 0$.
While it is possible to give formal solutions to equations \eqref{djzdaa}
and \eqref{djzdab} \textendash{} see
e.g. \cite{Seery:2012vj}, in
practice we will solve them numerically.  It is perhaps interesting to
consider how much work we will have to do.  Firstly, in terms of the
background dynamics we will have to solve $2M$ equations of motion for
the fields and their velocities.  Then, in solving for the
perturbations, we will have to solve for the $4M^2$ quantities
$\varphi^z_{,a}$ and for the $2M^2(2M+1)$ quantities $\varphi^z_{,ab}$,
where we have used the fact that $\varphi^z_{,ab}$ is symmetric in the
lower two indices.  Altogether we thus have $2M(2M^2 + 3M + 1)$
equations to solve, which for large $M$ goes as $\sim 4 M^3$.  We will
be able to compare this with the amount of work required when we make
the slow-roll approximation.

\subsubsection{Slow-roll approximation holds throughout inflation}

In the case that the leading or next-to-leading order slow-roll
approximation is valid throughout inflation, the field velocities are
not independent degrees of freedom as they are expressed in terms of the
field values via the slow-roll equations of motion.  Nevertheless, the
analysis goes through in exactly the same way as the non-slow-roll case
considered above, but with $\varphi_i\rightarrow \phi_I$ and $\mathbb
U^i \rightarrow U^{(i)}_{,I}$.  Explicitly, perturbing \eqref{EoM0} or \eqref{EoM1} up
to second order we have
\begin{equation}
{\delta\phi'_{Z}}(\bm x)=U^{(i)}_{,ZY}\delta\phi_Y(\bm x)+\frac{1}{2}U^{(i)}_{,ZYX}\delta\phi_{Y}(\bm x)\delta\phi_{X}(\bm x).\label{dZ}
\end{equation}
Then, in analogy with \eqref{jz}, we have
\begin{align}
\delta\phi_Z(\bm x) =  \phi_{Z,A}\delta\phi_A(\bm x) + \frac{1}{2}\phi_{Z,AB} \delta\phi_A(\bm x)\delta\phi_B(\bm x), 
\end{align}
where we have kept the convention that letters from the beginning and
end of the alphabet are used to denote quantities evaluated at the
initial and final times respectively.  On substituting this expansion
into \eqref{dZ} we obtain the equations
\begin{align}
{\phi_{Z,A}}' & =U^{(i)}_{Z,Y}\phi_{Y,A},\label{djzda-SR}\\
{\phi_{Z,AB}}' & =U^{(i)}_{,ZY}\phi_{Y,AB}+U^{(i)}_{,ZYX}\phi_{Y,A}\phi_{X,B},\label{djzdab-SR}
\end{align}
with the initial conditions $\phi_{Z,A}(N_0) = \delta_{ZA}$ and
$\phi_{Z,AB}(N_0) = 0$.  In this case, at background level we must solve
for just the $M$ scalar fields, and in solving for the perturbations we
must solve for the $M^2$ components of $\phi_{Z,A}$ and the $M^2(M+1)/2$
components of $\phi_{Z,AB}$, giving a total of $M(M^2+3M+2)/2$ equations
to solve.  For large $M$ this goes as $\sim M^3/2$, which is a factor of
$1/8$ fewer than the number of equations that need to be solved in the
case where no slow-roll approximation is made.

\subsubsection{Slow-roll approximation holds only around horizon crossing}

We finally consider the case where we only assume that the leading or
next-to-leading order slow-roll approximation holds around the time that
the scales of interest left the horizon, i.e. we allow for the
possibility that the slow-roll approximation breaks down during the
super-horizon evolution.

To allow for this possibility, we use the full equations of motion
\eqref{EoM-j} to solve for $\varphi^z(\bm x)$, but we assume that the
initial conditions are such that $\phi'_A(\bm x) = U^{(i)}_{,A}(\bm x)$.
As such, instead of \eqref{jz} we expand $\varphi^z(\bm x)$ as
 \begin{equation}
\varphi^{z}\left(\boldsymbol{x}\right)=\varphi^{z}+\varphi_{,A}^{z}\delta\phi_{A}\left(\boldsymbol{x}\right)+\frac{1}{2}\varphi_{,AB}^{z}\delta\phi_{A}\left(\boldsymbol{x}\right)\delta\phi_{B}\left(\boldsymbol{x}\right)+\ldots.\label{jz-SR}
\end{equation}
Note that the quantities $\varphi^z_{,A}$ and $\varphi^z_{,AB}$ have
mixed indices, in that $z$ runs from $1...2M$ whereas $A$ and $B$ only
run from $1...M$.  Substituting this expansion into \eqref{ddvphidN} we
obtain evolution equations for $\varphi^z_{,A}$ and $\varphi^z_{,AB}$ as
\begin{align}
{\varphi_{,A}^{z}}' &
=\mathbb{U}_{,y}^{z}\varphi_{,A}^{y},\label{djzda}\\
{\varphi_{,AB}^{z}}' &
=\mathbb{U}_{,y}^{z}\varphi_{,AB}^{y}+\mathbb{U}_{,yx}^{z}\varphi_{,A}^{y}\varphi_{,B}^{x},\label{djzdAB}
\end{align}
with the initial conditions
\begin{align}
 \varphi^z_{,A} = \delta^z_A,\quad \varphi^z_{,AB} = 0\qquad
 &\mbox{for}\quad 1\le z \le M,\\ 
\varphi^z_{,A} = \delta^{z-M}_B
 U^{(i)}_{BA},\quad \varphi^z_{,AB} = \delta^{z-M}_CU^{(i)}_{,CAB}\qquad
 &\mbox{for}\quad M+1\le z \le 2M.
\end{align}
In this case, at background level we have to solve the $2M$ equations
of motion for the fields and their velocities, and at the level of
perturbations we have to solve for the $2M^2$ components of
$\varphi^z_{,A}$ and $M^2(M+1)$ components of $\varphi^z_{,AB}$.  In
total this gives $M(M^2+3M+2)$ equations that we have to solve, which is
double the number we had to solve when we assumed that the slow-roll
approximation was valid throughout inflation.  For large $M$ the number
of equations scales as $\sim M^3$, which is a factor of four fewer than
the number of equations we have to solve when no slow-roll approximation
is made.

\subsection{The perturbations of initial conditions $\delta\varphi_{a}$\label{sec:initialdphi} }

The final ingredients in computing eq.~\eqref{z-phi} are the
perturbations $\delta\varphi_{a}\left(\boldsymbol{x}\right)$. They are
the field perturbations on the initial flat slice at time $N_{0}$. As
was already mentioned, $N_{0}$ can be set to be any moment, provided
that it is after the time at which the relevant scales exited the
horizon.  In practise, the standard procedure is to use linear
perturbation theory to solve for $\delta\varphi_{a}$ around the
horizon-crossing time by making use of the slow-roll approximation, and
to take $N_{0}=N_{\ast}$, where $N_{*}$ corresponds to a few e-foldings
after horizon crossing.  Such a computation was first performed up to
next-to-leading order in the slow-roll approximation by Nakamura \&
Stewart \cite{Nakamura:1996da}. Their method is based on
the assumption that slow-roll parameters change very little over the
course of a few e-foldings of expansion. This allows one to choose a
field basis in which the field equations of motion effectively become
decoupled for a few e-foldings before and after some given
instant in time. For our purposes, that instant in time is conveniently
chosen to be the moment of horizon exit. To show the relevant
assumptions of this method explicitly, we briefly recall the calculation
of ref.~\cite{Nakamura:1996da} below, specialising to
the case of a flat field space.

One can start by writing the equations of motion for the field
perturbation $\delta\phi_{I}$ on flat slices (see
e.g. ref.~\cite{lyth2009primordial})
\begin{equation}
\delta\ddot{\phi}_{k}^{I}+3H\delta\dot{\phi}_{k}^{I}+\left(\frac{k}{a}\right)^{2}\delta\phi_{k}^{I}+V_{,IJ}\delta\phi_{k}^{J}=\frac{\delta\phi_{k}^{J}}{a^{3}}\frac{\mathrm{d}}{\mathrm{d}t}\left(\frac{a^{3}}{H}\dot{\phi}_{I}\dot{\phi}_{J}\right),\label{EoM-dphik}
\end{equation}
where we work in Fourier space, hence the index $k$. Fundamentally
$\delta\phi_{k}^{I}$ are mode functions of a quantum field operator,
\mk{
$\hat{\phi}^I\left(\boldsymbol{x},t\right)=\int\left[\delta\phi_{\boldsymbol{k}}^{I}\left(t\right)\mathrm{e}^{i\boldsymbol{k}\cdot\boldsymbol{x}}\hat{a}\left(\boldsymbol{k}\right)+\delta\phi_{\boldsymbol{k}}^{I*}\left(t\right)\mathrm{e}^{-i\boldsymbol{k}\cdot\boldsymbol{x}}\hat{a}^{\dagger}\left(\boldsymbol{k}\right)\right]\mathrm{d}\boldsymbol{k}$,
}
where $\hat{a}^{\dagger}$ and $\hat{a}$ are creation and annihilation
operators.

Instead of using cosmic time it is more appropriate in this context
to switch to conformal time
\begin{equation}
\mathrm{d}\tau\equiv a^{-1}\mathrm{d}t.\label{tau-def}
\end{equation}
We then assume that over a few e-foldings of expansion the slow-roll
parameter $\epsilon$, defined in eq.~\eqref{eps-eta-def}, is constant,
which is consistent with the condition $\eta\ll 1$. In this case, we can
integrate eq.~\eqref{tau-def} to write
$\tau=-\left(1+\epsilon\right)/aH$.  Then, defining the field
\begin{equation}
\chi_{k}^{I}\equiv a\delta\phi_{k}^{I},
\end{equation}
eq.~\eqref{EoM-dphik} can be written as
\begin{equation}
\partial_{\tau}^{2}\chi_{k}^{I}+\left(k^{2}-\frac{2}{\tau^{2}}\right)\chi_{k}^{I}=\epsilon_{J}^{I}\frac{3\chi_{k}^{J}}{\tau^{2}},
\end{equation}
where
\begin{align}
\epsilon_{IJ} & \equiv\epsilon\delta_{IJ}+\delta_{IK}\delta_{JL}\phi_{K}'\phi_{L}'-\frac{V_{,IJ}}{3H^{2}}\\
 & \simeq\epsilon\delta_{IJ}+U_{,IJ}^{\left(0\right)}
\end{align}
and $\partial_{\tau}^{2}\chi\equiv\mathrm{d}^{2}\chi/\mathrm{d}\tau^{2}$.
To obtain the second line above we used the lowest-order slow-roll
approximations for $\phi'_K$ and $H^2$ given in eqs.~\eqref{EoM0} and \eqref{FrdmN-SR}

At any instant in time we can choose a field basis in which
$\epsilon_{IJ}$ is diagonalised, i.e. choose a basis in which the
fields, let us call them \mk{$\psi_{k}^{I}\left(\tau\right)$}, are
decoupled. Strictly speaking $\epsilon_{IJ}$ is diagonalised only
instantaneously, but if the slow-roll approximation is valid then the
off-diagonal components of $\epsilon_{IJ}$
should remain negligible for a few e-foldings before and after the instant at which $\epsilon_{IJ}$ is exactly diagonalised. Taking the instant in time to be that of horizon-crossing, the choice of initial conditions for \mk{$\psi_{k}^{I}\left(\tau\right)$}
becomes particularly simple. For $\left(\tau k\right)^{2}\gg1$
(the sub-horizon regime), the mode functions \mk{$\psi_{k}^{I}\left(\tau\right)$} coincide with those of a free massless field and we can use the Bunch-Davies vacuum state as our initial conditions. Later, when $\left(\tau k\right)^{2}\ll1$
(the super-horizon regime), the modes freeze and behave as classical
random fields. 

We should stop the integration of the equations for
\mk{$\psi_{k}^{I}\left(\tau\right)$} just a few e-foldings after horizon
crossing: late enough so that terms of order $(\tau k)^2$ can be
neglected and \mk{$\psi_{k}^{I}\left(\tau\right)$} are no longer oscillating,
but early enough so that the off-diagonal components of $\epsilon_{IJ}$
remain negligible. Denoting this time by $\tau_{*}$ and going back to
the original field basis we can write the final result as
\cite{Nakamura:1996da}
\mk{
\begin{equation}
\delta\phi_{k*}^{A}=\frac{iH_{*}}{\sqrt{2k^{3}}}\left\{ \left(1-\epsilon\right)\delta_{B}^{A}+\left[C+\ln\left(\frac{a_{*}H_{*}}{k}\right)\right]\epsilon_{B}^{A}\right\} b_{k}^{B},\label{dphi-star}
\end{equation}
}
where a subscript `$*$' is used to denote quantities evaluated at
$\tau_{*}$ and $C=2-\ln2-\gamma\simeq0.730$, where $\gamma\simeq0.577$ is the
Euler-Mascheroni constant. 
In the above equation we have used indices from the beginning of the alphabet, as the quantities $\delta\phi^A_{k\ast}$ correspond to the field perturbations on the initial flat slice that are used in evaluating eq.~\eqref{z-phi}.
The quantities $\epsilon$ and \mk{$\epsilon_{AB}$}
appearing in the above expression are evaluated at the horizon-crossing
time, and \mk{$b_{k}^{B}$} are effectively \emph{classical} random variables
satisfying
\mk{
\begin{equation}
\left\langle b_{\boldsymbol{k}}^{A}\right\rangle =0\text{ and }\left\langle b_{\boldsymbol{k}}^{A},b_{\boldsymbol{k}'}^{B*}\right\rangle =\left(2\pi\right)^{3}\delta^{AB}\delta^{\left(3\right)}\left(\boldsymbol{k}-\boldsymbol{k}'\right),
\end{equation}
}
where $\left\langle \ldots\right\rangle $ denotes ensemble averages
and \mk{$b_{\boldsymbol{k}'}^{B*}$} is the complex conjugate of \mk{$b_{\boldsymbol{k}'}^{B}$}.
Note that reality of \mk{$\delta\phi^{A}$} requires \mk{$b_{\boldsymbol{k}'}^{B*}=b_{-\boldsymbol{k}'}^{B}$}.  Also note that since gradient terms have been neglected, the solutions given in eq.~\eqref{dphi-star} coincide with solutions to the perturbed homogeneous background equations.

In principle one can go to higher order in the slow-roll approximation
using a Green's-function method \cite{Gong:2002cx},
but here we only make use of the next-to-leading order result. The fact
that we make use of the above next-to-leading order expression for
\mk{$\delta\phi_{\boldsymbol{k}}^{A}$} puts a limitation on the range of
models that we are able to consider. While our analysis can accommodate
a break down of slow-roll during the super-horizon evolution, it assumes
that the slow-roll approximation is reasonable around the time of
horizon crossing. If we wish to consider models in which slow-roll is
broken around the horizon-crossing time, then we could use the formalism
developed in \cite{Mulryne:2013uka}, where
perturbations are transported from sub-horizon scales and no assumptions
of slow-roll at horizon exit are necessary.  Before moving on, we note
that the appropriate expression for \mk{$\delta\phi^A_{k\ast}$} at leading
order in the slow-roll approximation is \mk{$\delta\phi^A_{k\ast} = iH_\ast
b^A_k/\sqrt{2k^3}$}.

\section{Correlation functions and observables}\label{Sec:corr}

In the preceding section we have given all the ingredients necessary to
use eq.~\eqref{z-phi} to determine the curvature perturbation $\zeta$
and its evolution during an epoch dominated by multiple scalar
fields. Ultimately we are interested in the statistics of the curvature
perturbation, and so in this section we bring all the pieces together
and consider the two- and three-point correlation functions of $\zeta$.

\subsection{Definitions}

We use the standard parametrisation of the two and three-point function,
defining the power spectrum $P_{\zeta}(k)$ and bispectrum
$B_{\zeta}(k_{1},k_{2},k_{3})$ as
\begin{align}
\left\langle \zeta\left(k_{1}\right),\zeta\left(k_{2}\right)\right\rangle  & =\left(2\pi\right)^{3}\delta^{(3)}\left(\boldsymbol{k}_{1}+\boldsymbol{k}_{2}\right)P_{\zeta}\left(k_{1}\right)\\
\left\langle \zeta\left(k_{1}\right),\zeta\left(k_{2}\right),\zeta\left(k_{3}\right)\right\rangle  & =\left(2\pi\right)^{3}\delta^{(3)}\left(\boldsymbol{k}_{1}+\boldsymbol{k}_{2}+\boldsymbol{k}_{3}\right)B_{\zeta}\left(k_{1},k_{2},k_{3}\right)
\end{align}
and the non-linearity parameter $f_{NL}$ as 
\begin{equation}
f_{NL}\equiv\frac{5}{6}\frac{B_{\zeta}\left(k_{1},k_{2},k_{3}\right)}{P_{\zeta}\left(k_{1}\right)P_{\zeta}\left(k_{2}\right)+\mathrm{c.p.}},
\end{equation}
where `$\mathrm{c.p.}$' stands for cyclic permutations of $k_{1}$,
$k_{2}$ and $k_{3}$. For an almost scale-invariant spectrum the reduced
power spectrum is further defined as
$\mathcal{P}_{\zeta}\left(k\right)\equiv\left(k^{3}/2\pi^{2}\right)P_{\zeta}\left(k\right)$,
and the tilt of the spectrum is defined such that
\begin{equation}
\mathcal{P}_{\zeta}\left(k\right)=\mathcal{P}_{\zeta}\left(k_{p}\right)\left(\frac{k}{k_{p}}\right)^{n_{s}-1},
\end{equation}
where $k_{p}$ is some arbitrary pivot scale.  Note that the quantity
$A_s$ appearing in the introduction coincides with $\mathcal
P_\zeta(k_p)$, and the pivot scale used by the {\it Planck}
collaboration is $k_p = 0.05\,{\rm Mpc ^{-1}}$.

In the case of the bispectrum, the different configurations of $k_{1}$,
$k_{2}$ and $k_{3}$ can be described in terms of the overall scale
$K=k_{1}+k_{2}+k_{3}$ and the relative magnitudes of the $k_{i}$'s,
where $i=1,2,3$. In our case we will be interested in the so-called
squeezed limit, where one momentum is much smaller than the other
two. Without loss of generality we take $k_{1}\ll k_{2}\simeq k_{3}$,
where the near equality of $k_{2}$ and $k_{3}$ follows from the
homogeneity requirement $\sum_{i}\boldsymbol{k}_{i}=0$. On introducing
$k_{\l}=k_{1}$ and $k_{\s}=k_{2}\simeq k_{3}$, where the subscripts
$\l$ and $\s$ label ``long'' and ``short'' respectively, we
can parametrise the bispectrum in the squeezed configuration in terms
of the two parameters $k_{\s}$ and 
\begin{equation}
R_{{\rm sq}}\equiv\frac{k_{\l}}{k_{\s}}
\end{equation}
where $R_{{\rm sq}}$ parameterises the level of squeezing. Note that
$K\simeq2k_{\s}$. Following \cite{Dias:2013rla},
we define the reduced bispectrum in the squeezed configuration as
\begin{equation}
B_{\zeta}(k_{\s},R_{{\rm sq}})=\frac{1}{R_{{\rm sq}}^{3}k_{\s}^{6}}\mathcal{B}_{\zeta}(k_{\s},R_{{\rm sq}}),\qquad\mathcal{B}_{\zeta}(k_{\s},R_{{\rm sq}})=\mathcal{B}_{\zeta}(k_{p})\left(\frac{k_{\s}}{k_{p}}\right)^{n_{B}}R_{{\rm sq}}^{n_{{\rm sq}}}.
\end{equation}
$n_{B}$ thus gives the dependence of the reduced bispectrum on the overall
scale, while $n_{{\rm sq}}$ gives the dependence on the squeezing
parameter. The parameter $n_{{\rm sq}}$ can be related to the tilt of
the scale dependent \mk{halo} bias, $n_{\delta b}$, as $n_{\delta b}=n_{{\rm
sq}}-n_{\zeta}$ \cite{Dias:2013rla}, where we have
introduced \mk{$n_{\zeta}\equiv n_{s}-1$}.

In the following we will also make reference to the power spectrum
of tensor fluctuations, which is parameterised in a similar way to
the scalar power spectrum. Focussing on the reduced power spectrum
we have 
\begin{equation}
\mathcal{P}_{T}(k)=\mathcal{P}_{T}\left(k_{p}\right)\left(\frac{k}{k_{p}}\right)^{n_{T}},
\end{equation}
and the tensor-to-scalar ratio is defined as
\begin{equation}
r \equiv \frac{P_T(k_p)}{P_\zeta(k_p)}.
\end{equation}

Having outlined the general parametrisation of the two- and three-point
correlation functions, we now turn to the specific case of the $\delta N$
expansion as discussed in Section~\ref{sec:sepuniv}. Given that we are assuming slow-roll to be satisfied at
horizon crossing, the $\delta N$ expansion can be written purely
in terms of the initial field values. Moving to Fourrier space we
have 
\mk{
\begin{equation}
\zeta\left(\boldsymbol{k}\right)=N_{,A}\delta\phi^{A}\left(\boldsymbol{k}\right)+\frac{1}{2}N_{,AB}\frac{1}{(2\pi)^{3}}\int d^{3}q\delta\phi^{A}\left(\boldsymbol{q}\right)\delta\phi^{B}\left(\boldsymbol{k}-\boldsymbol{q}\right)
\end{equation}
}
If we then introduce the parametrisation 
\mk{
\begin{align}
\langle\delta\phi^{A}\left(\boldsymbol{k}_{1}\right),\delta\phi^{B}\left(\boldsymbol{k}_{2}\right)\rangle & =\left(2\pi\right)^{3}\delta^{(3)}\left(\boldsymbol{k}_{1}+\boldsymbol{k}_{2}\right)\Sigma^{AB}\left(k_{1}\right),\label{SIJ-def}\\
\langle\delta\phi^{A}\left(\boldsymbol{k}_{1}\right),\delta\phi^{B}\left(\boldsymbol{k}_{2}\right),\delta\phi^{C}\left(\boldsymbol{k}_{3}\right)\rangle & =\left(2\pi\right)^{3}\delta^{(3)}\left(\boldsymbol{k}_{1}+\boldsymbol{k}_{2}+\boldsymbol{k}_{3}\right)B^{ABC}\left(k_{1},k_{2},k_{3}\right),
\end{align}
}
then we find that the power spectrum and bispectrum for $\zeta$ are
given as
\mk{ 
\begin{align}
P_{\zeta}\left(k\right) & =N_{,A}N_{,B}\Sigma^{AB}\left(k\right),\label{powgen}\\
B_{\zeta}\left(k_{1},k_{2},k_{3}\right) & =N_{,A}N_{,B}N_{,C}B^{ABC}\left(k_{1},k_{2},k_{3}\right)+N_{,AB}N_{,C}N_{,D}\left(\Sigma^{AC}\left(k_{1}\right)\Sigma^{BD}\left(k_{2}\right)+{\rm c.p.}\right).\label{bigen}
\end{align}
}
From this we see that there are essentially two contributions to the
bispectrum: one is the intrinsic non-Gaussianity of the initial field
perturbations just after horizon crossing, and the second comes from
the non-linear dependence of the local expansion on the initial field
values. In the squeezed limit, the first contribution is known to
be unobservably small \cite{Seery:2005gb,Lyth:2005qj}, so it
can be neglected compared to the second term in models where an observable
non-Gaussianity is generated in the squeezed limit.

As is evident in eqs.~\eqref{powgen} and \eqref{bigen}, the calculation
of the correlation functions of $\zeta$ can essentially be divided into
two parts: finding the dependence of the local expansion on the initial
conditions \textendash{} encoded in \mk{$N_{,A}$ and $N_{,AB}$} \textendash{}
and determining the correlation functions of perturbations in the
initial conditions themselves \textendash{} encoded in
\mk{$\Sigma^{AB}\left(k\right)$ and
$B^{ABC}\left(k_{1},k_{2},k_{3}\right)$}. In this sense, we can also
divide the slow-roll corrections to $P_{\zeta}\left(k\right)$ and
$B_{\zeta}\left(k_{1},k_{2},k_{3}\right)$ into two categories: the
slow-roll corrections to \mk{$N_{,A}$ and $N_{,AB}$} and the slow-roll
corrections to \mk{$\Sigma^{AB}\left(k\right)$ and
$B^{ABC}\left(k_{1},k_{2},k_{3}\right)$}.  The methods to determine
\mk{$N_{,A}$ and $N_{,AB}$} in terms of $N_{,z}$, $N_{,zy}$, $\varphi^z_{,A}$
and $\varphi^z_{,AB}$ under various levels of slow-roll approximation
have already been outlined in Section~\ref{sec:sepuniv}.  In the following we will
use eq.~\eqref{dphi-star} to determine leading and next-to-leading order
expressions for \mk{$\Sigma^{AB}(k)$}, and we will then show that the
corrections to the power spectrum and bispectrum of $\zeta$ associated
with the slow-roll corrections to \mk{$\Sigma^{AB}\left(k\right)$} can be
written in a relatively compact form.\footnote{As we neglect the
intrinsic contribution to the bispectrum, it is only necessary to
consider slow-roll corrections to \mk{$\Sigma^{AB}\left(k\right)$}.}

\subsection{Slow-roll corrections from the initial correlation functions}

\subsubsection{The power spectrum}\label{ICSRCpow}

Using eq.~\eqref{dphi-star} for \mk{$\delta\phi^A(\bm k_i)$} and the
definition of \mk{$\Sigma^{AB}$} in eq.~\eqref{SIJ-def} we find
\mk{
\begin{equation}
\Sigma^{AB}\left(N_{\ast},k\right)=\frac{H^{2}\left(N_{\ast}\right)}{2k^{3}}\left\{ \left(1-2\epsilon\right)\delta^{AB}+2\epsilon^{AB}\left[C+\ln\left(\frac{a\left(N_{\ast}\right)H\left(N_{\ast}\right)}{k}\right)\right]\right\} ,\label{sigma}
\end{equation}
}
where we temporarily write the argument $N_{*}$ explicitly to
emphasise that this expression is evaluated a few e-foldings after
the mode $k$ leaves the horizon. Using the above result we can read
off the tilt of the reduced power spectrum of $\zeta$ at leading order in slow-roll as
\mk{
\begin{equation}
\tilde n_{\zeta}\equiv\tilde n_{s}-1=-\frac{2\epsilon^{AB}N_{,A}N_{,B}}{N_{,C}N^{,C}} = -2\epsilon - \frac{2U^{(0)}_{AB}N^{,A}N^{,B}}{N_{,C}N^{,C}}.\label{nsleading}
\end{equation}
}
Note that here and in the following we mean ``leading order'' in the
sense that higher order slow-roll corrections to the initial correlation
functions \mk{$\Sigma^{AB}(k)$} have been neglected, which amounts to
assuming that the slow-roll approximation holds around horizon crossing.
The quantities \mk{$N_{,A}$} appearing in the above expression, however, can
still be calculated using the various levels of slow-roll approximation
outlined in Section~\ref{sec:sepuniv}.  In particular, this allows for the
possibility that the slow-roll approximation may break down at some point
during the super-horizon evolution.  We use a tilde to denote a quantity
that has been calculated to leading-order in the above sense.
 
Taking into account the fact that to leading order in slow-roll we have
$ n_{T}^{(0)}=-2\epsilon$, see e.g. \cite{Bassett:2005xm}, we can
thus express the power spectrum as
\mk{
\begin{equation}
\mathcal{P}_{\zeta}\left(N,k\right)=\left(\frac{H\left(N_{\ast}\right)}{2\pi}\right)^{2}N_{,A}\left(N,N_{\ast}\right)N^{,A}\left(N,N_{\ast}\right)\left\{ 1+ n_{T}^{(0)}-\tilde n_{\zeta}\left[C+\ln\left(\frac{a\left(N_{\ast}\right)H\left(N_{\ast}\right)}{k}\right)\right]\right\}, \label{pspec}
\end{equation}
}
where we have included the $N$- and $N_{\ast}$-dependencies explicitly
in order to aid the proceeding discussion.  As it stands, the right-hand
side of equation \eqref{pspec} appears to depend on $N_{\ast}$.  Indeed,
recall that the validity of eq.~\eqref{dphi-star}, which we have made
use of here, requires that $N_\ast$ be sufficiently late after horizon
crossing that gradient terms can be neglected, but not so late that the
assumption of constant $\epsilon$ and \mk{$\epsilon_{AB}$} breaks down.
However, as was discussed after eq.~\eqref{z-phi}, the value of $\zeta$
-- and hence $\mc P_\zeta$ -- does not depend on which flat slice we use
to evaluate \mk{$\delta\phi^{A}(\bm k)$}, i.e. it does not depend on the
exact choice of $N_\ast$.  Indeed, one can explicitly show that the
$N_\ast$-dependencies of the various terms on the right hand side of
\eqref{pspec} exactly cancel \cite{Nakamura:1996da}.  As such, we can
freely choose $N_{\ast}$, and so in order to remove the log term in the
square brackets we take $N_{\ast}=N_{k}$, where $N_{k}$ is the time at
which the mode $k$ left the horizon, namely
$k=a\left(N_{k}\right)H\left(N_{k}\right)$.  We thus obtain
\mk{
\begin{align}
\mathcal{P}_{\zeta}\left(N,k\right) &
=\tilde{\mathcal{P}}_{\zeta}\left(N,k\right)\left[1+n_{T}^{(0)}-\tilde
n_{\zeta}C\right],\label{powcompact}\\
\tilde{\mathcal{P}}_{\zeta}\left(N,k\right) &
=\left(\frac{H\left(N_{k}\right)}{2\pi}\right)^{2}N_{,A}\left(N,N_{k}\right)N^{,A}\left(N,N_{k}\right),\label{pow-leading}
\end{align}
}
where the terms in the square brackets correspond to the slow-roll
corrections arising from the corrections to the initial power spectra of
$\delta\phi^{A}$.  It is perhaps worth noting one subtlety, that
although we have taken $N_\ast = N_k$ in the above expression, we must
still take the final time $N$ to be at least a few e-foldings after
$N_k$ for the resulting expression to be valid.  This is because we must
wait for the decaying gradient terms that have been neglected in
deriving \eqref{powcompact} to become sufficiently small.\footnote{For
discussions regarding this issue in the single-field case see \cite{Nalson:2011gc}.}
The above form for the power spectrum and its slow-roll corrections has
also been derived using different methods in \cite{vanTent:2003mn,Byrnes:2006fr}.

We see from eq.~\eqref{powcompact} that slow-roll corrections to the
power spectrum arising from slow-roll corrections to \mk{$\Sigma^{AB}$} are
directly related to the observables $n_{T}$ and $n_{\zeta}$.  Moreover,
from eq.~\eqref{nsleading} we see that $\tilde n_\zeta = n_T^{(0)} +
... $, such that barring exact cancellation between terms, we expect
$|n_T^{(0)}|\lesssim |\tilde n_\zeta|$.  Given that $C\simeq 0.73$ and
$\tilde n_\zeta \simeq -0.032 $, this suggests that the slow-roll
corrections to $\mc P_\zeta$ should be $\lesssim \mc O(5\%)$ for any
model that gives an observationally viable value for $n_\zeta$.

\subsubsection{The spectral tilt}

One can now proceed to determine
the spectral tilt up to second order in slow-roll by using the relation 
\begin{equation}
 \frac{\mathrm{d}}{\mathrm{d}\ln k}
  \simeq (1+\epsilon)\frac{d}{dN_k},
\end{equation}
and one obtains 
\begin{align}
 n_s-1 = (1+\epsilon)\frac{d\ln \tilde{\mc P}_\zeta }{dN_k} -
 2\epsilon\eta - C\tilde \alpha_s,\label{tiltcompact}
\end{align}
where
\begin{align}
 \tilde \alpha_s &= \frac{d\tilde n_s}{d\ln k} \simeq
 \frac{d\tilde n_s}{dN_k}=
 -2\epsilon\eta-2U^{(0)}_{,C}U^{(0)}_{,CAB}M^{AB} + 4\epsilon^{AC}\epsilon^{BC}M_{AB} -
 \tilde n_\zeta^2\label{talpha}
\end{align}
and for brevity we have introduced
\begin{align}
  M_{AB} = \frac{N_{,A}(N,N_k)N_{,B}(N,N_k)}{N_{,C}(N,N_k)N_{,C}(N,N_k)}.
\end{align}
The above next-to-leading order expression for $n_s-1$ is again in
agreement with that given in \cite{vanTent:2003mn}. In eq.~\eqref{tiltcompact} the last two terms arise from slow-roll corrections to $\Sigma^{AB}$.  The term $-2\epsilon\eta$  actually coincides with the leading-order expression for the running of the tensor tilt, namely $dn_T/d\ln k = -2\epsilon \eta$ \cite{Ade:2015lrj}.  However, this term also appears in the leading-order expression for $\alpha_s$ given in eq.~\eqref{talpha}, where we have $\tilde\alpha_s = -2\epsilon \eta + ...$.  As such, barring exact cancellation amongst terms, we expect $|-2\epsilon\eta|\lesssim |\tilde\alpha_s|$, and the current $95\%$ confidence limit for $\alpha_s$ given in \cite{Ade:2015lrj} is
 \begin{equation}
  \alpha_s = -0.003 \pm 0.015.
 \end{equation}
This means that for models that satisfy observational constraints on $\alpha_s$, the last two terms in eq.~\eqref{tiltcompact} are expected to represent corrections to $n_s$ of order $10^{-2}$, which are thus likely to be important given that this is of the same order as the current $2\sigma$ bounds on $n_s$.

In computing $d\ln \tilde{\mc P}_\zeta/dN_k$ in the above expression, we
must be careful to work to next-to-leading order accuracy in the
slow-roll approximation.  Explicitly, one has
\begin{align}
 \frac{d\ln \tilde{\mc P}_\zeta}{dN_k}= -2\epsilon -2 M^{AB}\frac{\partial}{\partial\phi_A}(\phi'_B).
\end{align}
To leading order we usually use the fact that $\phi'_B = U^{(0)}_{,B}$,
which leads to the result given in eq.~\eqref{nsleading} for $\tilde
n_\zeta$, but to next-to-leading order we have $\phi'_B = U^{(1)}_{,B} =
U^{(0)}_{,B} - U^{(0)}_{,BA}U^{(0)}_{,A}/3$.  We thus obtain
\begin{align}
 n_s-1 = (1+\epsilon)\left[-2\epsilon - 2 M_{AB} U^{(1)}_{,AB}\right] -
 2\epsilon\eta - C\tilde \alpha_s.
\end{align}
If we expand everything in terms
of derivatives of $U^{(0)}$, we eventually obtain
\mk{
\begin{align}
 \nonumber n_s - 1 &= -U^{(0)}_{,A} U^{(0)}_{,A} - 2M^{AB}U^{(0)}_{,AB} \\
 &\quad - \frac{1}{2}(U^{(0)}_{,A}U^{(0)}_{,A})^2 +
 \frac{2}{3}U^{(0)}_{,A}U^{(0)}_{,AB}U^{(0)}_{,B}(3C-2) - U^{(0)}_{,C}U^{(0)}_{,C}M^{AB}U^{(0)}_{,AB} \\ 
\nonumber & \quad +
 4C(M^{AB}U^{(0)}_{,AB})^2+\frac{2}{3}M^{AB}U^{(0)}_{,AC}U^{(0)}_{,BC}(1-6C) + \frac{2}{3}M^{AB}U^{(0)}_{,ABC}U^{(0)}_{,C}(1+3C),
\end{align}
}
where terms on the first line correspond to the leading-order result and
those on the second and third lines represent next-to-leading order
corrections.  This result is in agreement with the expression of
Nakamura \& Stewart \cite{Nakamura:1996da}.

\subsubsection{Tensor to scalar ratio}

Although we will not derive the result here, the power spectrum for
tensor modes to next-to-leading order in slow-roll is given as \cite{Stewart:1993bc}
\begin{equation}
\mathcal{P}_{T}\left(N,k\right)=8\left(\frac{H\left(N_{k}\right)}{2\pi}\right)^{2}\left[1-n_{T}^{(0)}\left(C-1\right)\right],
\end{equation}
which on combining with the expression for $\mathcal{P}_{\zeta}\left(N,k\right)$
allows us to write the next-to-leading order expression for the tensor-to-scalar ratio as
\begin{align}
r &= \tilde r\left[1+C\left(\tilde n_{\zeta}-n_{T}^{(0)}\right)\right],\label{rel1}\\ 
\tilde r &= \frac{8}{N_{,I}\left(N,N_{k}\right)N^{,I}\left(N,N_{k}\right)}.\label{r-leading}
\end{align}
\mk{}
As in the case of the scalar power spectrum, the corrections to $r$ shown in eq.~\eqref{rel1} are given in terms of $n_\zeta$ d $n_T$.  As such, for models that give observationally allowed values of $n_s$, we expect slow-roll corrections to $r$ to be less than $5\%$.

\subsubsection{The bispectrum}

\mk{}

Having discussed the power spectrum, let us now turn to the bispectrum.
Focussing on the second term appearing in eq.~\eqref{bigen} and
making use of eq.~\eqref{sigma} we obtain 
\mk{
\begin{align}
B_{\zeta}(N,k_{L},k_{S}) &
 =N_{,AB}(N,N_{\ast})N_{,C}(N,N_{\ast})N_{,D}(N,N_{\ast})\frac{H^{4}(N_{\ast})}{4k_{S}^{3}k_{L}^{3}}\\\nonumber
 & \quad\times\left\{
 2\left[\delta^{AC}\delta^{BD}-2(\epsilon\delta^{AC}-2C\epsilon^{AC})\delta^{BD}+2\epsilon^{AC}\delta^{BD}\ln\left(\frac{a^{2}\left(N_{\ast}\right)H^{2}\left(N_{\ast}\right)}{k_{L}k_{S}}\right)\right]\right.\\\nonumber
 &
 \quad\left.+\frac{k_{L}^{3}}{k_{S}^{3}}\left[\delta^{AC}\delta^{BD}-2(\epsilon\delta^{AC}-2C\epsilon^{AC})\delta^{BD}+2\epsilon^{AC}\delta^{BD}\ln\left(\frac{a^{2}\left(N_{\ast}\right)H^{2}\left(N_{\ast}\right)}{k_{S}^{2}}\right)\right]\right\}
 ,
\end{align}
}
where we show the dependencies on $N$ and $N_{\ast}$ explicitly.
Given that $k_{L}/k_{S}\ll1$ we can neglect the terms on the third
line, and it is then possible to read off the quantities 
\mk{
\begin{equation}
\tilde n_{{\rm sq}}=\frac{\tilde n_{B}}{2}=-\frac{2N_{,AB}N_{,C}N_{,D}\epsilon^{AC}\delta^{BD}}{N_{,AB}N^{,A}N^{,B}},
\end{equation}
}
which in turn allows us to express the slow-roll corrections to $\mathcal{B}_{\zeta}\left(k_{S},R_{{\rm sq}}\right)$
concisely as 
\mk{
\begin{align}
\mathcal{B}_{\zeta}\left(N,R_{{\rm sq}},k_{S}\right) &
=\tilde{\mathcal{B}}_{\zeta}\left(N,N_\ast\right)\left\{
1+n_{T}^{(0)}+\tilde n_{{\rm sq}}\left[\ln\left(R_{{\rm
sq}}\right)+2\ln\left(\frac{k_{S}}{a\left(N_{\ast}\right)H\left(N_{\ast}\right)}\right)-2C\right]\right\}
\\ \tilde{\mathcal{B}}_{\zeta}\left(N,N_\ast\right) &
=\frac{1}{2}N_{,AB}(N,N_{\ast})N_{,C}(N,N_{\ast})N_{,D}(N,N_{\ast})H^{4}(N_{\ast})\delta^{AC}\delta^{BD}.
\end{align}
}
As with the power spectrum, in its current form the bispectrum looks to
be dependent on $N_{\ast}$. However, for exactly the same reason as with
the power spectrum, this apparent $N_\ast$-dependence is in fact
spurious, and we are free to choose $N_\ast$ as we like.\footnote{More
precisely, the total bispectrum is independent of $N_\ast$, whereas the
intrinsic and super-horizon contributions are not individually
$N_\ast$-independent.  See Appendix~\ref{Sec:Nastind} for more details.} Taking
$N_\ast$ to coincide with the horizon-crossing time of the short mode
$k_S$, namely $N_\ast = N_{k_S}$, where $k_S = a(N_{k_S})H(N_{k_S})$,
the above expression for $\mc B_\zeta$ simplifies to
\begin{align}
 \mathcal{B}_{\zeta}\left(N,R_{{\rm sq}},k_{S}\right) & =\tilde{\mathcal{B}}_{\zeta}\left(N,N_{k_S}\right)\left\{ 1+n_{T}^{(0)}+\tilde n_{{\rm sq}}\left[\ln\left(R_{{\rm sq}}\right)-2C\right]\right\}.\label{Bsimp}
\end{align}

Defining the quantity
$D_\zeta\left(k_{1},k_{2},k_{3}\right)=P_\zeta\left(k_{1}\right)P_\zeta\left(k_{2}\right)+{\rm c.p.}$, we find that it similarly can be expanded to
next-to-leading order as
\mk{
\begin{align}
D\left(N,R_{{\rm sq}},k_{S}\right) & =\tilde D\left(N,N_{k_S},R_{{\rm sq}},k_{S}\right)\left\{ 1+2\left(n_{T}^{(0)}-\tilde n_{\zeta}C\right)+\tilde n_{\zeta}\ln\left(R_{{\rm sq}}\right)\right\} \label{Dsimp},\\
\tilde D\left(N,N_{k_S},R_{{\rm sq}},k_{S}\right) & =\frac{H^{4}\left(N_{k_S}\right)\left[N_{,A}\left(N,N_{k_S}\right)N^{,A}\left(N,N_{k_S}\right)\right]^{2}}{2R_{{\rm sq}}^{3}k_{S}^{6}},
\end{align}
}
where we have once again exploited our freedom to choose $N_\ast =
N_{k_S}$.  As such, combining \eqref{Bsimp} and \eqref{Dsimp} we find
that $f_{NL}$ can be expanded as
\mk{
\begin{gather}\label{fNLcompact}
f_{NL}\left(N,R_{{\rm sq}},k_{S}\right)=\tilde
f_{NL}\left(N,N_{k_{S}}\right)\left\{ 1-n_{T}^{(0)}-\tilde n_{\delta
b}\left[2C-\ln\left(R_{{\rm sq}}\right)\right]\right\},\\ \tilde
f_{NL}\left(N,N_{k_S}\right)=\frac{5}{6}\frac{N_{,AB}\left(N,N_{k_S}\right)N^{,A}\left(N,N_{k_S}\right)N^{,B}\left(N,N_{k_S}\right)}{\left[N_{,C}\left(N,N_{k_S}\right)N^{,C}\left(N,N_{k_S}\right)\right]^{2}}.\label{fNL-leading}
\end{gather}
}
As with the power spectrum, we see that slow-roll corrections arising
from slow-roll corrections to the initial correlation functions
\mk{$\Sigma^{AB}$} are directly related to observables, and are thus in
principle constrained. However, we are not aware of any constraints on $n_{\delta b}$ at the present time, which means that 
the size of the slow-roll corrections to
$f_{NL}$ are not as tightly constrained as those to $\mc P_\zeta$, $n_s$ and $r$.

It is important to mention that there are some limits on the amount of
squeezing for which our expressions are valid. If $R_{{\rm sq}}$ is too
small then the time between horizon exit of the modes $k_{L}$ and
$k_{S}$ becomes too long, such that the assumption of constant
$\epsilon$ and \mk{$\epsilon_{AB}$} over this period breaks down. On the
other hand, if $R_{{\rm sq}}$ is not small enough, then the corrections
suppressed by $(k_{L}/k_{S})^{3}$ will become comparable to the
slow-roll corrections we are considering here and will thus need to be
properly taken into account. The case of extreme squeezing has been
considered by Kenton \& Mulryne~\cite{Kenton:2015lxa},
and they found that \mk{for highly squeezed configurations} corrections to the above formulae for
$f_{NL}$ and $n_{{\rm sq}}$ could be on the order of
20\%.

\section{Example Models}\label{Sec:examples}

In this section we consider some example models.  In doing so, there are
five different levels of slow-roll approximation that we consider when
calculating the curvature perturbation and its spectral properties, and
we label them SR0, SR1, HC0, HC1 and NUM.  The details of each are as
follows:
\begin{description}
 \item[SR0] In this case we assume that the leading-order slow-roll
	    approximation holds throughout inflation.  Correspondingly,
	    at background level we solve the equations of motion given
	    in eq.~\eqref{EoM0}.  For the quantities $N_{,Z}$ and
	    $N_{,ZY}$ we use the expressions given in eqs.~\eqref{Nz-SR}
	    and \eqref{Nzy-SR}, taking the index $i = 0$.  For the
	    quantities $\phi_{Z,A}$ and $\phi_{Z,AB}$ we solve
	    eqs.~\eqref{djzda-SR} and \eqref{djzdab-SR} numerically,
	    again taking the index $i = 0$.  Then, for $\mc P_\zeta$,
	    $n_s-1$, $r$ and $f_{NL}$ we use the expressions based on
	    the leading-order results for the correlation functions of
	    the initial field perturbations just after horizon crossing,
	    namely eqs.~\eqref{pow-leading}, \eqref{nsleading},
	    \eqref{r-leading} and \eqref{fNL-leading} respectively.
	    
 \item[SR1] In this case we assume that the next-to-leading order
	    slow-roll approximation holds throughout inflation.
	    Correspondingly, at background level we solve the equations
	    of motion given in eq.~\eqref{EoM1}.  For the quantities
	    $N_{,Z}$ and $N_{,ZY}$ we use the expressions given in
	    eqs.~\eqref{Nz-SR} and \eqref{Nzy-SR}, but now taking the
	    index $i = 1$.  For the quantities $\phi_{Z,A}$ and
	    $\phi_{Z,AB}$ we similarly solve eqs.~\eqref{djzda-SR} and
	    \eqref{djzdab-SR} numerically, now taking the index $i = 1$.
	    Then, for $\mc P_\zeta$, $n_s-1$, $r$ and $f_{NL}$ we use
	    the expressions based on the next-to-leading order results
	    for the correlation functions of the initial field
	    perturbations just after horizon crossing, namely
	    eqs.~\eqref{powcompact}, \eqref{tiltcompact}, \eqref{rel1}
	    and \eqref{fNLcompact} respectively.
	    
 \item[HC0] In this case we assume that the leading-order slow-roll
	    approximation holds around the time of horizon crossing, but
	    we allow for the possibility that slow-roll breaks down later
	    on during the super-horizon dynamics.  Correspondingly, at
	    background level we solve the full equations of motion given
	    in eq.~\eqref{EoMN}, but with the initial conditions for
	    $\phi_I'$ taken to be $\phi_I'(N_k) = U^{(0)}_{,I}(N_k)$.  For
	    $N_Z$, $N_{,Z'}$, $N_{,ZY}$, $N_{,ZY'}$ and $N_{,Z'Y'}$ we
	    use the results given in
	    \mk{eqs.~\eqref{NI-rez2}--\eqref{NdIdJ-rez2}}.  For the
	    quantities $\varphi^z_{,A}$ and $\varphi^z_{,AB}$ we
	    numerically solve eqs.~\eqref{djzda} and \eqref{djzdAB}.
	    Finally for $\mc P_\zeta$, $n_s-1$, $r$ and $f_{NL}$ we use
	    the expressions based on the leading-order results for the
	    correlation functions of the initial field perturbations
	    just after horizon crossing, namely
	    eqs.~\eqref{pow-leading}, \eqref{nsleading},
	    \eqref{r-leading} and \eqref{fNL-leading} respectively.
	    
 \item[HC1] In this case we assume that the next-to-leading order
	    slow-roll approximation holds around the time of horizon
	    crossing, but allow for the possibility that slow-roll breaks
	    down later on during the super-horizon dynamics.
	    Correspondingly, at background level we solve the full
	    equations of motion given in eq.~\eqref{EoMN}, but with the
	    initial conditions for $\phi_I'$ taken to be $\phi_I'(N_k) =
	    U^{(1)}_{,I}(N_k)$.  For $N_Z$, $N_{,Z'}$, $N_{,ZY}$, $N_{,ZY'}$
	    and $N_{,Z'Y'}$ we again use the results given in
	    \mk{eqs.~\eqref{NI-rez2}--\eqref{NdIdJ-rez2}}.  For the
	    quantities $\varphi^z_{,A}$ and $\varphi^z_{,AB}$ we
	    numerically solve eqs.~\eqref{djzda} and \eqref{djzdAB}.
	    Finally for $\mc P_\zeta$, $n_s-1$, $r$ and $f_{NL}$ we use
	    the expressions based on the next-to-leading order results
	    for the correlation functions of the initial field
	    perturbations just after horizon crossing, namely
	    eqs.~\eqref{powcompact}, \eqref{tiltcompact}, \eqref{rel1}
	    and \eqref{fNLcompact} respectively.

 \item[NUM] In this case we use the code {\it PyTransport} of Mulryne '16 \cite{Mulryne:2016mzv}
	    , which employs a transport method
	    that evolves perturbations from deep inside the horizon \cite{Mulryne:2013uka}.  In
	    doing so, it avoids the need to use the slow-roll
	    approximation around horizon crossing, and in this sense it
	    represents the ``exact'' result to which the other
	    approximations can be compared.  We only ever use this code
	    to calculate $\mc P_\zeta$ and $f_{NL}$.
	    
\end{description}

\subsection{Double quadratic potential}

We begin by considering the double quadratic potential 
\begin{equation}
 V = \frac{1}{2}m^2\left(\phi^2 + R^2\chi^2\right).
\end{equation}
The use of $R$ when writing down the potential allows us to
factor out the overall mass scale $m^2$.  Seeing as the equations of
motion for the fields, given in eq.~\eqref{EoMN}, only contain the ratio
$V_{,I}/V$, we find that the factor $m^2$ drops out, and so does not
affect the dynamics.  Consequently, the mass scale $m$ is only important
when it comes to determining the overall magnitude of the power
spectrum, as $\mc P_\zeta\propto H_k^2 = V_k/(3-\epsilon_k)\propto m^2$.
In particular, $n_s-1$, $r$ and $f_{NL}$ do not depend on $m$.

Following e.g. \cite{Dias:2011xy} we choose $R = 9$. The {\it
PyTransport} code used to determine the results labelled NUM requires us
to set initial conditions well before horizon crossing. In our case we
use the initial conditions $(\phi,\chi,\phi',\chi') = (12.94,9.9,0,0)$, and then consider the mode that leaves the horizon
$8.2$ e-foldings after these initial conditions are set.  When we
calculate $f_{NL}$, the mode $k_S$ is taken to leave at this time.  We
also choose $R_{\rm sq} = 0.1$, meaning that the mode $k_L$ left the
horizon approximately $5.9$ e-foldings after the initial conditions were
set.  Solving the full background equations, we determine that after
$8.2$ e-folds the field values are $(\phi_k,\chi_k) \simeq (12.91, 8.22)$,
and we use these values as the initial conditions for calculating the
results corresponding to approximations SR0, SR1, HC0 and HC1. 

\begin{figure}
\begin{subfigure}[t]{0.49\textwidth}
 \includegraphics[width = 0.9\columnwidth]{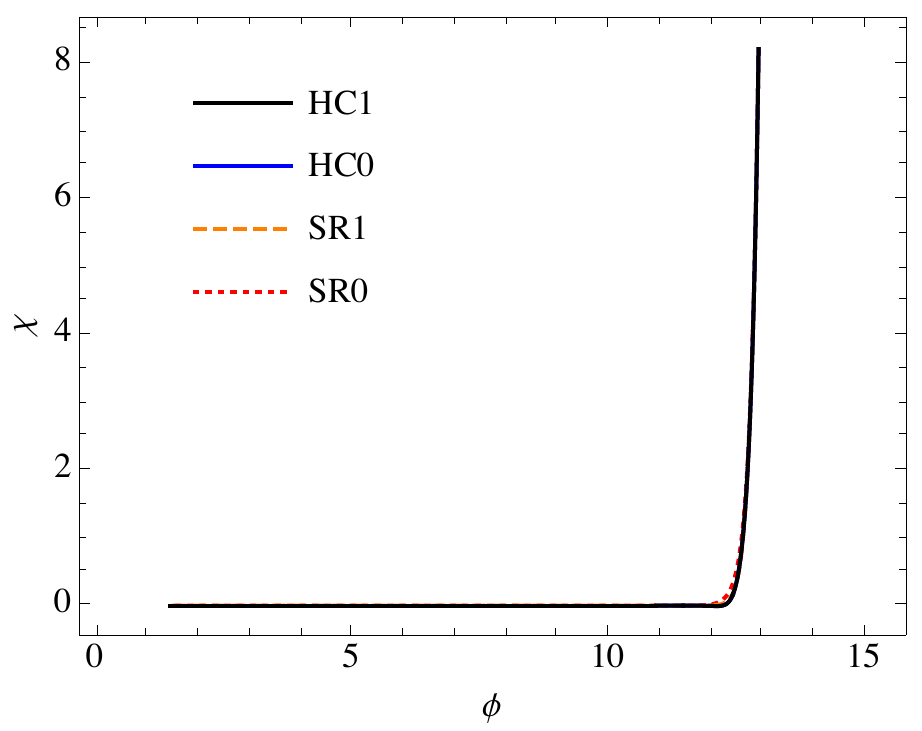}
 \caption{\label{dquad_bgtraj}}
\end{subfigure}
\begin{subfigure}[t]{0.49\textwidth}
 \includegraphics[width = 0.9\columnwidth]{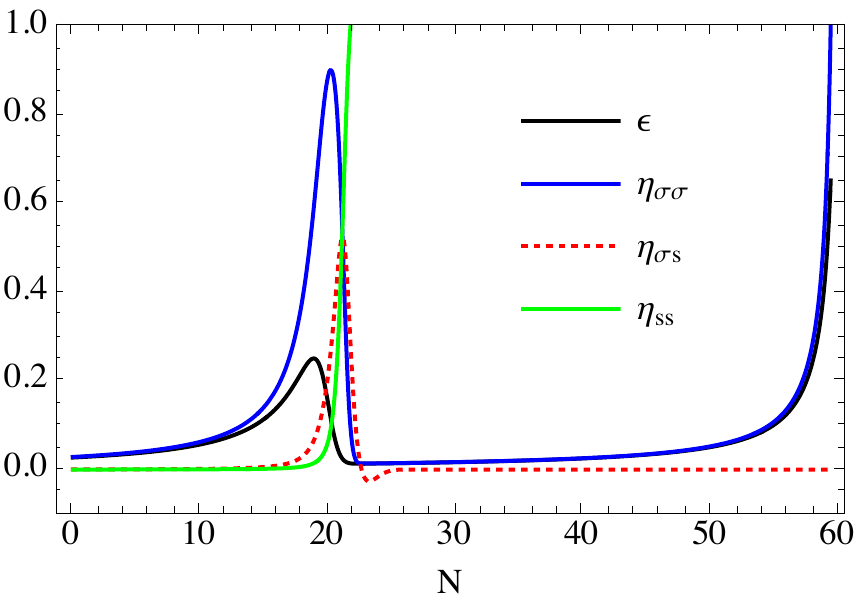}
 \caption{\label{dquad_srparams}}
\end{subfigure}
 \caption{\label{dquad_bg}(a) A plot of the background trajectory in
 field space for the double quadratic potential outlined in the text,
 with $R = 9$ and $(\phi_k,\chi_k)\simeq (12.91, 8.22)$.  Trajectories
 as determined under the four different levels of slow-roll
 approximation SR0, SR1, HC0 and HC1 are shown. (b) Evolution of the
 slow-roll parameters as a function of $N$, where $N$ is the number of
 e-foldings after the scale under consideration left the horizon.  Here
 the full background equations of motion are used, with the initial
 velocities taken to be $\phi'_I = U^{(1)}_{,I}$.}
\end{figure}

Plots of the background trajectory and the corresponding evolution of
the slow-roll parameters are given in Figure \ref{dquad_bg}.  The
e-folding number displayed in all figures is the number of e-foldings
after the horizon crossing time of the scale being considered, and for our choice of parameters we find that there are approximately $60$ e-foldings of inflation after horizon crossing. We always take the end of inflation to be defined as when $\epsilon^{(0)} = 1$. As can be seen from Figure~\ref{dquad_bgtraj}, the trajectory evolves to the minimum of the $\chi$ potential before the end
of inflation, and so we can expect that $\zeta$ becomes constant.

In Figure~\ref{dquad_srparams} we plot the evolution of the slow-roll
parameters for the trajectory under consideration.  Here $\epsilon$ is
as defined in eq.~\eqref{eps-eta-def}, while $\eta_{\sigma\sigma}$,
$\eta_{\sigma s}$ and $\eta_{ss}$ correspond to projections of
$\eta_{IJ}$ -- as defined in eq.~\eqref{etaIJdef} -- on to the
kinematic basis vectors $e^I_\sigma$ and $e^I_s$, which lie parallel and
perpendicular to the background trajectory respectively.  As an example,
we have $\eta_{\sigma s} = e^I_\sigma\eta_{IJ}e^J_s$.  It is helpful to
use these projections of $\eta_{IJ}$ as at lowest order in slow-roll
they are easy to interpret physically.  In particular, one finds that
$\eta\simeq -2\eta_{\sigma\sigma} + 4\epsilon^{(0)}$ and $\theta' \simeq
- \eta_{\sigma s}$, where $\theta$ is the angle between the tangent of
the trajectory and the $\phi$ axis \cite{Gordon:2000hv}.  As such,
$\eta_{\sigma\sigma}$ is related to the rate at which $\epsilon$
evolves, while $\eta_{\sigma s}$ is related to the rate at which the
trajectory turns.
Whether or not the trajectory turns is important when considering the temporal evolution of $\zeta$.  In the presence of so-called isocurvature perturbations, namely field fluctuations perpendicular to the background trajectory, it is known that the curvature perturbation $\zeta$ will be sourced by these isocurvature perturbations when the trajectory turns in field space.  This means that non-zero values of $\eta_{\sigma s}$ will in general be associated with a non-conservation of $\zeta$ \cite{Gordon:2000hv}.
  The final quantity $\eta_{ss}$ is related to the mass
of the direction in field space that is perpendicular to the trajectory.
If it is positive then we can expect neighbouring trajectories to
converge, while if it is negative we can expect them to diverge. In this
example we see that the quantities $\epsilon$, $\eta_{\sigma\sigma}$ and
$\eta_{\sigma s}$ remain less than unity right up until the end of
inflation.  The oscillatory feature in $\eta_{\sigma s}$ at around
$N\sim 20$ corresponds to the turning of the trajectory that we can see
in Figure~\ref{dquad_bgtraj}, and both $\epsilon$ and
$\eta_{\sigma\sigma}$ become temporarily large around the time of the
turn.  We also see that, as the trajectory evolves into the minimum of
the $\chi$-field potential, $\eta_{ss}$ evolves from being initially
small to a positive value greater than unity. The fact that $\eta_{ss}$
becomes large and positive reflects the fact that the trajectory has
reached an attractor where neighbouring trajectories converge.  If the
trajectory remains in the attractor for sufficiently long we expect a
so-called adiabatic limit to be reached, in which the dynamics is
effectively single-field in nature and so-called isocurvature
perturbations perpendicular to the trajectory have decayed away.  We
expect $\zeta$ to be conserved in this limit.

\begin{figure}
\begin{subfigure}[t]{0.49\textwidth}
 \includegraphics[width = 0.9\columnwidth]{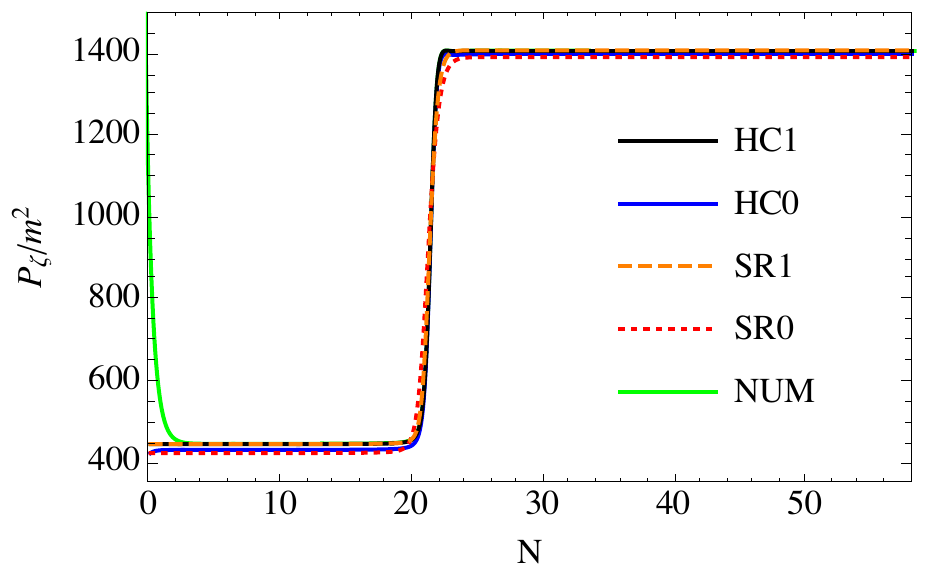}
 \caption{\label{dquad_pow}}
\end{subfigure}
\begin{subfigure}[t]{0.49\textwidth}
 \includegraphics[width = 0.9\columnwidth]{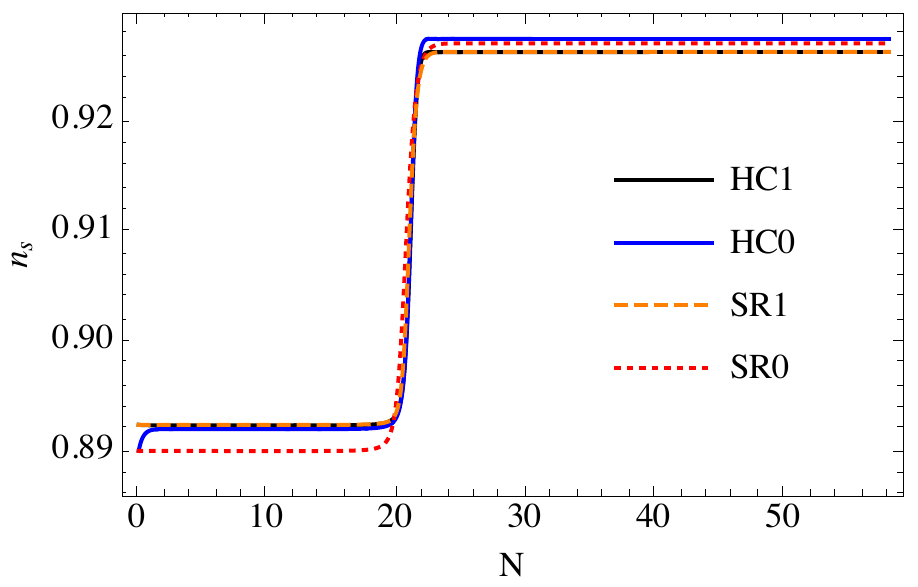}
 \caption{\label{dquad_ns}}
\end{subfigure}
\begin{subfigure}[t]{0.49\textwidth}
 \includegraphics[width = 0.9\columnwidth]{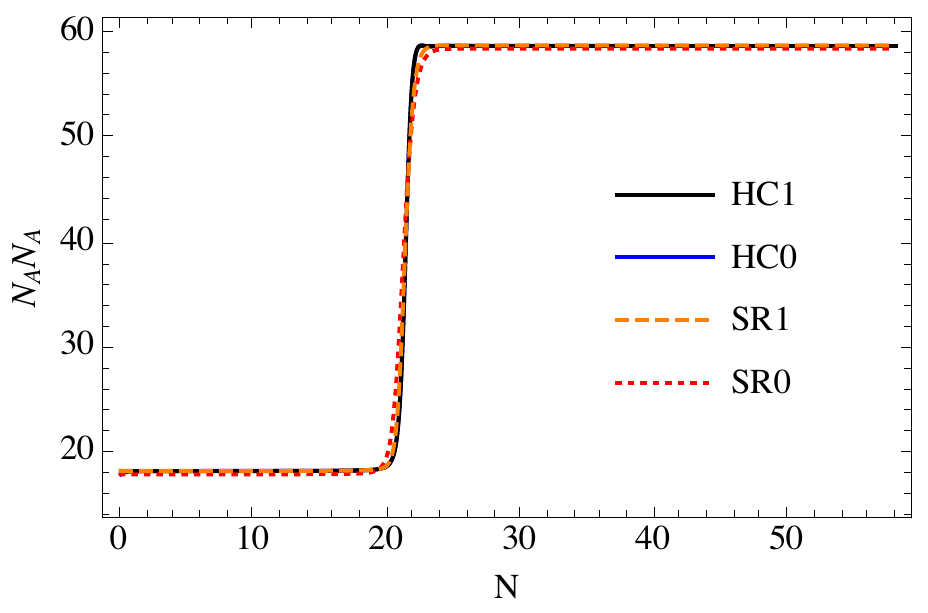}
 \caption{\label{dquad_NaNa}}
\end{subfigure}
 \begin{subfigure}[t]{0.49\textwidth}
 \includegraphics[width = 0.9\columnwidth]{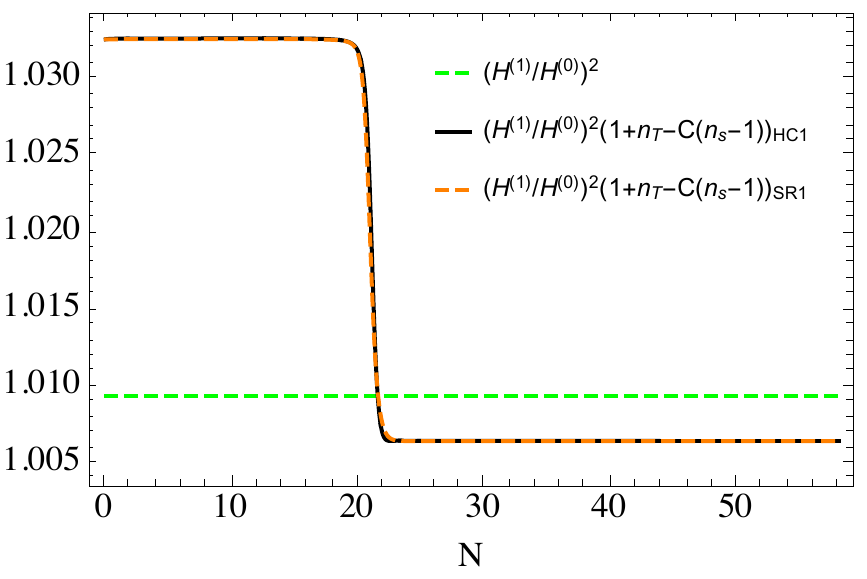}
 \caption{\label{dquad_HCcorr}}
\end{subfigure}
 \caption{\label{dquad_2pntf}(a) The evolution of $\mc P_\zeta/m^2$ for
 the same potential as considered in Figure~\ref{dquad_bg}.  Results are
 shown for the four different slow-roll approximations SR0, SR1, HC0 and
 HC1, and for comparison we also show results calculated using the
 {\it PyTransport} code of Mulryne '16 \cite{Mulryne:2016mzv} (NUM).  In (b) and (c) we show
 the evolution of $n_s$ and $N_{,A}N_{,A}$, respectively, as determined
 using the four different approximations SR0, SR1, HC0 and HC1. (d) The
 green dashed curve shows the ratio $(H^{(1)}/H^{(0)})^2$, where both
 quantities are evaluated at horizon crossing.  The dashed orange and
 black curves correspond to the slow-roll corrections associated with
 slow-roll corrections to $\Sigma^{AB}$ in approximations SR1 and HC1,
 respectively.}
\end{figure}

In Figure~\ref{dquad_2pntf} we plot quantities relating to the two-point
statistics of $\zeta$.  The first panel shows the evolution of $\mc
P_\zeta/m^2$ as determined using the five different levels of slow-roll
approximation outlined at the beginning of this section.  We see that
both before and after the turn $\zeta$ is conserved, as expected.  In
the very early stages there appears to be a large discrepancy between
the results of {\it PyTransport}, labeled NUM, and the other curves.
However, this simply reflects the fact that decaying modes have been
neglected in approximations SR0, SR1, \mk{HC0 and HC1}, and we should
therefore only expect them to coincide with the {\it PyTransport} result
a few e-foldings after horizon crossing, which they do to very good
approximation. At the end of inflation, all approximations are in
agreement to within just over $1\%$.  As we would naively expect, approximation SR0 deviates the most from the {\it PyTransport} result, while approximation $HC1$ provides the closest match.  The second best approximation is SR1 rather than HC0, which suggests that the slow-roll corrections to \mk{$\Sigma^{AB}$} are
important.  Indeed, this is confirmed by the results plotted in Figures
\ref{dquad_NaNa} \& \ref{dquad_HCcorr}.  In Figure~\ref{dquad_NaNa} we
have plotted the evolution of the quantity $N_{,A}N_{,A}$ for the four
different approximations SR0, SR1, HC0 and HC1.  This quantity
determines how the perturbations in the initial conditions are related
to $\delta N (= \zeta)$, and its evolution will depend on the
approximations made in solving the super-horizon dynamics.  Given that
approximations HC0 and HC1 both use the full equations of motion to
solve for the super-horizon evolution, we expect them to be in good
agreement, which they are.  Approximation SR1 is also in very good agreement with approximations HC0 and HC1, and even the leading-order approximation SR0 differs by less than $0.5\%$. In Figure \ref{dquad_HCcorr} the corrections arising from
slow-roll corrections to \mk{$\Sigma^{AB}$} are plotted for approximations
SR1 and HC1.  We see that these corrections are between $0.6$ and $0.7\%$.
Finally, Figure~\ref{dquad_ns} shows the evolution of $n_s$ as
determined under the four different approximation schemes SR0, SR1, HC0
and HC1.  The first thing to note is that the final value of $n_s$ is
not in agreement with observations.  The different approximations are in
agreement at the $\sim 0.1\%$ level, with approximations SR1 and HC1
being in very close agreement.  This again indicates that it is the
slow-roll corrections to perturbations on the initial flat hypersurface
that are dominant.

Overall, the very good agreement between the different approximations is
perhaps surprising, considering that intermediately the slow-roll
parameters become quite large.
A similar behaviour was demonstrated for the non-Gaussianity parameter $f_{NL}$ in ref.~\cite{Jung:2016kfd}, where they used a very different formalism.
Presumably it is because the slow-roll
parameters are only temporarily large that slow-roll corrections to
observables do not become significant.  It would be interesting to try
and gain a more precise understanding as to how large slow-roll
parameters must become and for how long in order for slow-roll
corrections to become significant.
\mk{}

For the double quadratic model we find that $f_{NL}$ is unobservably small, so we do not consider its evolution in detail here.

\subsection{Quadratic plus axion potential}

Next we consider the axion-quadratic model that is also
considered in \cite{Elliston:2011dr,Seery:2012vj}.  The potential takes
the form
\begin{align}
 V = \frac{1}{2}m^2\phi^2 +\Lambda^4\left(1-\cos\left(\frac{2\pi \chi}{f}\right)\right),
\end{align}
and we take
\begin{align}
 \Lambda^4 = \frac{R^2 m^2 f^2}{4\pi^2}.
\end{align}
Doing so once again allows us to factor out the overall mass
scale $m^2$, which is then only important when it comes to determining
the overall magnitude of the power spectrum, i.e. $n_s-1$, $r$, $f_{NL}$
do not depend on $m$.

For the particular example considered here, we take $f = 1$ and $R=5$.
Seeing as we are interested in slow-roll corrections to $f_{NL}$, we
choose initial conditions that are known to give rise to an observably
large $f_{NL}$.  Namely, $\chi$ is taken to start very close to the
maximum of its potential.  In using the {\it PyTransport} code we use the initial conditions
$(\phi,\chi,\phi',\chi') = (16.4,(1-5\times 10^{-4})/2,0,0)$, and then
consider the mode that leaves the horizon $8.2$ e-foldings after these
initial conditions are set.  When we calculate $f_{NL}$, the mode $k_S$
is again taken to leave at this time, and once again taking $R_{\rm sq} = 0.1$ means that the mode $k_L$ left the horizon approximately $5.9$
e-foldings after the initial conditions were set.  Solving the full
background equations, we determine that after $8.2$ e-folds the field
values are $(\phi_k,\chi_k) \simeq (15.42, 0.4989)$, and we use these values
as the initial conditions for calculating the results corresponding to
approximations SR0, SR1, HC0 and HC1.

\begin{figure}
\begin{subfigure}[t]{0.49\textwidth}
 \includegraphics[width = 0.9\columnwidth]{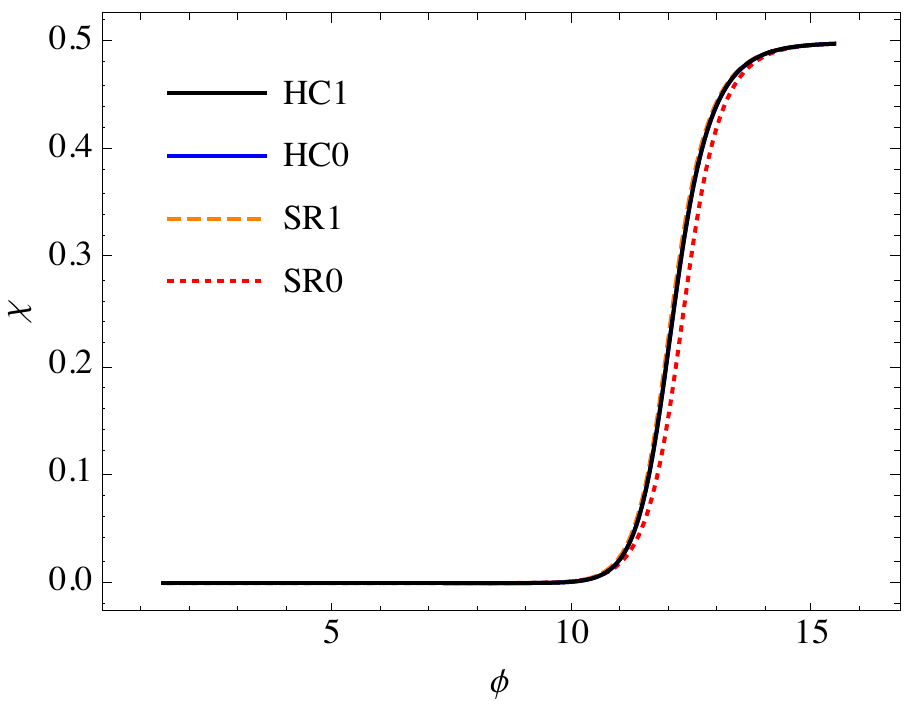}
 \caption{\label{quadAx_bgtraj}}
\end{subfigure}
\begin{subfigure}[t]{0.49\textwidth}
 \includegraphics[width = 0.9\columnwidth]{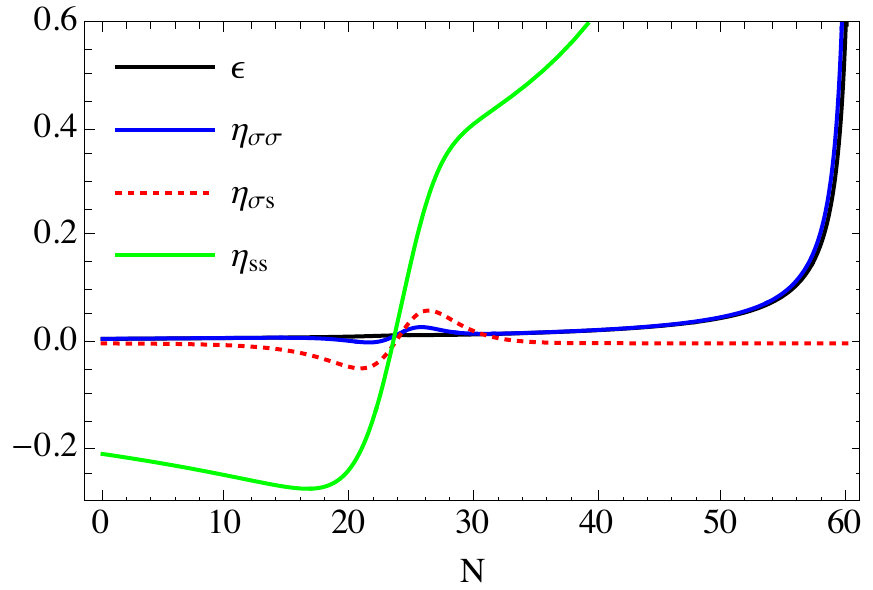}
 \caption{\label{quadAx_srparams}}
\end{subfigure}
 \caption{\label{quadAx_bg}As in Figure \ref{dquad_bg} but for the
 quadratic plus axion potential outlined in the text, with $f=1$, $R = 5$
 and $(\phi_k,\chi_k)\simeq (15.42, 0.4989)$.}
\end{figure}

Plots of the background trajectory and the corresponding evolution of
the slow-roll parameters are given in Figure \ref{quadAx_bg}, and we
find that there are approximately $60$ e-foldings of inflation after
horizon crossing.  As can be seen from Figure~\ref{quadAx_bgtraj}, the
trajectory evolves to the minimum of the $\chi$ potential before the end
of inflation, and so we can expect that $\zeta$ becomes constant.

Looking at Figure~\ref{quadAx_srparams}, we see that in
this example the quantities $\epsilon$,
$\eta_{\sigma\sigma}$ and $\eta_{\sigma s}$ remain less than $0.1$ right
up until the end of inflation.  The oscillatory feature in $\eta_{\sigma
s}$ at around $N\sim 25$ corresponds to the turning of the trajectory
that we can see in Figure~\ref{quadAx_bgtraj}.  The only relatively
large slow-roll parameter, then, is $\eta_{ss}$.  Initially it is
negative, corresponding to when the trajectory is along the peak of the
tachyonic $\chi$-field potential.  At around $N\sim 25$ the trajectory
then falls off the ridge and into the minimum of the $\chi$-field
potential, where $\eta_{ss}$ is positive and neighbouring trajectories
converge.  The fact that $\eta_{ss}$ becomes large and positive reflects
the fact that the trajectory has reached an attractor.  
\mk{
As was mentioned before, if the
trajectory remains in the attractor for sufficiently long we expect a
so-called adiabatic limit to be reached.
}

\begin{figure}
\begin{subfigure}[t]{0.49\textwidth}
 \includegraphics[width = 0.9\columnwidth]{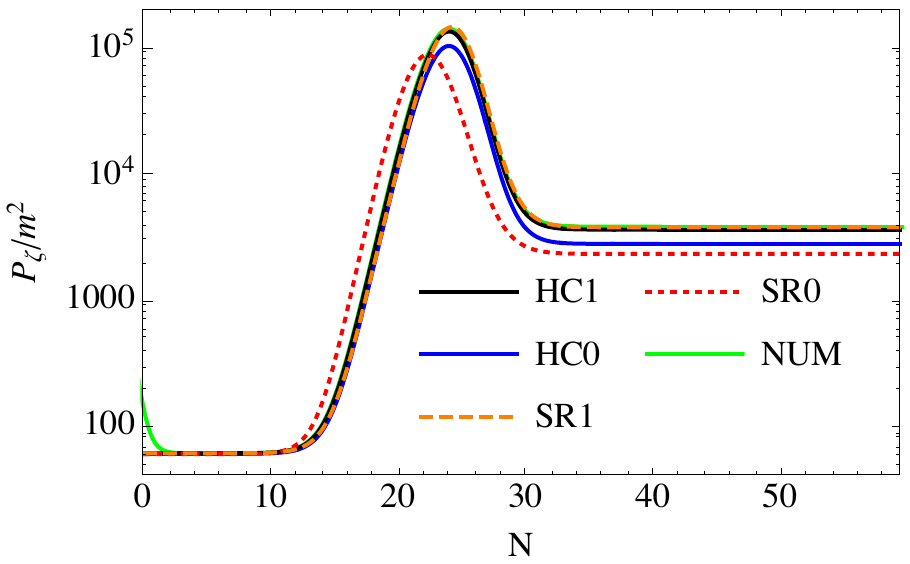}
 \caption{\label{quadAx_pow}}
\end{subfigure}
\begin{subfigure}[t]{0.49\textwidth}
 \includegraphics[width = 0.9\columnwidth]{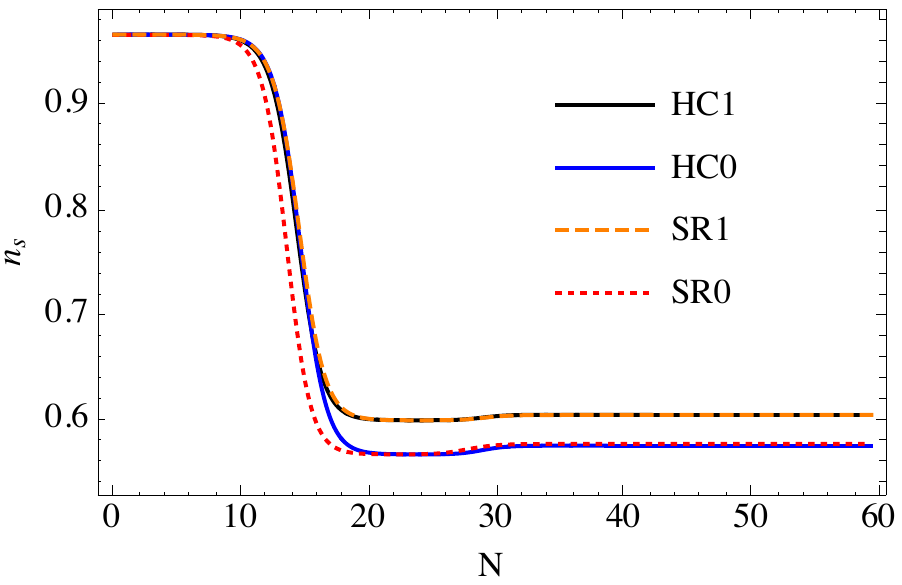}
 \caption{\label{quadAx_ns}}
\end{subfigure}
\begin{subfigure}[t]{0.49\textwidth}
 \includegraphics[width = 0.9\columnwidth]{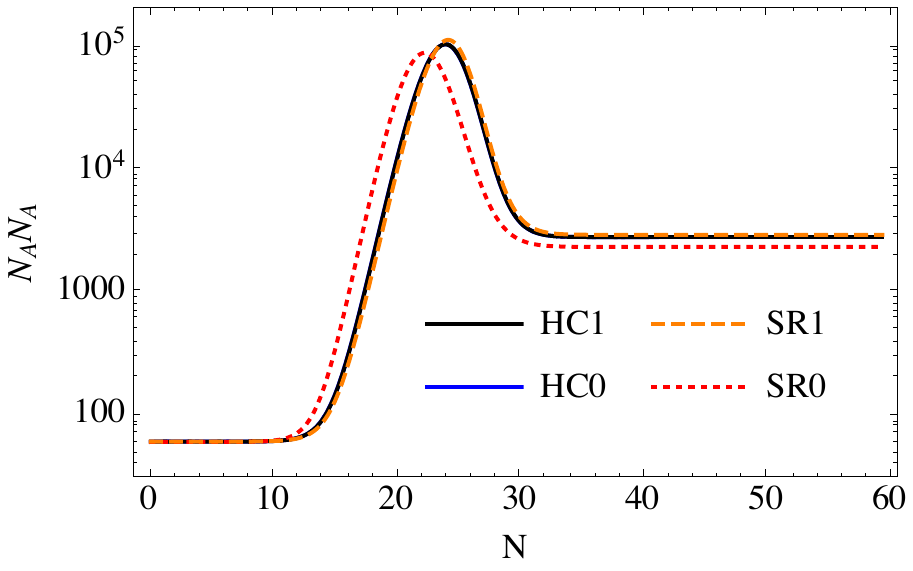}
 \caption{\label{quadAx_NaNa}}
\end{subfigure}
 \begin{subfigure}[t]{0.49\textwidth}
 \includegraphics[width = 0.9\columnwidth]{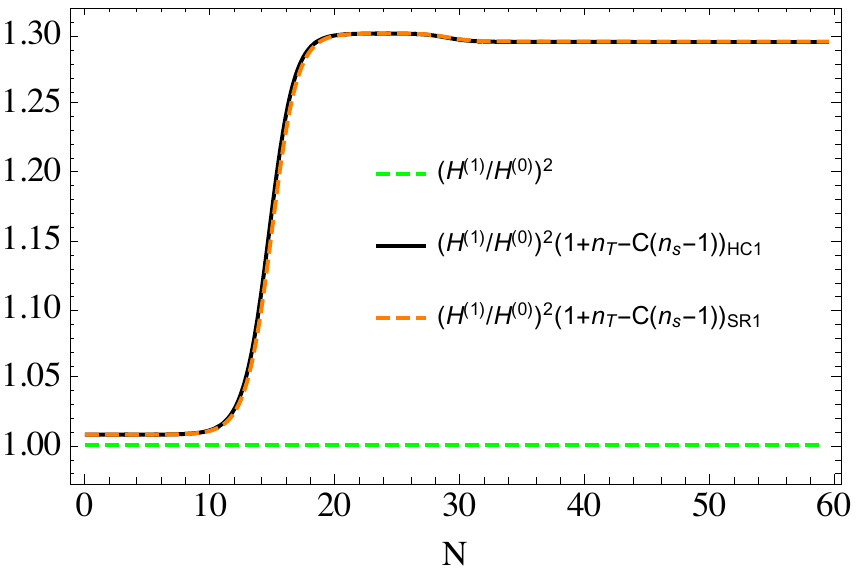}
 \caption{\label{quadAx_HCcorr}}
\end{subfigure}
 \caption{\label{quadAx_2pntf}As in Figure \ref{dquad_2pntf} but for the
 same quadratic plus axion potential as considered in Figure
 \ref{quadAx_bg}.}
\end{figure}

In Figure~\ref{quadAx_2pntf} we plot quantities relating to the
two-point statistics of $\zeta$.  The first panel shows the evolution of
$\mc P_\zeta/m^2$ as determined using the five different levels of
slow-roll approximation outlined at the beginning of this section.  In
the very early stages we see a large discrepancy between the NUM and
other curves.  However, \mk{as in the case of the double quadratic potential,} this simply reflects the fact that decaying
modes have been neglected in approximations SR0, SR1, \mk{HC0 and HC1}, and
we should therefore only expect them to coincide with the NUM curve a
few e-foldings after horizon crossing.  At later times, we see that
approximations HC1 and SR1 are in very good agreement with the {\it
PyTransport} results NUM.  For the first 10-efoldings the trajectory
remains straight and $\zeta$ is conserved.  The trajectory then starts
to bend, leading to a sourcing of $\zeta$, but after $N\sim 30$ the
trajectory reaches the minimum of the $\chi$ potential and $\zeta$ is
once again conserved.  Although qualitatively all five curves are in
good agreement, quantitatively we find a discrepancy of over $60\%$ in
the final value of $\mc P_\zeta$.  As approximations HC0 and HC1
essentially only differ in their assumptions for the power spectra of
the initial field fluctuations, i.e. $\Sigma^{AB}$, Figure
\ref{quadAx_pow} suggests that slow-roll corrections arising from the
slow-roll corrections to $\Sigma^{AB}$ must be important.  Indeed, in
Figure~\ref{quadAx_HCcorr} we plot the corrections due to slow-roll
corrections to \mk{$\Sigma^{AB}$}, and one can see that they are on the order
of $\sim 30\%$.  In Figure~\ref{quadAx_NaNa} we plot the evolution of
the quantity $N_{,A}N_{,A}$ for the four different approximations SR0,
SR1, HC0 and HC1.  
Given that approximations HC0 and HC1 both use
the full equations of motion to solve for the super-horizon evolution,
we see that they are in perfect agreement.  We also see that the
next-to-leading order approximation used in SR1 gives a significant
improvement over the leading-order approximation SR0.  The discrepancy
between SR0 and HC1 is on the order $20\%$.  Finally,
Figure~\ref{quadAx_ns} shows the evolution of $n_s$ as determined under
the four different approximation schemes.  The first thing to note is
that the final value of $n_s$ is not in agreement with observations.  As
discussed in Section~\ref{ICSRCpow}, a consequence of this will be that
slow-roll corrections from the initial conditions are large, which we
indeed see to be the case in Figure~\ref{quadAx_HCcorr}.  This is
further reflected in the large discrepancy between the HC0 and HC1
curves in Figure~\ref{quadAx_ns}, as essentially the only difference
between these approximation schemes is the choice of the initial
$\Sigma^{AB}$.

\begin{figure}
\centering
  \begin{subfigure}[t]{0.49\textwidth}
 \includegraphics[width = 0.9\columnwidth]{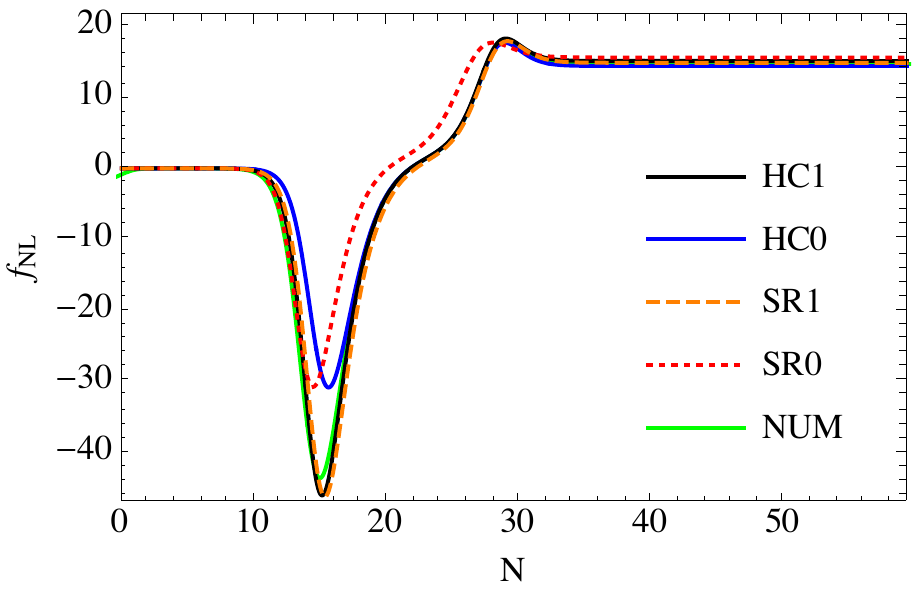}
 \caption{\label{quadAx_fNL}}
 \end{subfigure}
\\
 \begin{subfigure}[t]{0.49\textwidth}
 \includegraphics[width = 0.9\columnwidth]{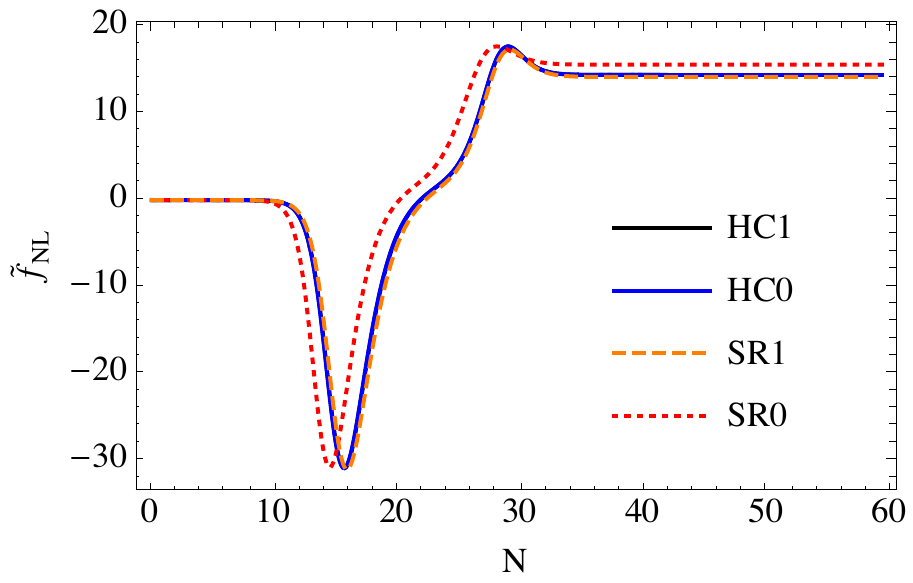}
 \caption{\label{quadAx_tildefNL}}
 \end{subfigure}
\begin{subfigure}[t]{0.49\textwidth}
 \includegraphics[width = 0.9\columnwidth]{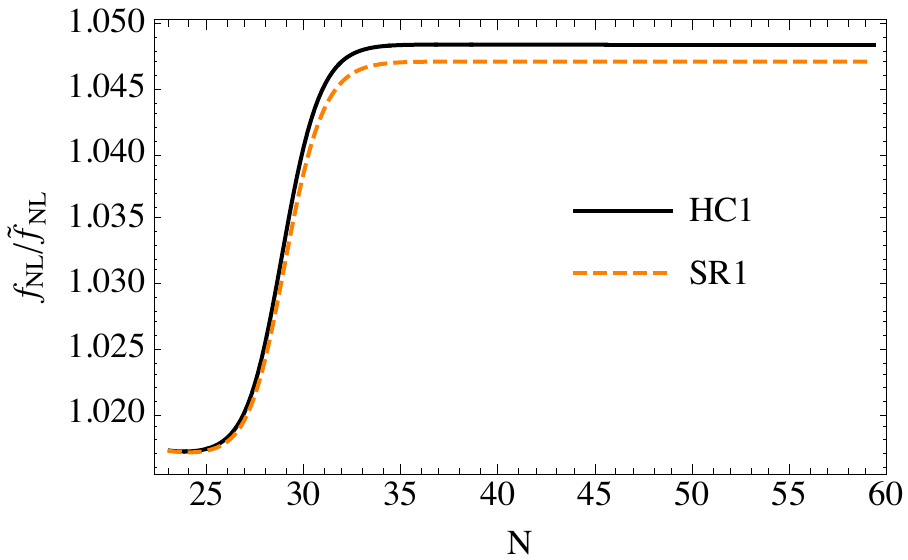}
 \caption{\label{quadAx_fNL_tildefNL}}
\end{subfigure}
 \caption{\label{quadAx_3pntf}Quantities relating to $f_{NL}$ for the quadratic plus axion potential considered in Figures \ref{quadAx_bg} \& \ref{quadAx_2pntf}:  (a) Evolution of $f_{NL}$ as determined
 using the four approximations SR0, SR1, HC0 and HC1, with numerical
 results calculated using the code {\it PyTransport}
 \cite{Mulryne:2016mzv} also given for comparison (NUM). (b) Evolution
 of $\tilde f_{NL}$ as determined using the four approximations SR0,
 SR1, HC0 and HC1.  (c) Evolution of the slow-roll corrections
 associated with slow-roll corrections to the initial $\Sigma^{AB}$ for
 the approximations SR1 and HC1.}
\end{figure}

In Figure~\ref{quadAx_3pntf} we plot quantities related to $f_{NL}$.
Looking at Figure~\ref{quadAx_fNL} we see that discrepancies between the
various approximation schemes can be large intermediately.  In
particular, the magnitudes of the minima at around $N\sim 15$ differ by
around $50\%$.  Interestingly, the magnitudes of the minima as
determined using approximations SR0 and HC0 are very similar, as are
those determined using approximations SR1 and HC1.  This suggests that
at this intermediate time it is the slow-roll corrections to
$\Sigma^{AB}$ that are important.  This is also reflected in
Figure~\ref{quadAx_tildefNL}, where $f_{NL}$ neglecting slow-roll
corrections to $\Sigma^{AB}$, namely $\tilde f_{NL}$, is plotted for the
four different approximation schemes.  In this plot we see that the
magnitudes of the minima at around $N\sim 15$ are very similar for all
the approximation schemes, while the location of the minimum is around
two e-foldings earlier for approximation SR0.  Turning to the final
value of $f_{NL}$, we find that the different approximations are in
better agreement, with differences being less than $ 10\%$.  From
Figure~\ref{quadAx_tildefNL} we see that $\tilde f_{NL}$ as determined
by HC1, HC0 and SR1 are approximately $10\%$ lower than the SR0 result.
In Figure~\ref{quadAx_fNL_tildefNL} we plot the slow-roll
corrections arising from slow-roll corrections to $\Sigma^{AB}$, namely
$f_{NL}/\tilde f_{NL}$, and for the cases HC1 and SR1 we see that they
give a positive correction of just under $5\%$.

In summary, compared to the double quadratic model considered in the preceding subsection, we have found slow-roll corrections to be much larger in this model with a quadratic plus axion potential.  This is perhaps to be expected given that the slow-roll parameter $\eta_{ss}$ in this model is relatively large and negative for an extended period, namely the first $\sim20$ e-foldings after horizon exit.  Interestingly, we found that corrections to the non-Gaussianity parameter $f_{NL}$ were significantly smaller than those to the power spectrum.

\subsection{Linear plus axion potential}

Next we consider a very similar model to that considered above, but
replace the quadratic potential of the $\phi$-field with a linear
potential.  It has been shown in \cite{McAllister:2008hb,Iso:2014gka} that such a
linear potential may be realised from axion monodromy. 
As we will see, with this
potential we are able to find a region in parameter space for which the
predictions for $n_s$ are in agreement with observations.  The potential
thus takes the form
\begin{align}
 V = \lambda\phi +\Lambda^4\left(1-\cos\left(\frac{2\pi \chi}{f}\right)\right),
\end{align}
and similar to the previous case we take
\begin{align}
 \Lambda^4 = \frac{\lambda f^2}{\mc M 4\pi^2}.
\end{align}
$\lambda$ can then be factored out and it only affects the overall
normalisation of $\mathcal P_\zeta$, with $\mc P_\zeta \propto H_k^2 =
V_k/(3-\epsilon_k)\propto \lambda$.

We take $f = 1/10$ and $\mc M=1$, and as in the previous case we choose
initial conditions such that an observably large $f_{NL}$ is generated.
In using the {\it PyTransport} code we set initial conditions as
$(\phi,\chi,\phi',\chi') = (11.6,(1-1.5\times 10^{-3})/20,0,0)$, and
consider a mode that leaves the horizon 8.2 e-foldings after these initial
conditions are set.  In calculations of $f_{NL}$ we once again take $k_S$ to leave the
horizon at this time, and we choose $R_{\rm sq} = 0.1$ such that $k_L$
left the horizon approximately $5.9$ e-foldings after the initial
conditions were set.  Using the full equations of motion we find that
after 8.2 e-folds the field values are given as $(\phi_k,\chi_k) \simeq
(10.90, 0.04985)$, and these are the values used as initial conditions
in approximations SR0, SR1, HC0 and HC1.  After horizon crossing we find
that there are approximately $60$ e-foldings of inflation.

\begin{figure}
\begin{subfigure}[t]{0.49\textwidth}
 \includegraphics[width = 0.9\columnwidth]{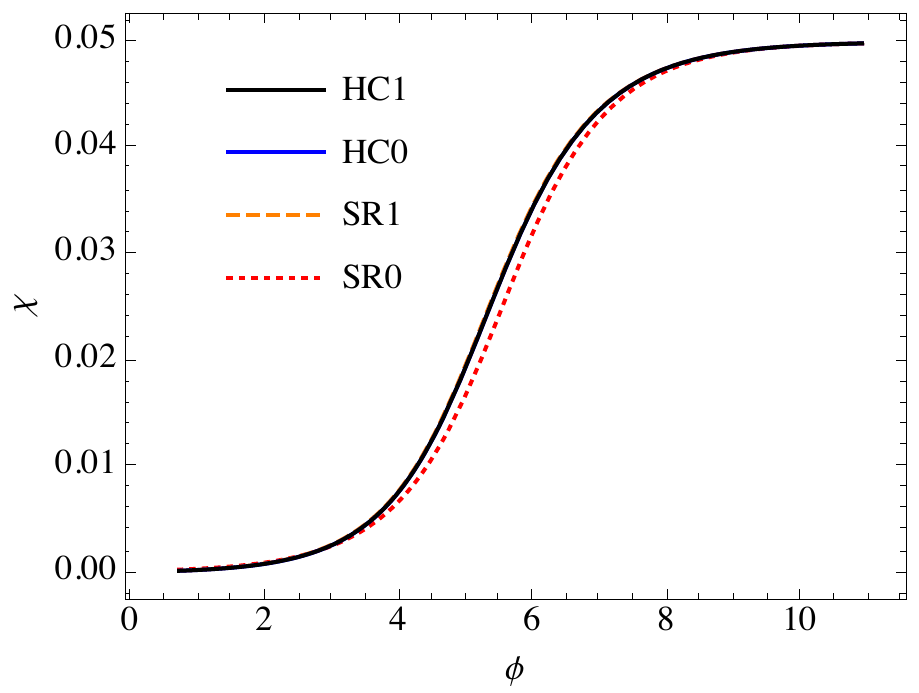}
 \caption{\label{linAx_bgtraj}}
\end{subfigure}
\begin{subfigure}[t]{0.49\textwidth}
 \includegraphics[width = 0.9\columnwidth]{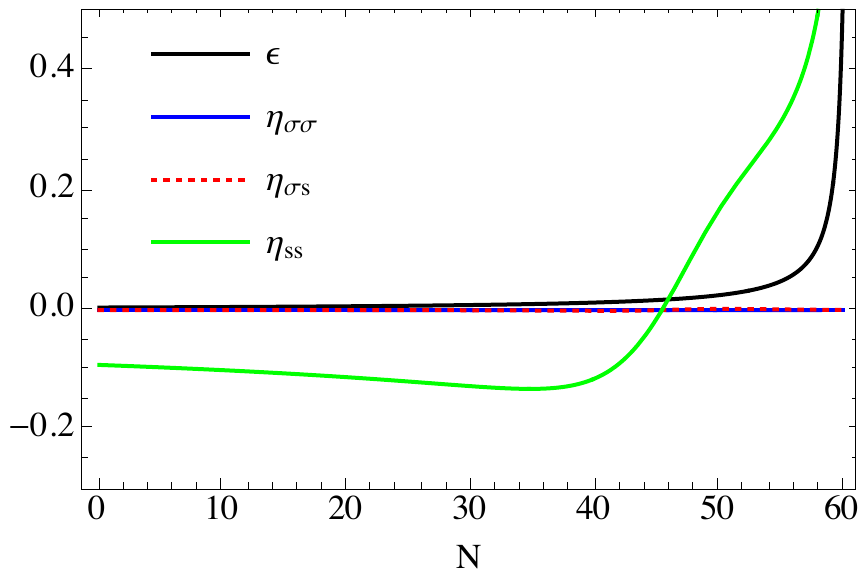}
 \caption{\label{linAx_srparams}}
\end{subfigure}
 \caption{\label{linAx_bg}As in Figure \ref{dquad_bg} but for the linear plus axion potential outlined in the text, with
 $f=1/10$, $\mc M = 1$ and $(\phi_k,\chi_k)\simeq (10.90, 0.04985)$.}
\end{figure}

In Figure~\ref{linAx_bgtraj} we give the background trajectory as
calculated using the different approximations SR0, SR1, HC0 and HC1.  As
in the previous example, the initial trajectory is along the maximum of
the $\chi$-field potential, but eventually it falls into the minimum at
$\chi = 0$ and continues to evolve in the $\phi$ direction.  Compared to
the trajectory shown in Figure~\ref{quadAx_bgtraj} we see that the
transition into the minimum of the $\chi$ potential is much more
gradual, but nevertheless the trajectory does reach the minimum before
the end of inflation.  We can thus expect that $\zeta$ will approach a
constant towards the end of inflation.

In Figure~\ref{linAx_srparams} we plot the evolution of the slow-roll
parameters described in the previous subsection.  We see that
$\eta_{\sigma\sigma}$ and $\eta_{\sigma s}$ remain very small for the
entire duration of inflation.  The fact that $\eta_{\sigma s}$ remains
much smaller in this example than in the previous example reflects the
fact that the turning of the trajectory in the transition to the minimum
of the $\chi$ potential is much more gradual.  Qualitatively we see that
the evolution of $\eta_{ss}$ is similar to that in the previous
example. Quantitatively, however, the initial tachyonic mass in the
$\chi$ direction is smaller in this case, and we also see that the
transition to positive $\eta_{ss}$ -- which indicates the time at which
the transition to the minimum of the $\chi$ potential occurs -- occurs
much later on in this example, at around $N\sim 45$.  This again is
consistent with our observation that evolution in the $\chi$ direction
occurs on a longer timescale in this example.

\begin{figure}
\begin{subfigure}[t]{0.49\textwidth}
 \includegraphics[width = 0.9\columnwidth]{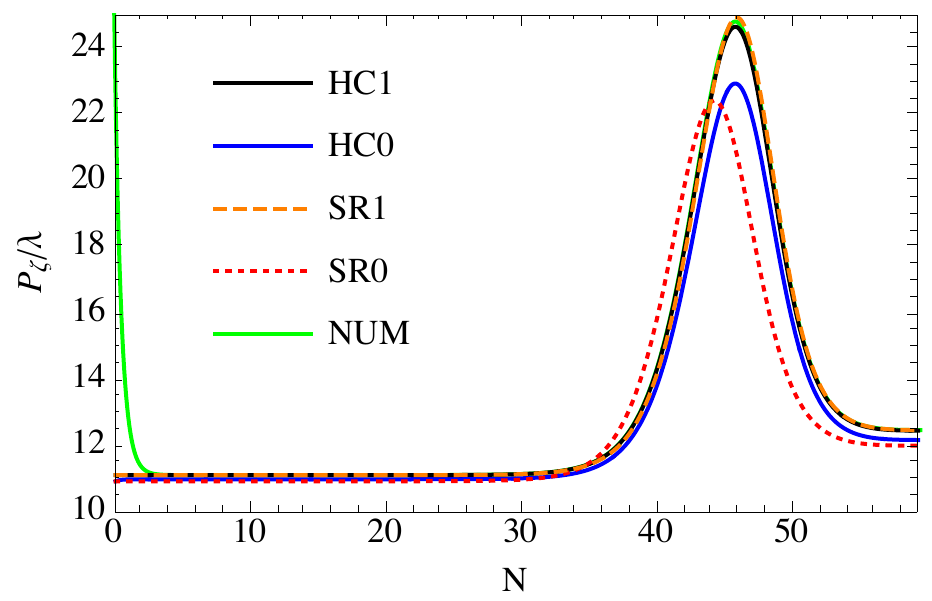}
 \caption{\label{linAx_pow}}
\end{subfigure}
\begin{subfigure}[t]{0.49\textwidth}
 \includegraphics[width = 0.9\columnwidth]{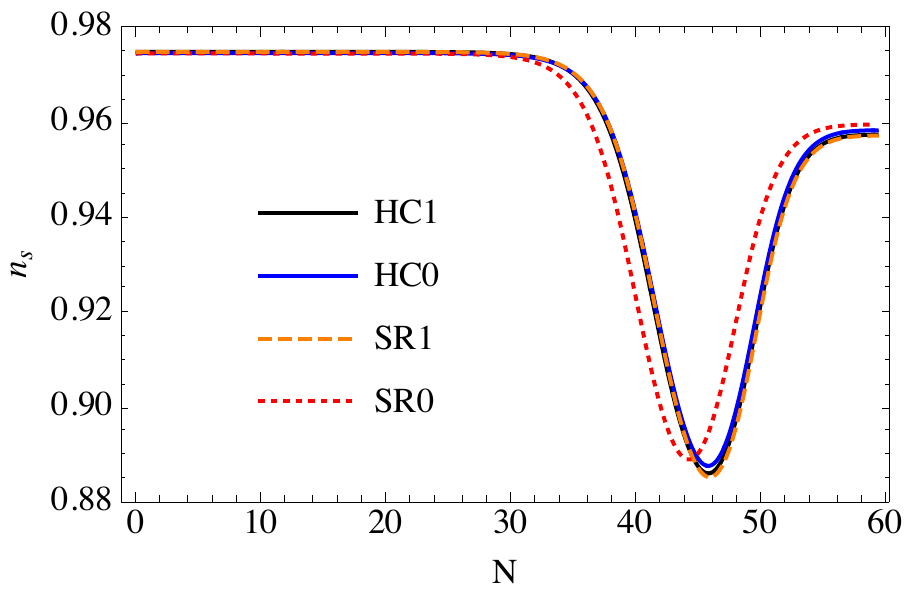}
 \caption{\label{linAx_ns}}
\end{subfigure}
\begin{subfigure}[t]{0.49\textwidth}
 \includegraphics[width = 0.9\columnwidth]{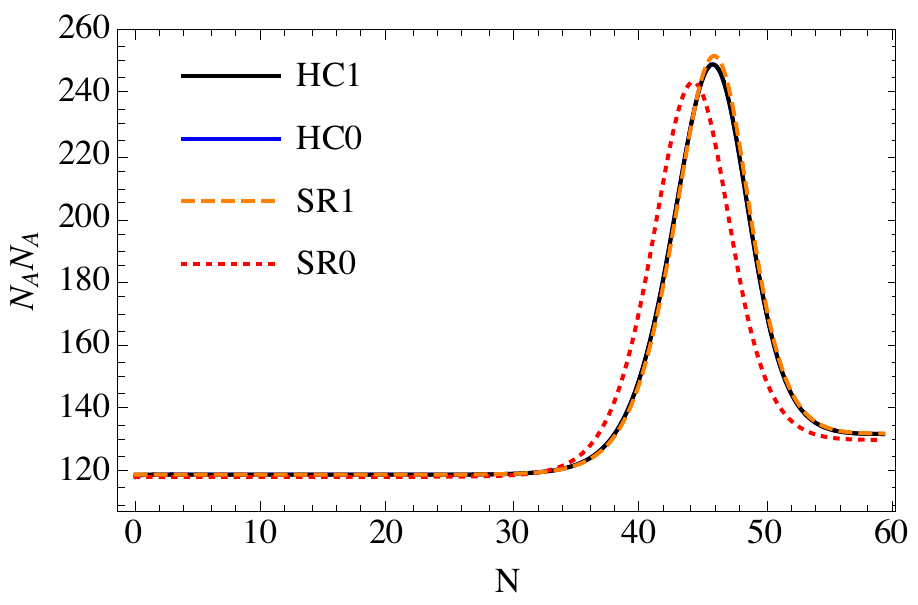}
 \caption{\label{linAx_NaNa}}
\end{subfigure}
 \begin{subfigure}[t]{0.49\textwidth}
 \includegraphics[width = 0.9\columnwidth]{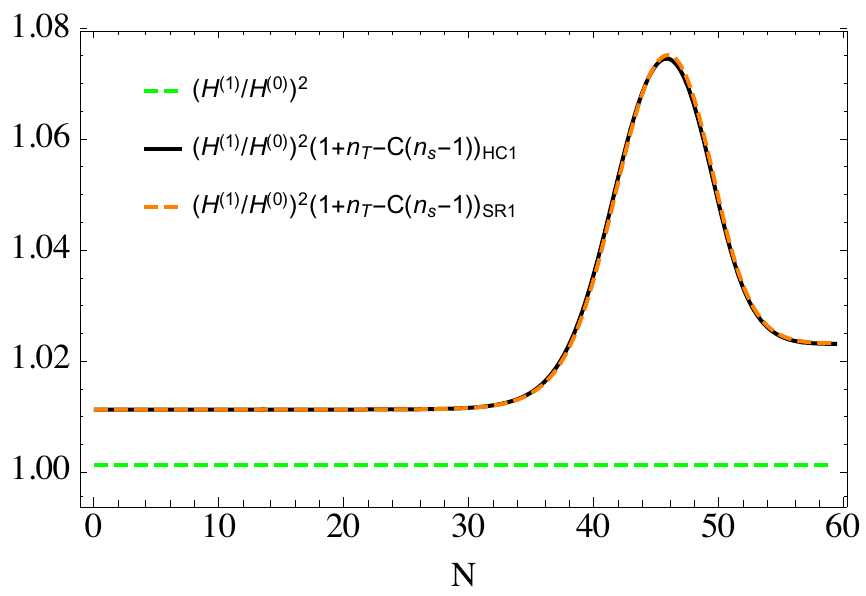}
 \caption{\label{linAx_HCcorr}}
\end{subfigure}
 \caption{\label{linAx_2pntf}Essentially as in Figure \ref{dquad_2pntf}
 but for the linear plus axion potential considered in Figure
 \ref{linAx_bg}.  The only difference is that the normalised power
 spectrum is given as $\mc P_\zeta/\lambda$ as opposed to $\mc
 P_\zeta/m^2$.}
\end{figure}

In Figure~\ref{linAx_pow} we plot the evolution of $\mc P_\zeta/\lambda$
as determined using the approximations SR0, SR1, HC0 and HC1, with
numerical results calculated using the {\it PyTransport} code
\cite{Mulryne:2016mzv} also given for comparison.  In the early stages
we see that it takes approximately four e-foldings before the decaying
modes become negligible and the approximations SR0, SR1, HC0 and HC1
come into good agreement with the NUM curve. Throughout the remaining
evolution we find that the SR1 and \mk{HC1} approximations are in very good
agreement with the results obtained using {\it PyTransport}.  As
expected after looking at the background trajectory in
Figure~\ref{linAx_bg}, we see that the curvature perturbation approaches
a constant right at the end of inflation, which is much later on than in
the previous example.  Qualitatively the SR0 approximation is in good
agreement with the {\it PyTransport} results NUM and the other
approximations.  Quantitatively, however, we see that the peak in $\mc
P_\zeta/\lambda$ is about $10\%$ lower than the NUM curve and also
occurs about two e-foldings earlier.  Looking at the final value of
$\mc P_\zeta/\lambda$, we see that corrections to approximation SR0 are
less than $5\%$.  The discrepancy between approximations HC0 and
HC1 highlights the importance of slow-roll corrections to the
initial power spectra $\Sigma^{AB}$, as this is essentially the only
difference between these two approximations.  We also note that, as in the
previous example, the next-to-leading order approximation SR1 offers a
significant improvement over the lowest order approximation SR0.

In Figure~\ref{linAx_ns} we show the evolution of $n_s$ as determined
using the four approximations SR0, SR1, HC0 and HC1.  Looking at the
final values we see that they are in very good agreement, with
corrections to the SR0 approximation being less than half of a percent.
Another important thing to note
is that the final value of $n_s$ is consistent with current
observations.  This is consistent with the fact that slow-roll
corrections to $\mc P_\zeta$ are relatively small.  As we saw in
Section~\ref{ICSRCpow}, slow-roll corrections to $\mc P_\zeta$ arising
from slow-roll corrections to $\Sigma^{AB}$ can be written in terms of
$n_s -1$, and if $n_s - 1$ is in agreement with observations then this
implies that the slow-roll corrections associated with corrections to
$\Sigma^{AB}$ should be small.  Indeed, this is confirmed explicitly in
Figure~\ref{linAx_HCcorr}, where the slow-roll corrections associated
with corrections to the initial power spectra $\Sigma^{AB}$ are plotted
as a function of time for the approximations SR1 (orange dashed curve) and
HC1 (solid black curve).  We see that at the end of inflation these
corrections are just over $2\%$.  For reference, the green dotted curve
also shows the ratio $(H^{(1)}/H^{(0)})^2$, with both quantities
evaluated at horizon crossing.  Given that this ratio is so close to
unity, we see that slow-roll corrections associated with $\Sigma^{AB}$
mostly come from the factor $(1+n_T-Cn_\zeta)$.

Given that slow-roll corrections arising from corrections to the initial
power spectra $\Sigma^{AB}$ are small, the only way to obtain large
corrections to $\mc P_\zeta$ would be if slow-roll corrections to the
super-horizon dynamics were large.  Such corrections are encoded in the
quantity $N_{,A}N_{,A}$, which is plotted for the four different
approximations in Figure~\ref{linAx_NaNa}, and we see that corrections
to the SR0 approximation are less than $2\%$ at the end of
inflation.  Also note that the HC0 and HC1 curves are in very good
agreement, which is to be expected given that they both use the full
equations of motion to solve for the super-horizon evolution.  The
approximation SR1 also offers a significant improvement over the SR0
approximation, being in very good agreement with approximations HC0
and HC1.

\begin{figure}
\centering
  \begin{subfigure}[h]{0.49\textwidth}
 \includegraphics[width = 0.9\columnwidth]{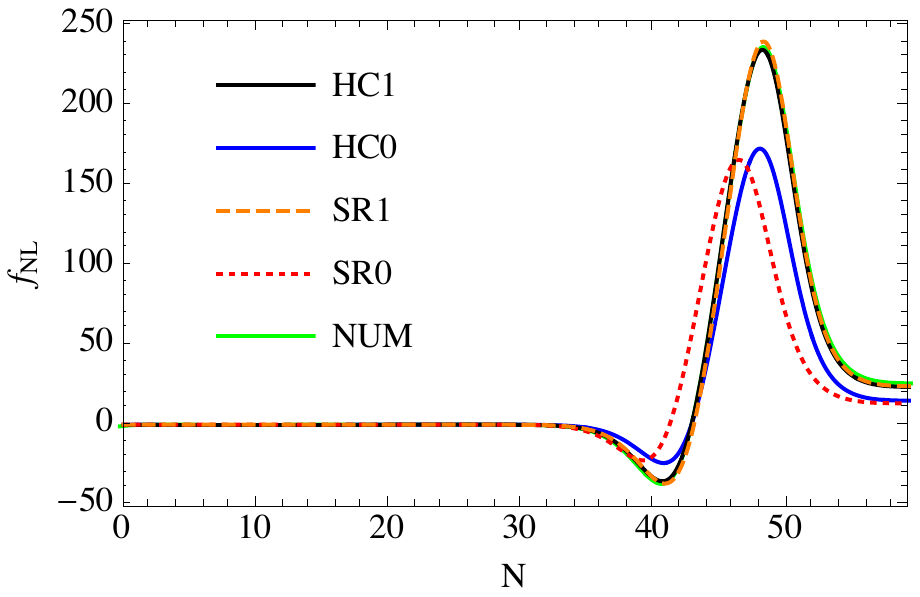}
 \caption{\label{linAx_fNL}}
 \end{subfigure}
\\
 \begin{subfigure}[h]{0.49\textwidth}
 \includegraphics[width = 0.9\columnwidth]{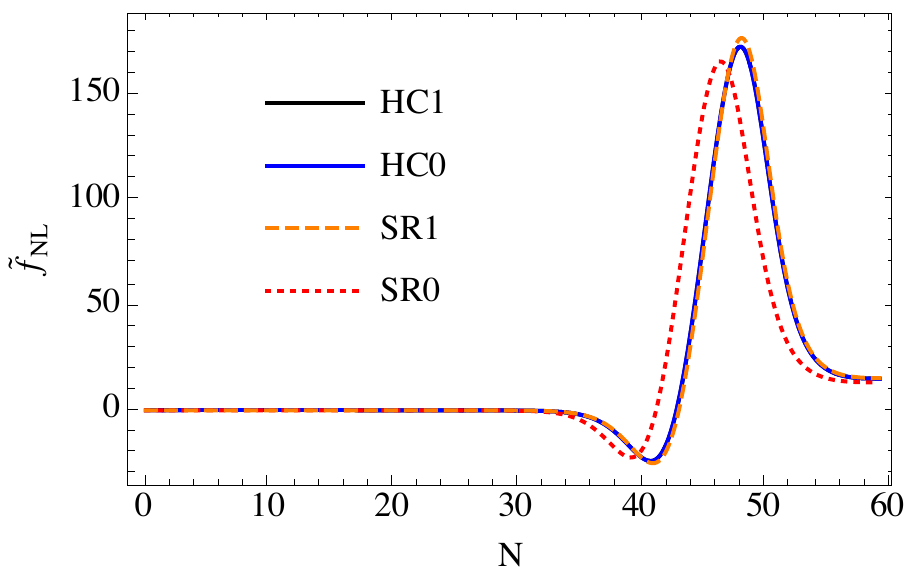}
 \caption{\label{linAx_tildefNL}}
 \end{subfigure}
\begin{subfigure}[h]{0.49\textwidth}
 \includegraphics[width = 0.9\columnwidth]{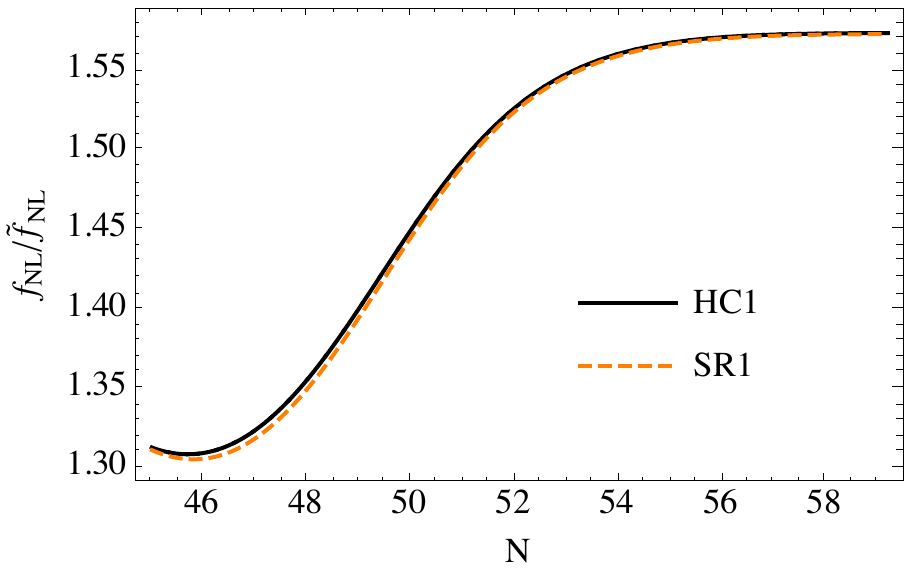}
 \caption{\label{linAx_fNL_tildefNL}}
\end{subfigure}
 \caption{\label{linAx_3pntf}As in Figure \ref{quadAx_3pntf} but for the linear plus axion potential considered in Figures \ref{linAx_bg} \& \ref{linAx_2pntf}.}
\end{figure}

Turning next to the non-Gaussianity, we plot quantities relating to
$f_{NL}$ in Figure~\ref{linAx_3pntf}.  As with the previous example, we
see that intermediately the peak values of $f_{NL}$ are quite different
for the different approximations, with approximations SR1 and HC1 giving
peak values $\sim 40$\% larger than approximations SR0 and HC0.  The
discrepancy between the peak values given by HC0 and HC1 suggests that
the difference can be attributed to the slow-roll corrections arising
from slow-roll corrections to $\Sigma^{AB}$, and this is confirmed in
Figure~\ref{linAx_tildefNL}, where we see that $\tilde f_{NL}$ essentially coincides
for approximations HC0 and HC1.  As with the power spectrum, we also
find that the peak in $f_{NL}$ as determined using approximation SR0
occurs a couple of e-foldings earlier than for the other approximations.
Focussing next on the final value of $f_{NL}$, we find that corrections
to the SR0 approximation are very large.  The approximation HC1 gives a
value almost $80$\% larger than the SR0 result, and the {\it PyTransport} result
NUM is then almost $10$\% larger than approximation HC1.  Corrections to
the SR0 result for $\tilde f_{NL}$ are less than $15$\%, so it is the
slow-roll corrections from $\Sigma^{AB}$ that are mostly responsible for
the large discrepancies.  This is confirmed in
Figure~\ref{linAx_fNL_tildefNL}, where we plot $f_{NL}/\tilde f_{NL}$
and see that corrections from the factor $1-n_T^{(0)}-\tilde n_{\delta
b}[2C-\ln(R_{\rm sq})]$ appearing in eq.~\eqref{fNLcompact} are close to
$60$\%.  Given that $n_T^{(0)}$ is small, this suggests that $n_{\delta
b}$ is relatively large in this model.

\begin{figure}
  \begin{subfigure}[t]{0.49\textwidth}
 \includegraphics[width = 0.9\columnwidth]{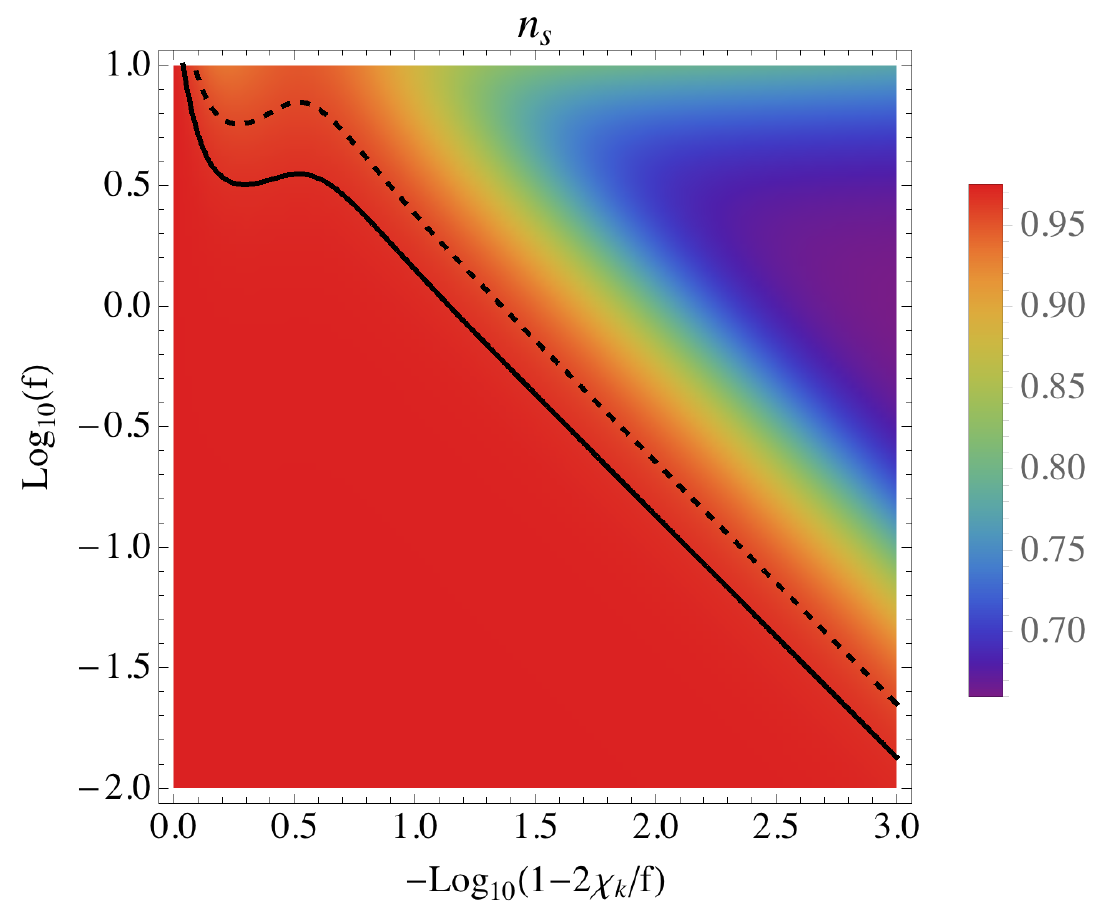}
 \caption{\label{linAx_nscon}}
 \end{subfigure}
 \begin{subfigure}[t]{0.49\textwidth}
 \includegraphics[width = 0.9\columnwidth]{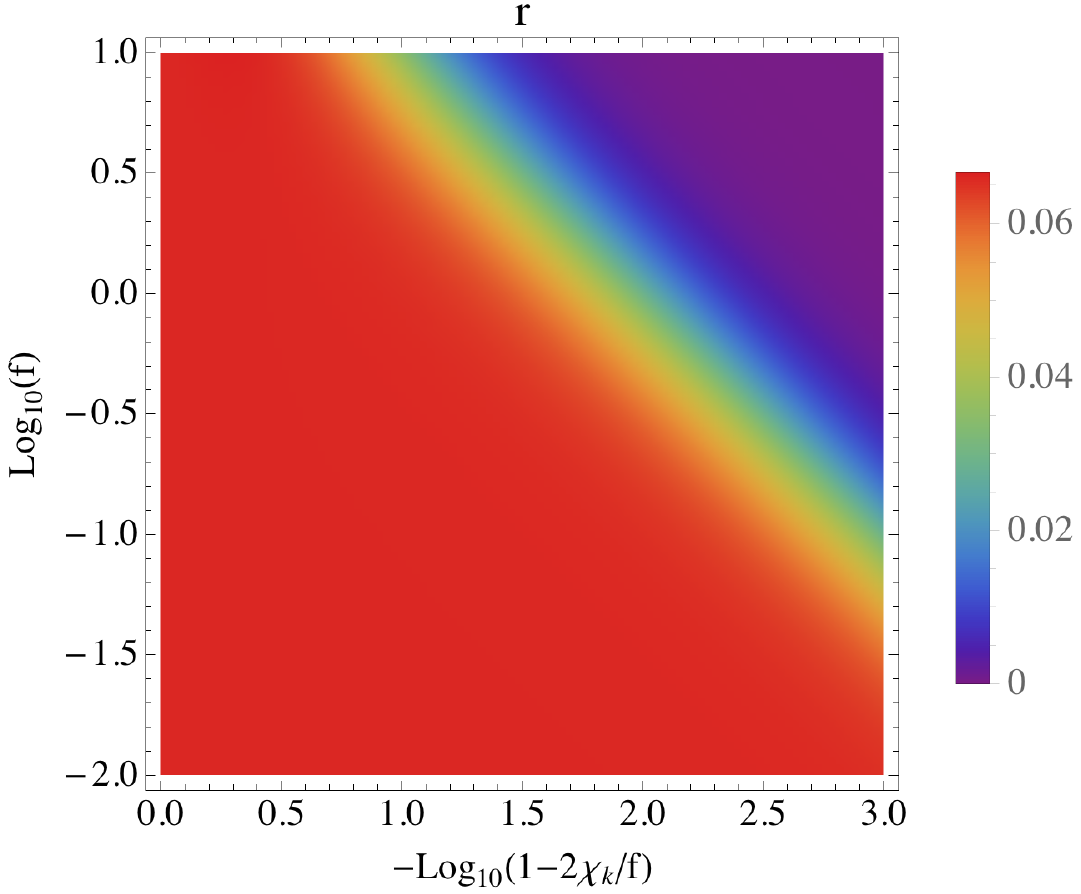}
 \caption{\label{linAx_rcon}}
 \end{subfigure}\\
 \centering
\begin{subfigure}[t]{0.49\textwidth}
 \includegraphics[width = 0.9\columnwidth]{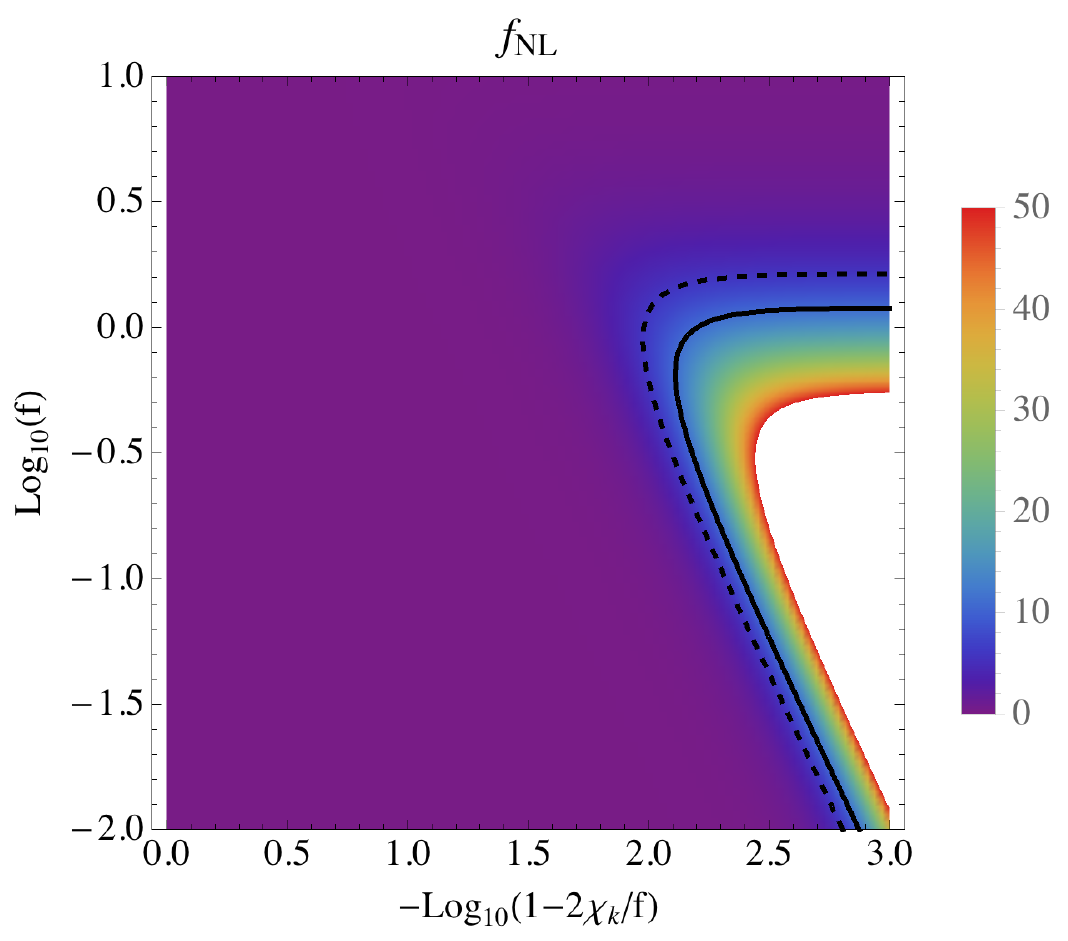}
 \caption{\label{linAx_fNLcon}}
\end{subfigure}
 \caption{\label{linAx_con}Predictions for the linear plus axion potential
 with $\mathcal M = 25/49$.  The decay constant is varied over the range
 $-2\le\log_{10}(f)\le 1$ and the initial value of $\chi$ is varied over
 the range $0\le -\log_{10}(1-2\chi_k/f)\le 3$.  The initial value of
 $\phi$ is determined by requiring a total of $60$ e-foldings of
 inflation, with the end of inflation being defined as when
 $\epsilon^{(0)} = 1$.  Approximation HC1 is used.  (a) Predictions for
 $n_s$.  The solid black line corresponds to the central observed value
 and the dashed line is the $2\sigma$ lower bound.  (b) Predictions for
 $r$.  (c) Predictions for $f_{NL}$.  The dashed and solid black lines
 correspond to the $1$- and $2$-$\sigma$ upper bounds respectively. The
 white region corresponds to where $f_{NL}>50$.}
\end{figure}

In summary, as in the previous example, we have found that slow-roll corrections in this model with a linear plus axion potential can be significant.  This is not surprising given that the slow-roll parameter $\eta_{ss}$ is relatively large and negative for an extended period of $\sim40$ e-foldings after horizon exit.  Interestingly, unlike the previous example, we have found that it is the slow-roll corrections to $f_{NL}$ that are most significant, with slow-roll corrections to the power spectrum being relatively small.

Given that predictions for $n_s-1$ appear to be in agreement with
observations in this model, it is perhaps interesting to take a closer
look at how the predictions depend on the model parameters.  In
Figure~\ref{linAx_con}, taking $\mathcal M = 25/49$ we plot the
predictions for $n_s$, $r$ and $f_{NL}$ as a function of the value of
$\chi$ at horizon crossing, $\chi_k$, and $f$.  All predictions are
calculated using approximation HC1.  We parametrise $\chi_k$ in terms
of how far it deviates from the maximum of the $\chi$-field potential as
$\chi_k = (f/2)(1-10^{-x})$.  We then vary $x$ from $0$ to $3$.  For $f$
we take values in the range $10^{-2}\le f \le10$.  In each case, the
initial value of $\phi$ is determined by requiring that $60$ e-foldings
of inflation are obtained, with the condition for the end of inflation
being $\epsilon^{(0)} = 1$.  This range of parameters has been chosen
such that inflation is essentially driven by the $\phi$ potential and so
that all trajectories converge to the minimum of the $\chi$ potential
before the end of inflation, in order that the final value of $\zeta$
approaches a constant.  In Figure \ref{linAx_nscon} the solid black line
represents the central observed value \mk{$n_s = 0.968$}, with the dashed line
being the lower $2\sigma$ bound \mk{\cite{Ade:2015lrj}}.  As such, we see that just under half
of the parameter space considered is ruled out by observations.  In
Figure \ref{linAx_rcon} we plot the predictions for $r$. At $2\sigma$
$r=0$ is still not ruled out, and the constraint $r<0.1$ does not rule
out any of the parameter space considered here.  Finally, in
Figure~\ref{linAx_fNLcon} we plot the predictions for $f_{NL}$.  Here we
can clearly see that observably large $f_{NL}$ is only obtained when
$\chi$ is tuned to start very close to the top of its potential.  The
black solid line corresponds to the $2\sigma$ upper bound of $f_{NL} =
10.8$ while the dashed line is the $1\sigma$ upper bound of $f_{NL} =
5.8$ \mk{\cite{Ade:2015ava}}.

\begin{figure}
  \centering \includegraphics[width =
  0.6\columnwidth]{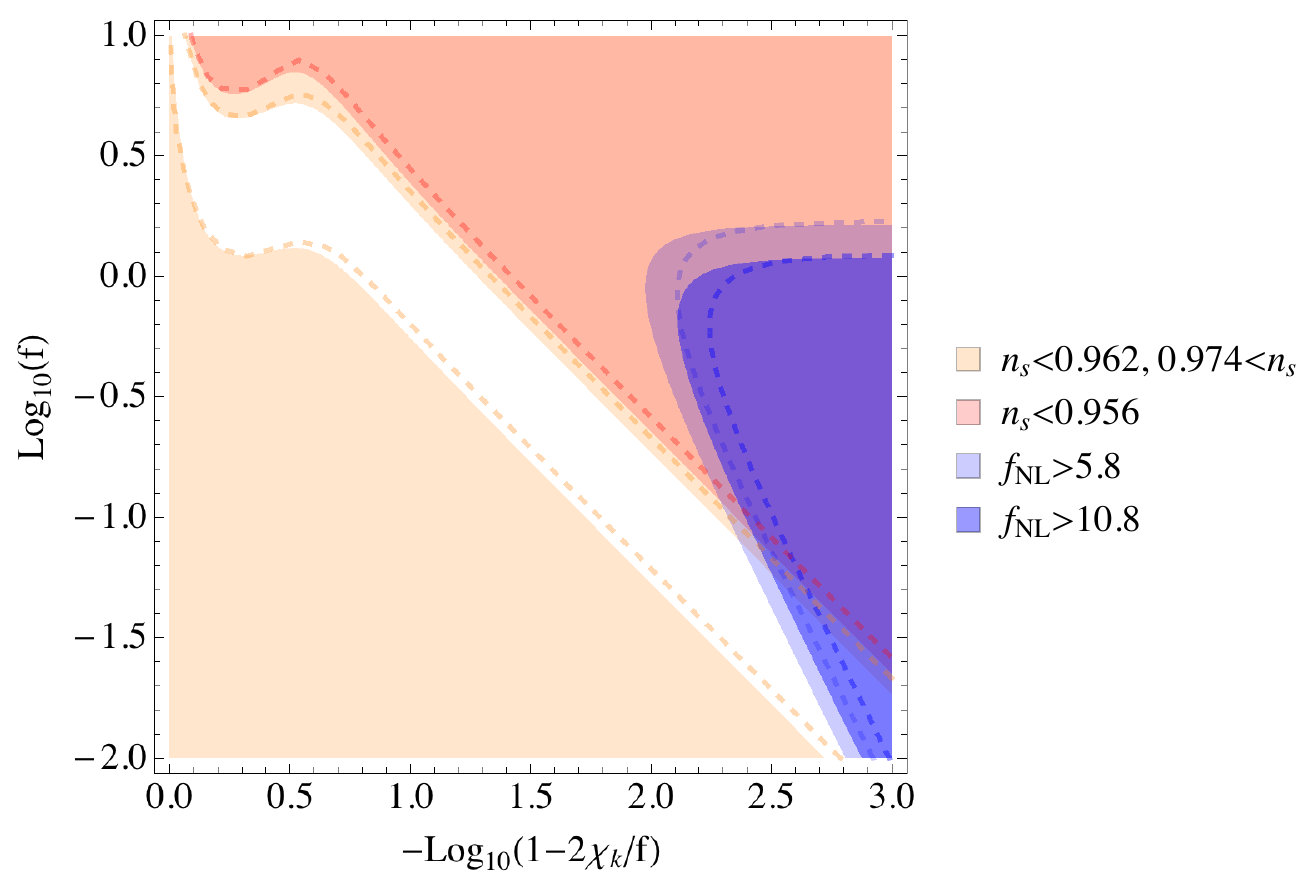}
  \caption{\label{linAx_comb_con}Combined $1$- and $2$-$\sigma$
  constraints in the $\chi_k$--$f$ plane for the linear plus axion model
  with $\mathcal M = 25/49$.  Solid shaded regions correspond to
  excluded regions as determined using approximation HC1, while the
  dashed lines indicate how the excluded regions change if one uses
  approximation SR0.}
\end{figure}

In Figure~\ref{linAx_comb_con} we combine the constraints coming from
$n_s$, $r$ and $f_{NL}$.  The light- and dark-blue regions are the $1$-
and $2$-$\sigma$ excluded regions associated with constraints on
$f_{NL}$.  The orange and red shaded regions are the $1$- and
$2$-$\sigma$ excluded regions associated with constraints on $n_s$.
$2\sigma$ constraints on $r$ do not rule out any of the parameter space.
The dashed curves with matching colours indicate how the excluded
regions change if approximation SR0 is used instead of HC1 to determine
the predictions. If we focus on the $2\sigma$ constraints on $n_s$ and
$f_{NL}$, we see that the allowed region is larger if we use
approximation SR0.

\begin{figure}
  \centering \includegraphics[width = 0.8\columnwidth]{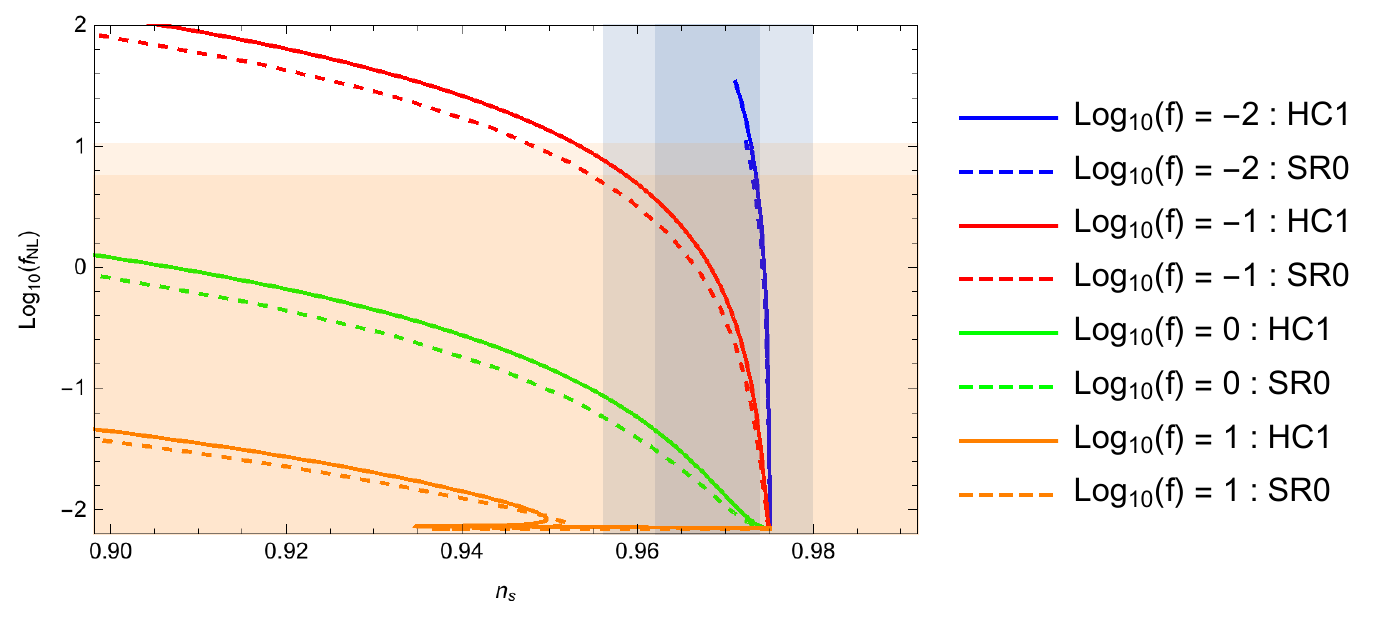}
  \caption{\label{linAx_ns_fNL}Predictions in the $n_s$--$f_{NL}$ plane
  for the linear plus axion model with $\mathcal M = 25/49$.  The
  different coloured lines correspond to different choices of $f$, as
  given in the plot legend.  Solid lines are the predictions as
  determined using approximation HC1, while dashed lines are those
  determined using approximation SR0.  In all cases, the ends of the
  curves located to the bottom right-hand side of the figure correspond
  to $\chi_k = 0$, where predictions coincide with linear chaotic
  inflation.  The initial value of $\chi$ then increases as we move
  left and upwards along the curves. The blue shaded regions correspond
  to the $1$- and $2$-$\sigma$ constraints on $n_s$, while the orange shaded
  regions correspond to the $1$- and $2$-$\sigma$ constraints on $f_{NL}$.}
\end{figure}

In Figure~\ref{linAx_ns_fNL} we plot the predictions for $f_{NL}$ as a
function of the predictions for $n_s$, taking $\log_{10}(f) =
-2,-1,0,1$.  The solid and dashed lines correspond predictions as
calculated using approximations HC1 and SR0, respectively.  Different
points along any given line correspond to different initial conditions
for $\chi$.  For $\chi_k = 0$ the predictions for any $f$ coincide with
those of linear chaotic inflation, and this is the point at which the
lines converge in the bottom right-hand corner of the plot.  Moving left
and upwards along any given line then corresponds to increasing the
initial value of $\chi$.  In the case of $\log_{10}(f)=-2$ we can see
the other end of the line, which corresponds to $-\log_{10}(1-2\chi_k/f)
= 3$.  For observationally relevant values of $n_s$ we see that only the
curves corresponding to $\log_{10}(f)=-2$ and $\log_{10}(f)=-1$ predict
observable values of $f_{NL}$, and these values do differ
depending on whether approximation SR0 or HC1 is used.

\section{Summary and Conclusions}\label{Sec:conclusions}

Motivated by the need for high-precision theoretical predictions to
compare with current and future data, in this paper we have investigated
the importance of slow-roll corrections in determining predictions for
the curvature perturbation $\zeta$ in multi-field models of inflation.
In our analysis we made use of the separate universes approach and
$\delta N$ formalism, which allowed us to consider slow-roll corrections
to the non-Gaussianity of $\zeta$ as well as corrections to its
two-point statistics.  In the context of the $\delta N$ formalism,
slow-roll corrections to the correlation functions of $\zeta$ can
essentially be divided into two categories.  First one has corrections
associated with slow-roll corrections to the correlation functions $\Sigma^{AB}$ of
the field perturbations on the initial flat hypersurface at horizon
crossing.  Second one has slow-roll corrections to the
derivatives of $N$ with respect to the field values on the initial flat slice, $N_{,A}$ and
$N_{,AB}$.

Slow-roll corrections to the correlation functions of the field
perturbations on the initial flat hypersuface can be calculated by
solving the perturbation equations of motion around the horizon-crossing
time, and we have made use of the next-to-leading order results derived
by Nakamura \& Stewart \cite{Nakamura:1996da}.  We find that the
corrections to $\zeta$ arising from these slow-roll corrections can be
written in a compact form, with expressions for $\mathcal P_\zeta$,
$n_s$, $r$ and $f_{NL}$ being given in eqs.~\eqref{powcompact},
\eqref{tiltcompact}, \eqref{rel1} and \eqref{fNLcompact} respectively.
The expressions for corrections to $\mathcal P_\zeta$, $r$ and $n_s$ are
in agreement with those derived using different methods in
\cite{vanTent:2003mn,Byrnes:2006fr}.  To our knowledge, the compact
expression for corrections to $f_{NL}$ has not previously been given
explicitly in the literature.  An appealing feature of the compact
expressions is that the slow-roll corrections associated with
$\Sigma^{AB}$ are written in terms of quantities that in principle are
observable.  In particular, slow-roll corrections to $\mathcal P_\zeta$
and $r$ contain terms proportional to $n_s-1$ and \mk{$n_T$}.  The fact that
$n_s-1$ is now tightly constrained by observations, and that barring an
exact cancellation amongst terms we have \mk{$|n_T|\lesssim |n_s - 1|$},
means that slow-roll corrections to $\mc P_\zeta$ and $r$ for any
observationally viable model are constrained to be small.  Similarly,
slow-roll corrections to $n_s$ and $f_{NL}$ include terms proportional
to $\alpha_s$ and $n_{\delta b}$ respectively. Observational constraints on $\alpha_s$ imply that for observationally viable models these slow-roll corrections to $n_s$ are relatively small.  We are unaware, however, of any similar constraints on $n_{\delta b}$ at the present time, meaning that these slow-roll corrections to $f_{NL}$ are not necessarily constrained to be small, even for observationally viable models.

Slow-roll corrections to the quantities $N_{,A}$ and $N_{,AB}$ arise
from using different levels of slow-roll approximation in solving for
the super-horizon evolution, which in turn corresponds to using
different levels of slow-roll approximation in the background equations
of motion. 
\mk{We have considered
four different levels of approximation,} which we labelled SR0, SR1, HC0
and HC1.  SR0 corresponds to assuming that the lowest order slow-roll
equations of motion -- given in eq.~\eqref{EoM0} -- hold throughout the
super-horizon evolution.  Similarly, SR1 corresponds to assuming that
the next-to-leading order slow-roll equations of motion -- given in
eq.~\eqref{EoM1} -- hold throughout inflation.  Approximations HC0 and
HC1, on the other hand, use the full equations of motion on
super-horizon scales, only assuming that slow-roll is a good
approximation around horizon crossing.  The fact that slow-roll is
assumed to hold around horizon crossing means that only initial field
values need to be specified, with the initial field velocities being
given by the leading and next-to-leading order relations \eqref{EoM0}
and \eqref{EoM1} for HC0 and HC1 respectively.  Note that we were forced
to assume that the slow-roll approximation holds around the horizon
crossing time due to the fact that we made use of the results of
Nakamura \& Stewart for $\Sigma^{AB}$.  Although we expect
approximations HC0 and HC1 to more correctly reproduce the super-horizon
evolution, the advantage of approximations SR0 and SR1 is that fewer
equations have to be solved.  If we wish to expand $\delta N$ up to
second order in the initial field perturbations we find that for
approximations HC0 and HC1 the number of equations we must solve goes as
$M^3$ for large $M$, where $M$ is the number of fields.  For
approximations SR0 and SR1, however, the number is half that, going as
$M^3/2$.  As such, if we are considering models with many fields, SR0
and SR1 will be numerically advantageous provided they give sufficiently
accurate results.

To our knowledge, the next-to-leading order slow-roll equations of
motion used in approximation SR1 have not received much attention in the
literature up to now.  For the examples considered in
Section.~\ref{Sec:examples}, however, we found that approximation SR1 offers
a significant improvement over SR0, often being in very good agreement
with approximation HC1 and the results obtained using {\it PyTransport} \cite{Mulryne:2016mzv}.
In two of the examples considered, we found that corrections to the leading order approximation
SR0 could be sizeable, on the order of tens of percent.  The quadratic
plus axion model gave an example where predictions for $n_s-1$ deviated
significantly from the observed value, which in turn meant that
corrections to $\mc P_\zeta$ coming from slow-roll corrections to
$\Sigma^{AB}$ were large.  Corrections to $f_{NL}$ in this model,
however, were relatively small, being less than $10\%$.  The
linear plus axion model offered a good example where corrections to the
power spectrum were relatively small, at less than $5\%$, but
corrections to the non-Gaussianity were large, being around $95\%$.  In
this case, even corrections to the approximation HC1 as compared to the
results obtained using {\it PyTransport} were found to be on the order of $10\%$, which
suggests that it may be necessary to go to next-to-next-to-leading
order in the slow-roll expansion at horizon crossing to obtain
predictions of the desired precision.
  
Finally, in the model with a double quadratic potential we found that all slow-roll corrections remained small.  This was perhaps somewhat surprising given that intermediately some of the slow-roll parameters became quite large.  We take this to indicate that the slow-roll parameters were not large enough for long enough to give rise to significant corrections.  Indeed, in the other two models considered one of the slow-roll parameters was relatively large for an extended period after the horizon crossing time.

In terms of future directions, it would be desirable to determine
compact expressions for slow-roll corrections to the quantities $N_{,A}$
and $N_{,AB}$, in order that the conditions under which they become
large could be better understood.  
This would help us to better understand why slow-roll corrections remained small in the model with a double quadratic potential, despite the slow-roll parameters intermediately becoming large.
Also, for models that give rise to slow-roll corrections on the order of tens of percent, it will
likely be necessary to consider next-to-next-to-leading order
corrections in order to obtain predictions of the desired precision. 
The correlation functions of the initial field perturbations up to this
order in slow-roll have been determined using a Green's function method
in \cite{Gong:2002cx}, so these results could be used.  It would then be
necessary to determine the background field equations at
next-to-next-to-leading order in slow roll.
Finally, when using the lowest-order background equations of motion it is known that for some classes of potentials the derivatives of $N$ can be determined analytically.  It would be interesting to explore whether analytical results can also be obtained when using the next-to-leading order background equations.

\section*{Acknowledgements}

The work of M.K. for this project was supported by the Academy of Finland project 278722 and by JSPS as an International Research Fellow of the Japan Society for the Promotion of Science. M.K. would also like to thank the Theory Center of KEK in Tsukuba for hospitality during the JSPS fellowship. The work of K.K. was partially supported by JSPS KAKENHI Grant Nos.~26247042, JP17H01131, and MEXT KAKENHI Grant Nos.~JP15H05889, and JP16H0877. J.W. would like to thank David Mulryne for many useful discussions and for help with using the code PyTransport introduced in Ref.~\cite{Mulryne:2016mzv}.  He would also like to thank the Cosmology Group at the University of Jyvaskyla and the Astronomy Unit at Queen Mary University of London for their kind hospitality during visits made while this work was in progress.

\appendix
\section{Background equations of motion at next-to-leading order in slow-roll\label{sec:SR-aprx}}

In order to expand the equations of motion in eq.~\eqref{EoMN} up to
higher orders in the slow-roll approximation we first write them
as\footnote{In this equation we assumed that $\phi''$ and $\phi'^{2}$
terms vanish equally fast as $\xi\rightarrow0$. This can be checked to be
the case by applying the dominant balance condition (see
e.g. ref.~\cite{zwillinger1997handbook}).}
\begin{equation}
\phi_{I}'=U_{,I}-\frac{\xi\phi_{I}''}{3-\frac{\xi}{2}\phi_{I}'\phi_{I}'},\label{Aeom}
\end{equation}
where $U$ is defined as
\begin{equation}
U\equiv-\ln V\label{U-def}
\end{equation}
In the above expression we also used eq.~\eqref{eps-eta-dphidN} and
introduced a parameter $\xi$ (not to be confused with the parameter $\xi
= 1/(LH)$ used in Section~\ref{sec:sepuniv}) to help us organise small
terms in the computations that follow. Initially we take $\xi\ll1$ and
expand eq.~\eqref{Aeom} and fields $\phi_{I}$ in terms of $\xi$. The
latter expansion can be written as
\begin{equation}
\phi_{I}=\phi_{I}^{\idx 0}+\xi\phi_{I}^{\left(1\right)}+\mathcal{O}\left(\xi^{2}\right).\label{Aphi-exp}
\end{equation}
At the end of the computations we take $\xi=1$.

Plugging eq.~\eqref{Aphi-exp} into eq.~\eqref{Aeom} and expanding in
terms of $\xi$ we find
\begin{equation}
{\phi_{I}^{\idx 0}}'+\xi{\phi_{I}^{\idx 1}}'+\mathcal{O}\left(\xi^{2}\right)=U_{,I}^{\idx 0}+\xi\left(U_{,IJ}^{\idx 0}\phi_{J}^{\idx 1}-\frac{1}{3}{\phi_{I}^{\idx 0}}''\right)+\mathcal{O}\left(\xi^{2}\right),\label{AEoM-prtb}
\end{equation}
where $U_{,I}^{\idx 0}$ and $U_{,IJ}^{\idx 0}$ are functions of
fields $\phi_{I}^{\idx 0}$, for example
\begin{equation}
U_{,I}^{\idx 0}\equiv\frac{\partial U\left(\vec{\phi}^{\idx 0}\right)}{\partial\phi_{I}^{\idx 0}}
\end{equation}
and similarly
\begin{equation}
U_{,IJ}^{\idx 0}\equiv\frac{\partial^{2}U\left(\vec{\phi}^{\idx 0}\right)}{\partial\phi_{I}^{\idx 0}\partial\phi_{J}^{\idx 0}}.
\end{equation}

At the zeroth order in $\xi$ eq.~\eqref{AEoM-prtb} becomes
\begin{equation}
{\phi_{I}^{\idx 0}}'\simeq U_{,I}^{\idx 0}.
\end{equation}
Taking the time derivative of this expression and plugging back into
eq.~\eqref{AEoM-prtb} we get
\begin{equation}
{\phi_{I}^{\idx 0}}'+\xi{\phi_{I}^{\idx 1}}'\simeq U_{,I}^{\idx 0}+\xi\left[U_{,IJ}^{\idx 0}\phi_{J}^{\idx 1}-\frac{1}{3}U_{,IJ}^{\idx 0}U_{,J}^{\idx 0}\right],\label{AEoM-prtb2}
\end{equation}
where we kept only terms up to the first order in $\xi$.

In the last expression derivatives of $U^{\idx 0}$ are functions
of the fields $\phi_{I}^{\idx 0}$, but for our purpose the above equation
is more useful if expressed as a closed system of equations for the fields
\begin{equation}
\phi_{1I}\equiv\phi_{I}^{\idx 0}+\xi\phi_{I}^{\idx 1}.
\end{equation}
To do that, we can compute
\begin{equation}
U_{,I}\left(\vec{\phi}_{1}\right)=U_{,I}^{\idx 0}+\xi U_{,IJ}^{\idx 0}\phi_{J}^{\idx 1}+\mathcal{O}\left(\xi^{2}\right),\label{AUd1}
\end{equation}
where we defined
\begin{equation}
U_{,I}\left(\vec{\phi}_{1}\right)\equiv\frac{\partial U\left(\vec{\phi}_{1}\right)}{\partial\phi_{1I}}.\label{AUd1-def}
\end{equation}
Similarly the second derivative of $U$ for fields $\phi_{1I}$ can
be easily obtained
\begin{equation}
U_{,IJ}\left(\vec{\phi}_{1}\right)=U_{,IJ}^{\idx 0}+\xi U_{,IJK}^{\idx 0}\phi_{K}^{\idx 1}+\mathcal{O}\left(\xi^{2}\right),\label{AUdd1}
\end{equation}
where the definition of $U_{,IJ}\left(\vec{\phi}_{1}\right)$ is analogous
to the one of $U_{,I}\left(\vec{\phi}_{1}\right)$ in eq.~\eqref{AUd1-def}.

Finding next the expression for $\xi U_{,IJ}^{\idx 0}\phi_{J}^{\idx
1}$from eq.~\eqref{AUd1} and $U_{,IJ}^{\idx 0}$ from eq.~\eqref{AUdd1}
and plugging the result into eq.~\eqref{AEoM-prtb2} we compute
\begin{equation}
\phi_{1I}'\simeq U_{,I}\left(\vec{\phi}_{1}\right)-\frac{\xi}{3}U_{,IJ}\left(\vec{\phi}_{1}\right)U_{,J}\left(\vec{\phi}_{1}\right),
\end{equation}
where the result is given up to terms proportional to $\xi^{2}$.
Taking $\xi=1$, it agrees with the expression given in
ref.~\cite{Nakamura:1996da}.  The final result is obtained using the
definition of $U$ in eq.~\eqref{U-def}.  It is given by
\begin{equation}
\phi_{1I}'\simeq-\frac{V_{,I}}{V}-\frac{1}{3}\left(\frac{V_{,IJ}}{V}-\frac{V,_{I}V_{,J}}{V^{2}}\right)\frac{V_{,J}}{V},\label{AEoM1st}
\end{equation}
where $V$ is a function of the fields $\phi_{1I}$.

\section{$N_\ast$-independence of $\zeta$ and its correlation functions}\label{Sec:Nastind}

\mk{}

As discussed in Section~\ref{sec:sepuniv}, in the context of the
\mk{separate universes} approach and the $\delta N$ formalism, $\zeta$ can be
expressed in the following form
\mk{
\begin{align}\label{dNexp}
 \zeta(N,\vec x) = \delta N (N,\vec x) =
 N_{,i}(N,N_\ast)\delta\varphi^{i}(N_\ast,\vec x)
 +\frac{1}{2}N_{,ij}(N,N_\ast)\delta\varphi^{i}(N_\ast,\vec
 x)\delta\varphi^{j}(N_\ast,\vec x) + ...\, ,
\end{align}
}
where here we have given the arguments of the various elements
explicitly in order to aid our discussion.  The right-hand side of this
expression appears to depend on $N_\ast$, but in fact we know that we
are free to take $N_\ast$ to be any time after the relevant scales have
left the horizon.  The reason is that the number of e-foldings
between any two flat hypersurfaces is equal to the unperturbed number of
e-foldings, such that evolution between the two surfaces makes no
contribution to $\delta N$.  Here let us show explicitly that the
right-hand side of \ref{dNexp} is independent of $N_\ast$.  First, for
the derivatives of \mk{$N_{,i}$ and $N_{,ij}$} we have
\mk{
\begin{align}\label{Nderivs}
 \frac{d}{dN_\ast}N_{,i} &= \varphi'_{j} N_{,ij} =
 \frac{\partial}{\partial\varphi^{i}}\left(\varphi'_{j}N_{,j}\right) -
 \frac{\partial}{\partial\varphi^{i}}\left(\varphi'_{j}\right)N_{,j} =
 -\mathbb U^j{}_{,i}N_{j},\\ \frac{d}{dN_\ast}N_{,ij} &= \varphi'_{k}
 N_{,ijk} =
 \frac{\partial}{\partial\varphi^{j}}\left(\varphi'_{k}N_{,ik}\right) -
 \frac{\partial}{\partial\varphi^{j}}\left(\varphi'_{k}\right)N_{,ik} \\
 &=
 \frac{\partial}{\partial\varphi^{j}}\left(\frac{\partial}{\partial\varphi^{i}}\left(\varphi'_{k}N_{,k}\right)
 -
 \frac{\partial}{\partial\varphi^{i}}\left(\varphi'_{k}\right)N_{,k}\right)
 - \mathbb U^{k}{}_{,j} N_{,ik} \\ &= -\mathbb U^{k}{}_{,ij}N_{,k} -
 \mathbb U^{k}{}_{,i}N_{,kj} - \mathbb U^{k}{}_{,j} N_{,ik},
\end{align}
}
where we have used \mk{$\varphi'_{i}N_{i} = -1$ and $\varphi'_{i} = \mathbb
U^{i}$}.  Next, for the derivatives of \mk{$\delta\varphi^i$}, namely
\mk{$d\delta\varphi^i(N_\ast,\bm x)/dN_\ast$}, we use eq.~\eqref{ddvphidN}. 
Putting everything together we thus have
\mk{
\begin{align}
 \frac{d}{dN_\ast}\delta N(N,\vec x) &= -\mathbb
 U^{j}{}_{,i}N_{,j}\delta\varphi^{i} + N_{,i}\left(\mathbb
 U^{i}{}_{,j}\delta\varphi^{j} + \frac{1}{2}\mathbb
 U^{i}{}_{,jk}\delta\varphi^{j}\delta\varphi^{k}\right)\\&\qquad +
 \frac{1}{2}\left(-\mathbb U^{k}{}_{,ij}N_{,k} - 2\mathbb
 U^{k}{}_{,i}N_{,jk}\right)\delta\varphi^{i}\delta\varphi^{j}+ N_{,ij}
 \mathbb U^{i}{}_{,k}\delta\varphi^{k}\delta\varphi^{j}\\ &=0,
\end{align}
}
as expected.

Given that $\zeta(N,\vec x)$ is independent of $N_\ast$, it follows that
any correlation functions of $\zeta$ should also be independent of
$N_\ast$.  In relation to the bispectrum, however, there is perhaps one
subtlety worth pointing out.  While the full bispectrum $B_\zeta(k_1,
k_2, k_3)$ is of course independent of $N_\ast$, the so-called ``intrinsic'' and
``super-horizon'' contributions (see below) individually are not.  As such, given that
the intrinsic contribution is often neglected, strictly speaking the
remaining approximation for the bispectrum, namely the super-horizon
contribution, is not independent of $N_\ast$.  Practically speaking,
however, as long as we do not take $N_\ast$ to be too long after
horizon-crossing, then it should remain true that the super-horizon
contribution dominates over the intrinsic one for any model in which
non-Gaussianity is observably large.  As such, we can effectively
consider the super-horizon contribution to be $N_\ast$-independent.

For completeness, let us show the above more explicitly.  In proceeding
we move to Fourier space, where correlation functions are most
commonly considered.  Making use of \eqref{ddvphidN} we determine that the
equations of motion for the perturbations in Fourier space take the form
\mk{
\begin{align}\label{Fperteom}
 \frac{d}{dN}\delta\varphi^{i}(N,\vec k) = \mathbb U^{i}{}_{,j}\delta\varphi^{j}(N,\vec k) + \frac{1}{2}\mathbb U^{i}{}_{,jl}\int\frac{d^3\vec q}{(2\pi)^3}\delta\varphi^{j}(N,\vec k - \vec q)\delta\varphi^{l}(N,\vec q),
\end{align}
}
which can in turn be used to determine the evolution equations for \mk{$\Sigma^{ij}$ and $B^{ijk}$} as
\mk{
\begin{align}\label{Sevol}
 \frac{d}{dN}\Sigma^{ij}(N,k) &= \mathbb U^{i}{}_{,k}\Sigma^{jk}(N,k)+\mathbb U^{j}{}_{,k}\Sigma^{ik}(N,k) + \mathcal O(\delta\varphi^3),\\\nonumber
 \frac{d}{dN}B^{ijk}(N,k_1,k_2,k_3) &= \mathbb U^{i}{}_{,l}B^{jkl}(N,k_1,k_2,k_3)+\mathbb U^{j}{}_{,l}B^{ikl}(N,k_1,k_2,k_3)+\mathbb U^{k}{}_{,l}B^{ijl}(N,k_1,k_2,k_3)\\\nonumber
 &\quad + \mathbb U^{i}{}_{,lm}\Sigma^{jl}(N,k_2)\Sigma^{km}(N,k_3)+ \mathbb U^{j}{}_{,lm}\Sigma^{il}(N,k_1)\Sigma^{km}(N,k_3) \\ &\quad+ \mathbb U^{k}{}_{,lm}\Sigma^{il}(N,k_1)\Sigma^{jm}(N,k_2) +\mathcal O(\delta\varphi^5)\label{Bevol},
\end{align}
}
where we have assumed all $\vec k_i$ to be non-zero, \mk{the
intrinsic 4-point correlation function to be negligible} and made use of the
fact that \mk{$\mathbb U^{i}{}_{,jk}$} is symmetric in its lower two indices.

The first of the above relations, combined with the first relation in
\eqref{Nderivs}, can be used to show that \mk{$P_\zeta(N,k) =
N_{,i}(N,N_\ast)N_{,j}(N,N_\ast)\Sigma^{ij}(N_\ast,k)$} is indeed
independent of $N_\ast$.  Turning to the bispectrum, we decompose it
into \mk{``intrinsic'' and ``super-horizon''} contributions as
\mk{
\begin{align}
 B_\zeta(N,k_1,k_2,k_3) &= B^{\rm int}_\zeta(N,N_\ast,k_1,k_2,k_3) + B^{{\rm sh}}_{\zeta}(N,N_\ast,k_1,k_2,k_3),\\
 B^{\rm int}_\zeta(N,N_\ast,k_1,k_2,k_3) &= N_{,i}(N,N_\ast)N_{,j}(N,N_\ast)N_{,k}(N,N_\ast)B^{ijk}(N_\ast,k_1,k_2,k_3),\\
\nonumber B^{\rm sh}_\zeta(N,N_\ast,k_1,k_2,k_3) &= N_{,ik}(N,N_\ast)N_{,j}(N,N_\ast)N_{,l}(N,N_\ast)\times\\
&\qquad\left(\Sigma^{ij}(N_\ast,k_1)\Sigma^{kl}(N_\ast,k_2) + {\rm 2\, perms.}\right).
\end{align}
}
This labelling makes sense if we take $N_\ast$ to be shortly after
horizon-crossing, as the first term then represents the contribution
arising due to the intrinsic non-gaussianity of the field perturbations
shortly after horizon-crossing -- which is often taken to be negligible
-- while the second term is the contribution resulting from the
non-linear dynamics on super-horizon scales.  However, the split is not
independent of $N_\ast$.  Indeed, if we imagine that $B^{\rm int}_\zeta$
does vanish at some initial time $N_{\rm i}$, namely \mk{$B^{ijk}(N_{\rm i},
k_1,k_2,k_3) = 0$}, due to the non-linear evolution of the
$\delta\varphi^{a}$, giving rise to the terms on the second and third
lines of \eqref{Bevol}, \mk{$B^{ijk}$} and hence $B^{\rm int}_\zeta$ will not
remain zero at some later time $N_{\rm f}$.  Explicitly, we find that
\mk{
\begin{align}\nonumber
 \frac{d}{dN_\ast} B_\zeta^{\rm int}(N,N_\ast,k_1,k_2,k_3) &= N_{,i}(N,N_\ast)N_{,j}(N,N_\ast)N_{,k}(N,N_\ast)\mathbb U^{k}{}_{,lm}\Sigma^{li}(N_\ast,k_1)\Sigma^{mj}(N_\ast,k_2)\\ &\quad + {\rm c.p.}\\&=-
 \frac{d}{dN_\ast} B_\zeta^{\rm sh}(N,N_\ast,k_1,k_2,k_3).
\end{align}
}

While \eqref{Sevol} and \eqref{Bevol} give the relevant evolution
equations for the field correlation functions on super-horizon scales,
we still need to determine the initial conditions by evolving field
perturbations from far inside the horizon up until just after horizon
exit.  In our case we have used the results of Nakamura \& Stewart \cite{Nakamura:1996da}, as
given in eq.~\eqref{sigma}.  Here let us confirm that \mk{$\Sigma^{AB}$} as
given in eq.~\eqref{sigma} does indeed satisfy the super-horizon equations
of motion \eqref{Sevol}, which will in turn ensure the
$N_\ast$-independence of $P_\zeta$.\footnote{“Note that in \cite{Nakamura:1996da} and in our paper we have made the assumption that the slow-roll approximation holds around horizon crossing, which means that we only ever consider the correlation functions of the field perturbations $\Sigma^{AB}$ and $B^{ABC}$.  However, the preceding analysis in this appendix is more general, with $\Sigma^{ij}$ and $B^{ijk}$ representing the correlation functions of both field and field-velocity perturbations.” }  Recall that in deriving
\eqref{sigma} $\epsilon$ and \mk{$\epsilon^{AB}$} were assumed to be
constant, such that
\begin{align}
 \frac{d}{dN_\ast}\Sigma^{AB}(N_\ast,k) \simeq 2 U^{(0)}_{,AB}\frac{H^2(N_\ast)}{2k^3} \simeq 2 U^{(0)}_{,AC} \Sigma^{CB}(N_\ast,k). 
\end{align}
This is in agreement with \eqref{Sevol} once we take into account that
at leading order in slow-roll $\mathbb U\rightarrow U^{(0)}$.  

\bibliographystyle{JHEP}
\bibliography{squeezed}

\end{document}